\documentclass{JHEP3}

\usepackage{graphicx}
\usepackage{subfigure}
\usepackage{dcolumn}
\usepackage{bm,bbm,epsfig,multicol}
\usepackage{amssymb}
\usepackage{amsmath}

\title{Lepton Flavor Violation in Models with $A_4$ and $S_4$ Flavor Symmetries}
\author{Gui-Jun Ding\thanks{dinggj@ustc.edu.cn}, Jia-Feng Liu
\\
Department of Modern Physics, University of Science and Technology
of China, Hefei, Anhui 230026, China}

\abstract{The lepton flavor violation $\mu\rightarrow e\gamma$,
$\tau\rightarrow e\gamma$, $\tau\rightarrow\mu\gamma$,
$\mu\rightarrow3e$, $\tau\rightarrow3e$, $\tau\rightarrow3\mu$ and
$\mu-e$ conversion in Al and Ti are studied in both the
Altarelli-Feruglio $A_4$ model and the $S_4$ model of Ding. The
rates of these lepton flavor violation process for the inverted
hierarchy neutrino mass spectrum are enhanced considerably by the
next to leading order contributions. For both models, we find that
the $\mu-e$ conversion in Ti is within the precision of next
generation experiments in all the parameter space considered, the
$\mu-e$ conversion in Al should be observable at least in a very
significant part of the parameter space, whereas $\tau\rightarrow
e\gamma$, $\tau\rightarrow\mu\gamma$, $\tau\rightarrow3e$ and
$\tau\rightarrow3\mu$ are below the expected future sensitivity. The
detectability of $\mu\rightarrow e\gamma$ and $\mu\rightarrow3e$ in
near future depends on the models and the neutrino mass spectrum. We
suggest that a comprehensive consideration of the lepton flavor
violation processes is important to test and distinguish both
discrete flavor symmetry models. }


\begin{document}

\section{Introduction\label{sec:intr}}

In the past decades, one of the most striking developments in
particle physics beyond the standard model (SM) is the experimental
establishment of neutrino masses and the large mixing property,
which is quite different from the small mixing in the quark sector.
A global fit to the current neutrino oscillation data from the
solar, atmospheric, reactor and long baseline neutrino experiments
gives the following $3\sigma$ interval for the mixing parameters
\cite{Schwetz:2008er,GonzalezGarcia:2007ib}
\begin{eqnarray}
\nonumber&&\sin^2\theta_{12}=0.304^{+0.066}_{-0.054},~~\sin^2\theta_{23}=0.50^{+0.17}_{-0.13},~~\sin^2\theta_{13}<0.056\\
\label{1}&&{\rm \Delta
m^2_{21}=7.65^{+0.69}_{-0.50}\times10^{-5}\,eV^2,~~\Delta
m^2_{31}=\pm2.40^{+0.35}_{-0.33}\times10^{-3}\,eV^2}
\end{eqnarray}
For the overall scale of the neutrino and the mixing angle
$\theta_{13}$ currently only upper limit exists. But considerable
progress is expected from future double beta decay
\cite{Avignone:2007fu} and reactor neutrino oscillation experiments
\cite{Ardellier:2006mn,Wang:2006ca}. It is remarkable that such a
mixing pattern is very close to the so-called tri-bimaximal (TB)
mixing \cite{TBmix}, which leads to $\sin^2\theta_{12}=1/3$,
$\sin^2\theta_{23}=1/2$ and $\sin^2\theta_{13}=0$. Thus it is a
important subject to realize the TB mixing from the theoretical
viewpoint.

Recently it is found that the TB mixing appears naturally in models
with discrete flavor symmetry such as $A_4$, $T'$ and $S_4$, and
many models based on these flavor symmetries have been proposed
\cite{
A4,Altarelli:2005yp,Altarelli:2005yx,Altarelli:2009kr,Tprime,Ding:2009iy,S4,S4_previous}.
We note that the $T'$ model usually has the same structure as that
of the $A_4$ model in the lepton sector. In these models, the flavor
symmetry is realized at a higher energy scale, the lepton fields
transform nontrivially under the corresponding flavor group, and the
flavor symmetry is spontaneously broken by a set of flavons along
appropriate directions to provide a realistic description of the
lepton masses and mixing angles. To be concrete, in this work we
will concentrate on the  Altarelli-Feruglio $A_4$ model
\cite{Altarelli:2005yx} and the $S_4$ model of Ding
\cite{Ding:2009iy}, which represent a typical class of flavor
models. The structure of the model is tightly constrained by the
flavor symmetry, as a result only few parameters are involved at
leading order (LO), and the models are rather predictive. The common
features of both models are as follows: they are supersymmetry
(SUSY) models and the neutrino masses are generated via the type-I
Seesaw mechanism. At LO the charged lepton mass matrix is diagonal,
the light neutrino mass matrix is controlled only by two complex
parameters, and it is diagonalized by the TB mixing matrix exactly.
Despite of the tight constraints on the model parameters, and in
particular the fact that there is only one real parameter left after
taking into account $\Delta m^2_{21}$ and $\Delta m^2_{31}$ measured
from the neutrino oscillation experiments, it has been shown that
the $A_4$ models can reproduce the observed matter-anti matter
asymmetry for natural values of the parameters in a quite
satisfactory way
\cite{Jenkins:2008rb,Bertuzzo:2009im,Hagedorn:2009jy,AristizabalSierra:2009ex,Felipe:2009rr}.

Since this kind of $A_4$ and $S_4$ models are so attractive and
predictive, it is highly desirable to verify or exclude these model
experimentally. There is no double that the precise measurement of
$\theta_{13}$ is a crucial test of the models. One of the
implications of the observation of neutrino oscillation is the
possibility of measurable branching ratio for charged lepton flavor
violating (LFV) decays. While the LFV processes are still highly
suppressed in the SM, predictions of the SUSY Seesaw for these rare
decays are much more enhanced, as these processes are suppressed by
the SUSY scale rather than the Planck scale. Furthermore, as
different models obtain large neutrino mixing angles through
different mechanisms, their predictions for the LFV decays can be
very distinct. In the past years, LFV processes have been explored
extensively from experiments, the upper bound of the branch ratio
would be improved considerably in future, which could strongly
constrain the new physics beyond the SM. Consequently, LFV rare
decays may provide a way to test the $A_4$ and $S_4$ flavor
symmetries.

In this work, we shall investigate the predictions for the various
LFV processes in the Altarelli-Feruglio model and the Ding's $S_4$
model, assuming the so-called minimal supergravity (mSUGRA) boundary
conditions. Although our analysis is performed for the two specific
models, it has generic features which are universal to the models
based on the discrete flavor symmetry. We find that these models are
so predictive that the branching ratios of the LFV processes are
closely related to light neutrino mass, and there are some
characteristic relations between various LFV branching ratios.
Present work is different from the previous studies of the LFV decay
branching ratio within the SUSY Seesaw framework. Since the low
energy data allow to reconstruct only partially the high energy
couplings of the theory, they generally inversed the Seesaw formula
following the procedure of Casas and Ibarra \cite{Casas:2001sr},
then they carried out the Monte Carlo studies by scanning the
unknown right handed neutrino mass spectrum and the angles and
phases of the inversion matrix in order to present scatter plots of
the rare branching ratios
\cite{Arganda:2005ji,Ibarra:2008uv,Hirsch:2008dy}. Another method is
the top-down approach, where the neutrino Yukawa coupling matrix and
the right handed neutrino mass matrix is determined by a specific
SUSY grand unified theory \cite{GUT_LFV}.

The paper is organized as follows: in the second section we briefly
review the LFV within the framework of the SUSY Seesaw. Then we give
a concise introduction to the Altarelli-Feruglio model and the
Ding's $S_4$ model in section 3. Our analysis of LFV decay branching
ratios for these two interesting models is presented in section 4
and section 5 respectively. Finally we summarize our results and
draw the conclusions.

\section{LFV in SUSY Seesaw}

SUSY is a well motivated possibility for new physics
\cite{Drees:2004jm,Martin:1997ns}, and the supersymmetric Seesaw
mechanism can accommodate simultaneously tiny neutrino masses and
large hierarchy between the electroweak scale and the high Seesaw
scale without serious fine-tunings. In this framework the particle
content of the minimal supersymmetric standard model (MSSM) is
extended by including three gauge singlet right handed neutrino
superfields $\nu^{c}_i$ ($i$=1, 2, 3). Imposing $R-$ parity
conservation, the leptonic part of the superpotential is thus given
by
\begin{equation}
\label{1}W_{lep}=e^{c}_i{\mathbf Y}_{eij}L_jH_d+\nu^{c}_i{\mathbf
Y}_{\nu ij}L_jH_u+\frac{1}{2}\;\nu^{c}_i{\mathbf M}_{\nu
ij}\nu^{c}_{j}
\end{equation}
where $L_i$ denotes the left handed lepton doublet, $e^{c}_i$ is the
right handed charged lepton, $H_u$ and $H_d$ are the hypercharge
$+1/2$ and $-1/2$ Higgs doublets respectively. ${\mathbf Y}_{e}$ and
${\mathbf Y}_{\nu}$ are the charged lepton and neutrino Yukawa
coupling matrices respectively, and ${\mathbf M}_{\nu}$ represents
the $3\times3$ majorana mass matrix for the heavy right handed
neutrinos. In general, the overall scale of $M_{\nu}$, which we
shall denote by $M_{maj}$, is much larger than the electroweak
scale. At energies below $M_{maj}$, the theory will be well
described by the following effective superpotential
\begin{equation}
\label{2}W^{eff}_{lep}=e^{c}_i{\mathbf
Y}_{eij}L_jH_d+\frac{1}{2}{\mathcal M}_{\nu ij}(L_iH_u)(L_jH_u)
\end{equation}
where the light neutrino effective mass matrix ${\mathcal M}_{\nu}$
is given by the well-known Seesaw formula
\begin{equation}
\label{3}{\mathcal M}_{\nu}=-{\mathbf Y}^{T}_{\nu}{\mathbf
M}^{-1}_{\nu}{\mathbf Y}_{\nu}\,v^2_u
\end{equation}
where $v_u$ is the vacuum expectation value of the neutral component
of the Higgs doublet $H_u$. In nature SUSY must be a broken
symmetry, we thus introduce the soft SUSY breaking (SSB) Lagrangian
${\mathcal L}_{soft}$, which could be new sources of flavor
violation. The leptonic part of ${\mathcal L}_{soft}$ has the
following form,
\begin{eqnarray}
\nonumber-{\mathcal L}_{soft}=&&({\mathbf
m}^2_{\widetilde{L}})_{ij}\widetilde{L}^{\dagger}_i\widetilde{L}_j+({\mathbf
m}^2_{\widetilde{e}})_{ij}\widetilde{e}^{*}_i\widetilde{e}_{j}+({\mathbf
m}^2_{\widetilde{\nu}})_{ij}\widetilde{\nu}^{*}_i\widetilde{\nu}_j+\Big[({\mathbf
A}_e)_{ij}\widetilde{e}^{*}_i\widetilde{L}_jH_d\\
\label{4}&&+({\mathbf
A}_{\nu})_{ij}\widetilde{\nu}^{*}_i\widetilde{L}_jH_u+h.c.\Big]
\end{eqnarray}
where $\widetilde{L}_i$, $\widetilde{e}_i$ and $\widetilde{\nu}_i$
are the supersymmetric partners of the left handed lepton doublet,
right handed charged lepton and right handed neutrinos respectively.
${\mathbf m}^2_{\widetilde{L}}$, ${\mathbf m}^2_{\widetilde{e}}$ and
${\mathbf m}^2_{\widetilde{\nu}}$ are the corresponding soft mass
matrices squared, ${\mathbf A}_{e}$ and ${\mathbf A}_{\nu}$ are the
charged lepton and neutrino soft trilinear couplings respectively.
In general, above SSB terms can have arbitrary flavor structures
which would induce unacceptably large flavor violating effects. The
simplest solution to this SUSY flavor problem is to assume a
flavor-blind SUSY breaking mediation mechanism, which will generate
flavor universal SSB terms at some high scale. In the present work,
we will restrict ourselves to the so-called minimal supergravity
scenario (mSUGRA). It assumes that, at the GUT scale, the slepton
mass matrices are diagonal and universal in flavor, and that the
trilinear couplings are proportional to the Yukawa couplings
\begin{eqnarray}
\nonumber&& ({\mathbf m}^2_{\widetilde{L}})_{ij}=({\mathbf m}^2_{\widetilde{e}})_{ij}=({\mathbf m}^2_{\widetilde{\nu}})_{ij}=m^2_0\delta_{ij}\;,~~~m^2_{H_u}=m^2_{H_d}=m^2_0\\
\label{5}&&({\mathbf A}_{e})_{ij}=A_0{\mathbf Y}_{eij}\;,~~~({\mathbf A}_{\nu})_{ij}=A_0{\mathbf Y}_{\nu ij}
\end{eqnarray}
where $m_0$ is the common scalar mass, and $A_0$ is the common
trilinear parameter. In addition, there are still three parameters
characterizing the mSUGRA: the common gaugino mass $M_{1/2}$,
$\tan\beta\equiv v_u/v_d$ and the sign of the Higgs mixing
parameters $\mu$, where $v_d$ is the vacuum expectation value of the
neutral component of the Higgs doublet $H_d$. However, these SSB
terms will not be universal at the weak scale. Since the Yukawa
coupling matrices ${\mathbf Y}_e$ and ${\mathbf Y}_{\nu}$ can not be
simultaneously diagonalized, non-vanishing off-diagonal elements
would be generated in the SSB mass-squared matrices and the
trilinear couplings due to renormalization effect. The one-loop
renormalization group equations for ${\mathbf m}^2_{\widetilde{L}}$,
${\mathbf m}^2_{\widetilde{e}}$ and ${\mathbf A}_{e}$ are as follows
\cite{Ibarra:2008uv,Petcov:2003zb},
\begin{eqnarray}
\nonumber\mu\frac{d\mathbf m^2_{\widetilde{L}}}{d\mu}&&=\frac{1}{16\pi^2}\Big[({\mathbf m}^2_{\widetilde{L}}+2m^2_{H_u}){\mathbf Y}^{\dagger}_{\nu}{\mathbf Y}_{\nu}+({\mathbf m}^2_{\widetilde{L}}+2m^2_{H_d}){\mathbf Y}^{\dagger}_e{\mathbf Y}_e+2{\mathbf Y}^{\dagger}_{\nu}{\mathbf m}^2_{\widetilde{\nu}}{\mathbf Y}_{\nu}+2{\mathbf Y}^{\dagger}_{e}{\mathbf m}^2_{\widetilde{e}}{\mathbf Y}_e\\
\nonumber&&+({\mathbf Y}^{\dagger}_{\nu}{\mathbf Y}_{\nu}+{\mathbf Y}^{\dagger}_e{\mathbf Y}_e){\mathbf m}^2_{\widetilde{L}}+2{\mathbf A}^{\dagger}_{\nu}{\mathbf A}_{\nu}+2{\mathbf A}^{\dagger}_e{\mathbf A}_e-6g^2_2|M_2|^2-\frac{6}{5}g^2_1|M_1|^2-\frac{3}{5}g^2_1{\mathcal S}\Big]\\
\nonumber\mu\frac{d{\mathbf m}^2_{\widetilde{e}}}{d\mu}&&=\frac{1}{16\pi^2}\Big[(2{\mathbf m}^2_{\widetilde{e}}+4m^2_{H_d}){\mathbf Y}_{e}{\mathbf Y}^{\dagger}_e+4{\mathbf Y}_{e}{\mathbf m}^{2}_{\widetilde{L}}{\mathbf Y}^{\dagger}_e+2{\mathbf Y}_e{\mathbf Y}^{\dagger}_e{\mathbf m}^2_{\widetilde{e}}+4{\mathbf A}_e{\mathbf A}^{\dagger}_e\\
\nonumber&&-\frac{24}{5}g^2_1|M_1|^2+\frac{6}{5}g^2_1{\mathcal S}\Big]\\
\nonumber\mu\frac{d{\mathbf A}_e}{d\mu}&&=\frac{1}{16\pi^2}\Big\{{\mathbf A}_e[{\rm Tr}(3{\mathbf Y}_d{\mathbf Y}^{\dagger}_d+{\mathbf Y}_e{\mathbf Y}^{\dagger}_e)+5{\mathbf Y}^{\dagger}_e{\mathbf Y}_e+{\mathbf Y}^{\dagger}_{\nu}{\mathbf Y}_{\nu}-3g^2_2-\frac{9}{5}g^2_1]\\
\label{6}&&+{\mathbf Y}_e[{\rm Tr}(6{\mathbf A}_d{\mathbf
Y}^{\dagger}_d+2{\mathbf A}_e{\mathbf Y}^{\dagger}_e)+4{\mathbf
Y}^{\dagger}_e{\mathbf A}_e+2{\mathbf Y}^{\dagger}_{\nu}{\mathbf
A}_{\nu}+6g^2_2M_2+\frac{18}{5}g^2_1M_1]\Big\}
\end{eqnarray}
where $\mu$ is the renormalization point, and we have defined
\begin{equation}
\label{7}{\mathcal S}=m^2_{H_u}-m^2_{H_d}+{\rm Tr}[{\mathbf
m}^2_{\widetilde{Q}}-{\mathbf m}^2_{\widetilde{L}}-2{\mathbf
m}^2_{\widetilde{u}}+{\mathbf m}^2_{\widetilde{d}}+{\mathbf
m}^2_{\widetilde{e}}]
\end{equation}
Here ${\mathbf m}^2_{\widetilde{Q}}$, ${\mathbf
m}^2_{\widetilde{u}}$ and ${\mathbf m}^2_{\widetilde{d}}$ are
respectively the SSB mass-squared matrices for the supersymmetric
partner of the left handed quark doublet, right handed up type quark
and right handed down type quark. In the phenomenological studies it
is convenient to work in the leptonic basis where both the charged
lepton mass matrix and the mass matrix of the right handed neutrino
are real and diagonal. In the leading-logarithmic approximation with
universal boundary conditions Eq.(\ref{5}), the off-diagonal
elements of the above SSB slepton mass matrices and trilinear
coupling are given by \cite{Hisano:1995cp,Hisano:1998fj}
\begin{eqnarray}
\nonumber&&({\mathbf m}^2_{\widetilde{L}})_{i\neq j}\simeq-\frac{1}{8\pi^2}\,(3m^2_0+A^2_0)(\hat{{\mathbf Y}}^{\dagger}_{\nu}{\mathbf L}\hat{{\mathbf Y}}_{\nu})_{ij}\\
\nonumber&&({\mathbf A}_e)_{i\neq j}\simeq-\frac{3A_0}{16\pi^2}\,(\hat{{\mathbf Y}}_{e}\hat{{\mathbf Y}}^{\dagger}_{\nu}{\mathbf L}\hat{{\mathbf Y}}_{\nu})_{ij}\\
\label{8}&&({\mathbf m}^2_{\widetilde{e}})_{i\neq j}\simeq0
\end{eqnarray}
where the hat denotes the basis in which the right handed neutrino
and the charged lepton mass matrix are real and diagonal, and the
factor ${\mathbf L}$ is defined as
\begin{equation}
\label{9}{\mathbf L}_{ij}=\log(\frac{M_G}{M_i})\,\delta_{ij}
\end{equation}
Here $M_i$ is the ith heavy right handed neutrino mass, and $M_G$ is
the GUT scale typically equal to $2\times10^{16}$ GeV. Eq.(\ref{8})
shows that, within the type-I Seesaw mechanism the off-diagonal
elements of the right handed charged slepton matrix approximately
don't run in the leading-log approximation, and the running of the
trilinear parameters ${\mathbf A}_e$ is suppressed by the charged
lepton masses. At low energy, the flavor off-diagonal correction
$({\mathbf m}^2_{\widetilde{L}})_{ij}$ induces the LFV processes
such as $l_i\rightarrow l_j\gamma$, $l_i\rightarrow3l_j$ and
$l_i-l_j$ conversion in nuclei, with $i>j=1,\,2,\,3$. A very useful
tool to treat analytically the complicated expression for the
branching ratio is the mass insertion approximation, where the small
off-diagonal elements of the soft mass matrix are treated as
insertions in the sfermion propagators in the loop. Then we obtain
the compact expression for the branching ratio of the charged lepton
LFV radiative decay
\begin{equation}
\label{10}Br(l_i\rightarrow l_j\gamma)\simeq Br(l_i\rightarrow l_j\bar{\nu}_j\nu_i)\frac{\alpha^3}{G^2_Fm^8_s}|({\mathbf m}^2_{\widetilde{L}})_{ij}|^2\tan^2\beta
\end{equation}
where $G_F$ is the Fermi coupling constant and $m_s$ is the
characteristic mass scale of the SUSY particle in the loop. It has
been show that the expression (\ref{10}) with $({\mathbf
m}^2_{\widetilde{L}})_{ij}$ given by Eq.(\ref{8}) represents an
excellent approximation to the exact renormalization group result if
one sets \cite{Petcov:2003zb}
\begin{equation}
\label{11}m^8_s\simeq0.5m^2_0M^2_{1/2}(m^2_0+0.6M^2_{1/2})^2
\end{equation}
The LFV processes $l_i\rightarrow l_j\gamma$, $l_i\rightarrow3l_j$
and $l_i-l_j$ conversion in nuclei occur via $\gamma-$, $Z-$ and
Higgs-penguins as well as squark/slepton box diagrams. It has been
shown that the contribution from the $\gamma-$ penguin diagrams are
almost undistinguishable from the total rates in the universal
mSUGRA scenario, although the contribution of Higgs-penguins becomes
large in the region of large $\tan\beta\sim60$ and light Higgs boson
mass$\sim100$ GeV, which is not allowed by the current experimental
lower bounds on the MSSM particle masses \cite{Amsler:2008zzb}. In
this approximation, the branching ratio for the trilepton decays is
approximately given by \cite{Arganda:2005ji,Arganda:2007jw}
\begin{equation}
\label{12}Br(l_i\rightarrow 3l_j)\simeq\frac{\alpha}{3\pi}(\log\frac{m^2_{l_i}}{m^2_{l_j}}-\frac{11}{4})Br(l_i\rightarrow l_j\gamma)
\end{equation}
For $\mu-e$ conversion, the $\gamma-$penguins dominance gives
\begin{equation}
\label{13}CR(\mu N\rightarrow eN)=\frac{\Gamma(\mu N\rightarrow eN)}{\Gamma_{cap}}\simeq\frac{\alpha^4m^5_{\mu}G^2_{F}}{12\pi^3\Gamma_{cap}}\,ZZ^4_{eff}|F(q^2)|^2Br(\mu\rightarrow e\gamma)
\end{equation}
where $\Gamma_{cap}$ is the measured total muon capture rate, $Z$ is
the proton number of the nucleus, $Z_{eff}$ is the effective atomic
charge obtained by averaging the muon wavefunction over the nuclear
density, and $F(q^2)$ denotes the nuclear form factor at momentum
transfer $q$ \cite{mueconversion}. In this work, we will consider
$\mu-e$ conversion in two nuclei ${\rm ^{27}_{13}Al}$ and ${\rm
^{48}_{22}Ti}$. ${\rm ^{27}_{13}Al}$ would be used by the proposed
Mu2e experiment at Fermilab \cite{mu2e}, $Z_{eff}=11.5$, the values
of the relevant parameters are $F(q^2\simeq-m^2_{\mu})=0.64$ and
$\Gamma_{cap}=4.64079\times10^{-19}$ GeV. For the nucleus ${\rm
^{48}_{22}Ti}$ used by the proposed PRIME experiment at J-PARC
\cite{prime}, $Z_{eff}=17.6$ , $F(q^2\simeq-m^2_{\mu})\simeq0.54$
and $\Gamma_{cap}=1.70422\times10^{-18}$ GeV.

Experimental discovery of LFV is one of smoking gun signatures of
new physics beyond the SM, thus many experiments have been developed
to detect the rare LFV processes. The present and projected bounds
on these processes are summarized in Table \ref{tab:lfv}. In the
following, we shall investigate in details the predictions for the
rare processes $l_i\rightarrow l_j\gamma$, $l_i\rightarrow3l_j$ and
$l_i-l_j$ conversion in ${\rm ^{27}_{13}Al}$ and ${\rm
^{48}_{22}Ti}$ in the AF model and the $S_4$ model of Ding, and
discuss the possibility of testing these discrete flavor symmetry
models by LFV.

\begin{table}
\begin{center}
\begin{tabular}{|l|c|c|}
\hline\hline
  & ~~Present~~ & ~~Future~~  \\
\hline
BR($\mu \rightarrow e\gamma$)    &~~ $1.2 \times 10^{-11}$ ~~  &~~  $10^{-13}$~~ \\
BR($\tau \rightarrow \mu\gamma $)&~~ $4.5 \times 10^{-8}$~~  &~~  $10^{-9}$~~ \\
BR($\tau \rightarrow e\gamma$)   &~~ $3.3 \times 10^{-8}$~~  &~~  $10^{-9}$ ~~ \\
BR($\mu \rightarrow eee$)        &~~ $1.0 \times 10^{-12}$~~  &~~  $10^{-14}$~~ \\
BR($\tau \rightarrow \mu\mu\mu $)&~~ $3.2 \times 10^{-8}$~~  &~~  $10^{-9}$~~  \\
BR($\tau \rightarrow eee$)       &~~$3.6 \times 10^{-8}$~~  &~~  $10^{-9}$~~  \\
CR($\mu\,\textrm{Ti}\rightarrow e\,\textrm{Ti}$) &~~ $4.3 \times
10^{-12}$~~  &~~
$10^{-18}$~~ \\
CR($\mu\,\textrm{Al}\rightarrow e\,\textrm{Al}$) &~~  -- ~~ &~~  $10^{-16}$~~  \\
\hline\hline
\end{tabular}
\caption{Present and upcoming experimental limits on various lepton
flavor violation processes \cite{Amsler:2008zzb}. \label{tab:lfv}}
\end{center}
\end{table}

\section{Flavor models with $A_4$ and $S_4$ symmetries}

In this section, we will briefly present the Altarelli-Feruglio
$A_4$ model \cite{Altarelli:2005yx} and the $S_4$ model of Ding
\cite{Ding:2009iy}, the next to leading order (NLO) corrections to
the lepton mixing parameters will be discussed in detail, especially
the allowed region of the reactor neutrino angle $\theta_{13}$ and
the Jarlskog invariant $J$ in the lepton sector
\cite{Jarlskog:1985cw}.

\subsection{AF $A_4$ model\label{sec:A4_model}}

The AF model is a typical supersymmetric flavor model with $A_4$
symmetry, and the auxiliary symmetries are $Z_3$ and the
Froggatt-Nielsen symmetry $U(1)_{FN}$. The $U(1)_{FN}$ is to
reproduce the observed charged lepton mass hierarchies, and the
$Z_3$ symmetry is to guarantee that the neutrino and charged lepton
couple with different flavons at LO, so that the $A_4$ symmetry is
broken down to $Z_3$ and $Z_2$ subgroups in the charged lepton and
neutrino sectors respectively at LO, this misalignment of symmetry
breaking is exactly the origin of TB mixing. $A_4$ is the group of
the even permutation of four objects, geometrically it is the
invariant group of the regular tetrahedron in 3-dimensional space.
It has three independent one-dimensional representations ${\mathbf
1}$, ${\mathbf{1'}}$ and ${\mathbf 1''}$ and one three-dimensional
representations ${\mathbf{3}}$. The multiplication rule for two
triplet representations is the following
\begin{equation}
\label{14}{\mathbf 3}\otimes {\mathbf 3}={\mathbf 1}\oplus{\mathbf
1'}\oplus{\mathbf 1''}\oplus{\mathbf 3}_S\oplus{\mathbf 3}_A
\end{equation}
If we denote the two triplets as $(a_1,a_2,a_3)$ and
$(b_1,b_2,b_3)$, the singlets and triplets contained in the product
are given by
\begin{eqnarray}
\nonumber&&{\mathbf 1}\sim a_1b_1+a_2b_3+a_3b_2\\
\nonumber&&{\mathbf 1'}\sim a_3b_3+a_1b_2+a_2b_1\\
\nonumber&&{\mathbf 1''}\sim a_2b_2+a_1b_3+a_3b_1\\
\nonumber&&{\mathbf
3}_S\sim(2a_1b_1-a_2b_3-a_3b_2,2a_3b_3-a_1b_2-a_2b_1,2a_2b_2-a_1b_3-a_3b_1)\\
\label{15}&&{\mathbf 3}_A\sim
(a_2b_3-a_3b_2,a_1b_2-a_2b_1,a_3b_1-a_1b_3)
\end{eqnarray}
The fields in the model and their transformation properties under
the flavor group are listed in Table \ref{tb:A4rappresentazioni}. At
LO the flavon fields develop the following vacuum expectation values
(VEVs),
\begin{eqnarray}
\nonumber&&\langle\varphi_{T}\rangle=(V_T,0,0),~~~\langle\theta\rangle=V_{\theta},\\
\nonumber&&\langle\xi\rangle=V_{\xi},~~~\langle\tilde{\xi}\rangle=0,\\
\label{16}&& \langle\varphi_S\rangle=(V_S,V_S,V_S)
\end{eqnarray}

\begin{table}[h]
\begin{center}
\begin{tabular}{|c|c|c|c|c|c|c|c|c|c|c|c|c|c|c|c|c|}
 \hline\hline
 & $\ell$ & $e^c$ & $\mu^c$ & $\tau^c$ & $\nu^c$ & $h_{u,d}$ & $\varphi_T$&
$\varphi_S$&$\xi$&$\tilde\xi$& $\theta$ &$\varphi_0^T$&$\varphi_0^S$&$\xi_0$\\
\hline
$A_4$ & 3&1&$1''$&$1'$&3&1&3&3&1&1& 1&3&3&1\\
\hline
$Z_3$&$\omega$&$\omega^2$&$\omega^2$&$\omega^2$&$\omega^2$&1&1&$\omega^2$&$\omega^2$&$\omega^2$& 1&1&$\omega^2$&$\omega^2$\\
\hline
$U(1)_{FN}$ & 0&2&1&0&0&0&0&0&0& 0&-1&0&0&0\\
\hline
$U(1)_{R}$ & 1&1&1&1&1&0&0&0&0&0&0&2&2&2\\
\hline\hline
\end{tabular}\caption{The fields of the $A_4\times Z_3\times U(1)_{FN}$ model and their
 representations, where $\omega$ is the third root of unity $\omega=e^{i2\pi/3}$.}
\label{tb:A4rappresentazioni}
\end{center}
\end{table}

\begin{figure}[h]
\begin{center}
\begin{tabular}{c}
\includegraphics[scale=1,width=5.0cm]{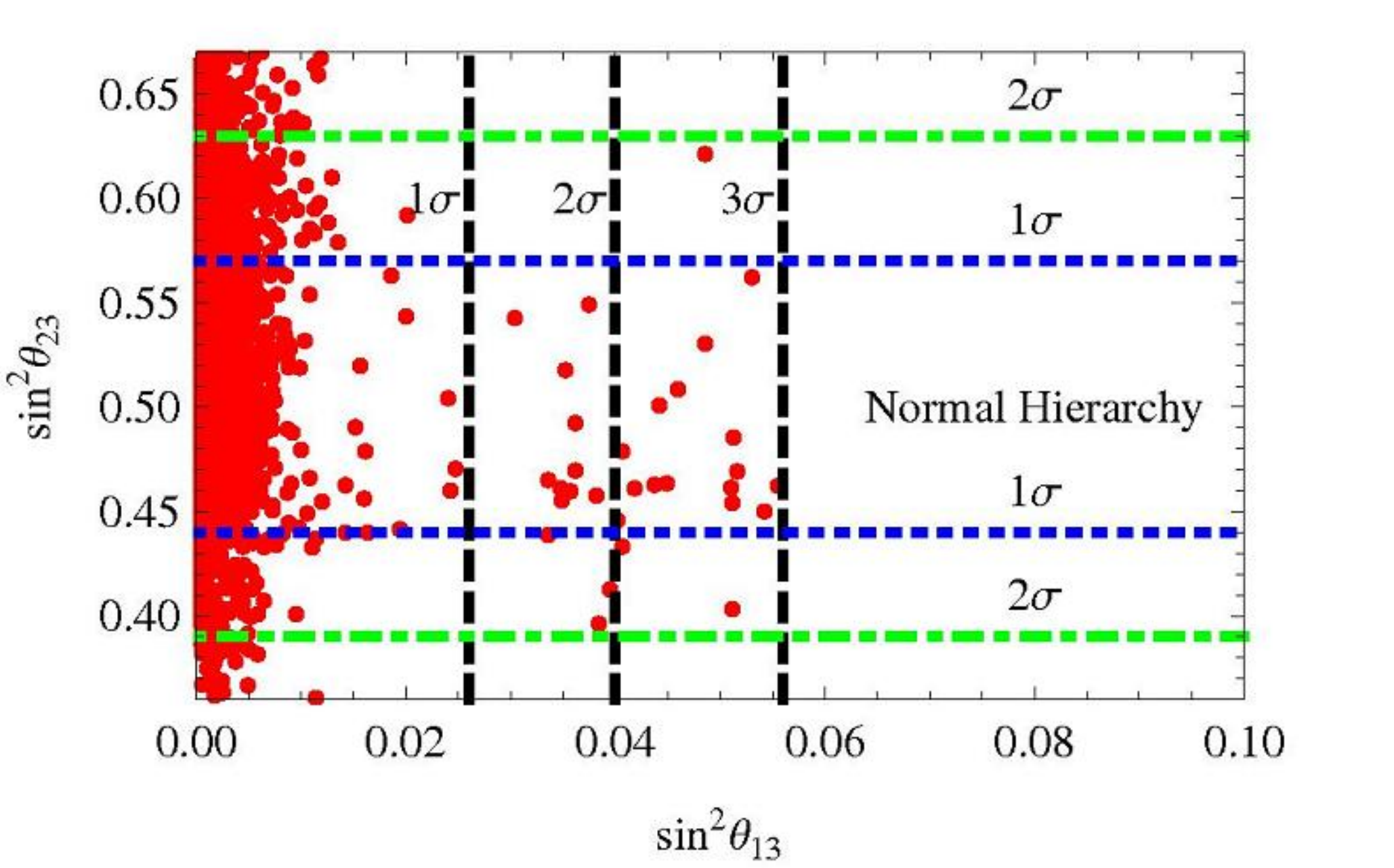}
\hspace{1cm}
\includegraphics[scale=1,width=5.0cm]{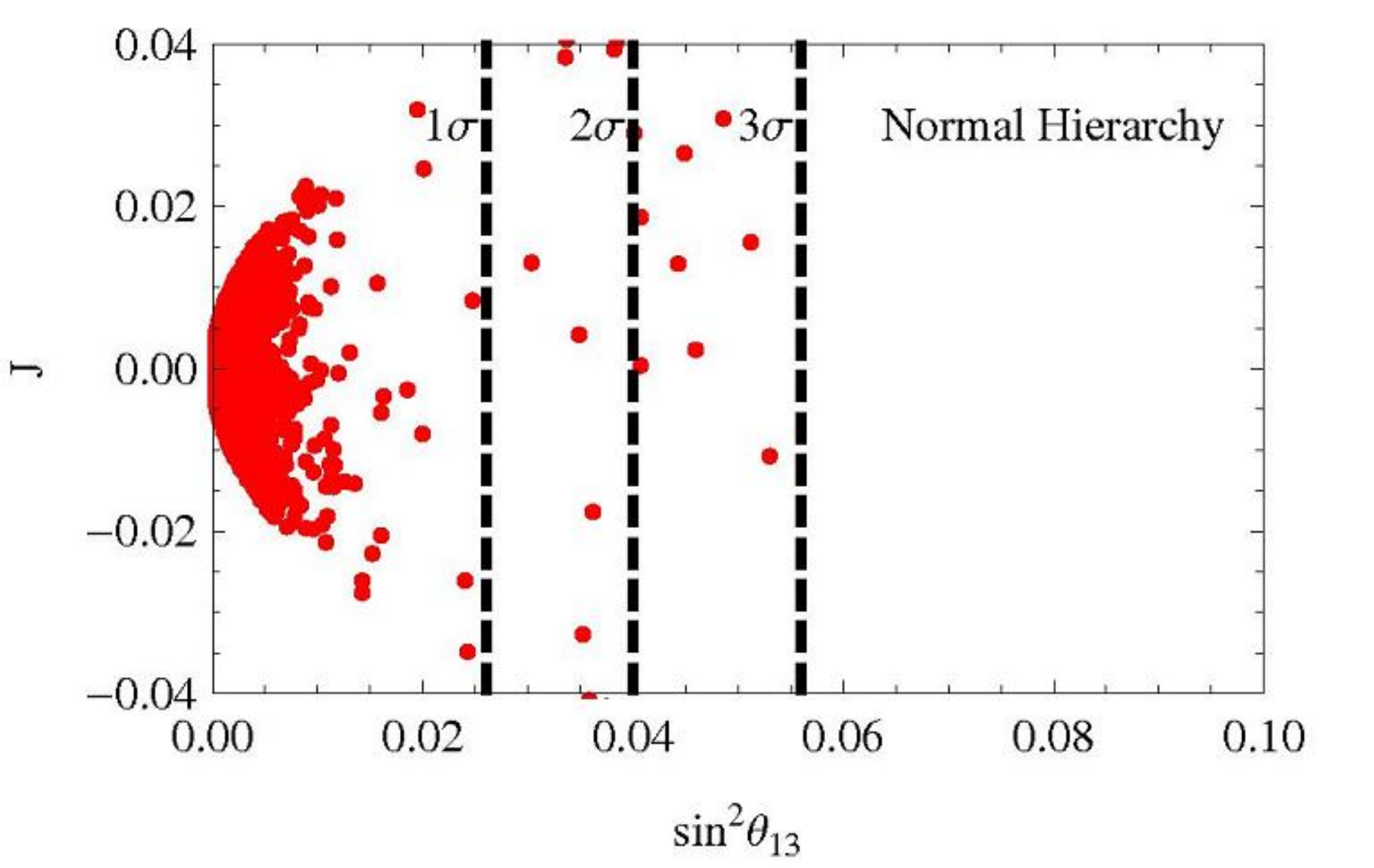}\\\\
\includegraphics[scale=1,width=5.0cm]{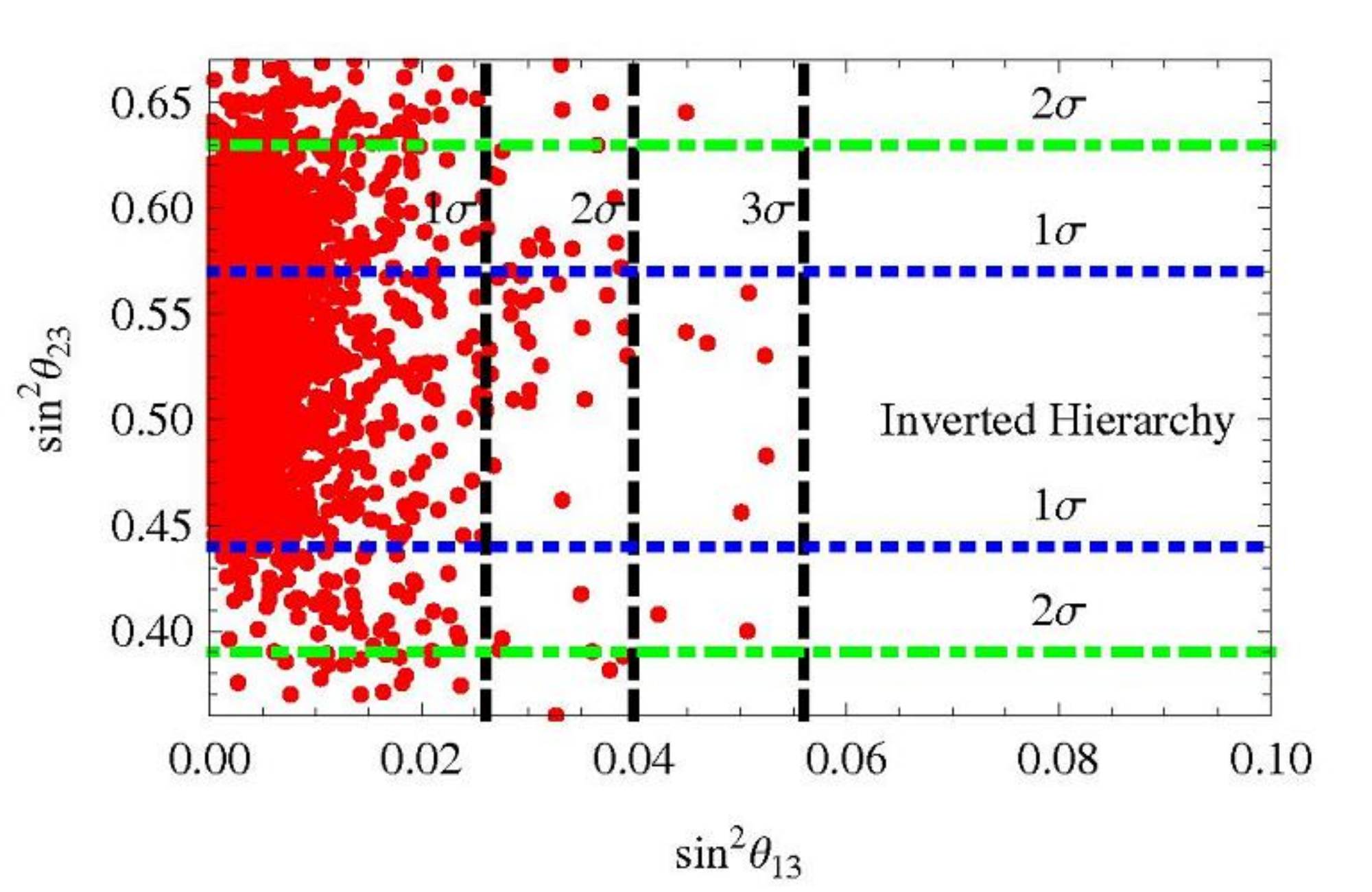}
\hspace{1cm}
\includegraphics[scale=1,width=5.0cm]{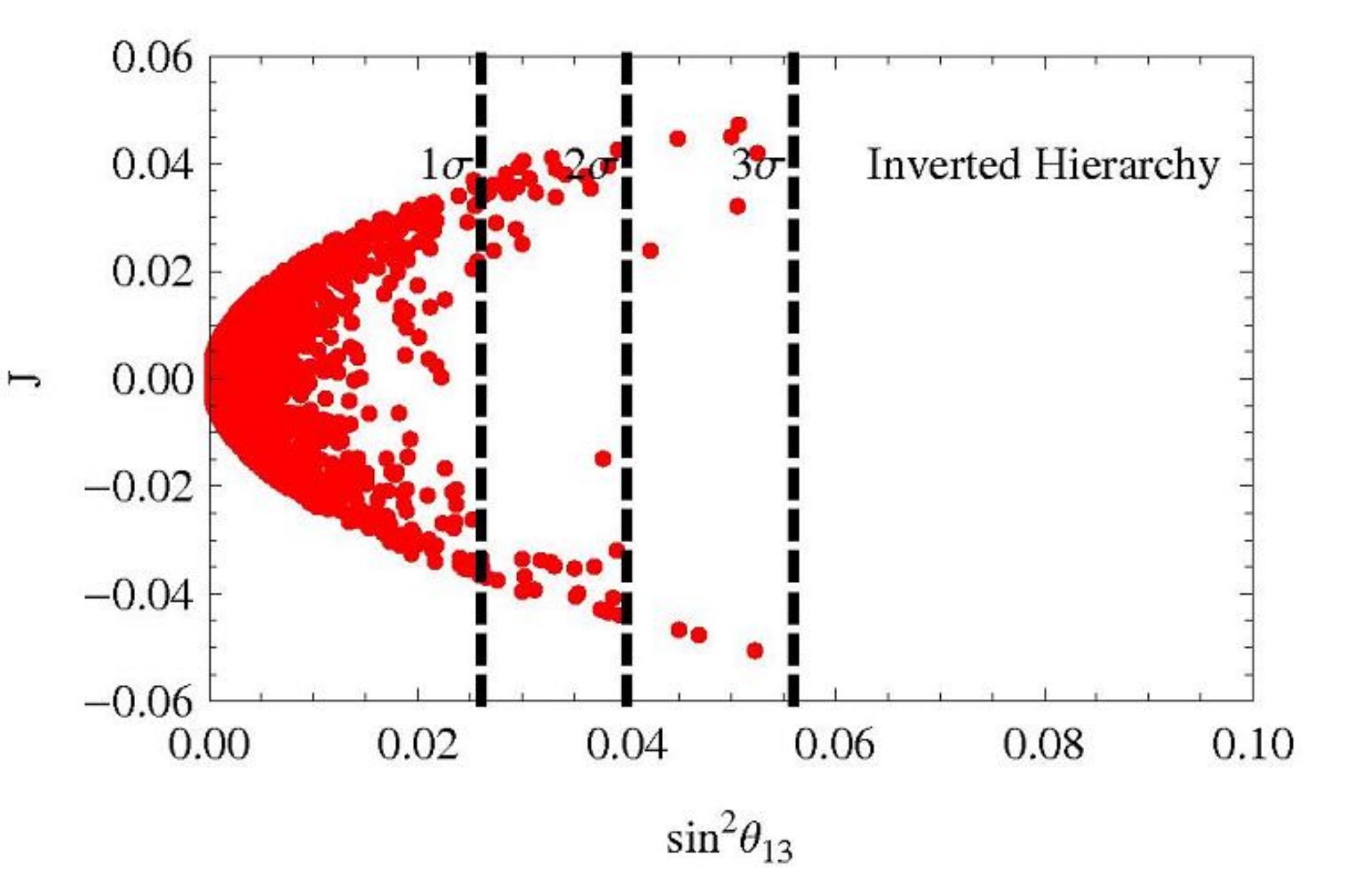}
\end{tabular}
\caption{\label{fig:AF_NLO}Scatter plot of $\sin^2\theta_{23}$ and
the Jarlskog invariant $J$ against $\sin^2\theta_{13}$ for both
normal hierarchy and inverted hierarchy neutrino mass spectrum in AF
model.}
\end{center}
\end{figure}

At LO the neutrino masses are generated by the following
superpotential
\begin{equation}
\label{17}w_{\nu}=y_{\nu}(\nu^c\ell)h_u+(x_A\xi+\tilde{x}_A\tilde{\xi})(\nu^c\nu^c)+x_{B}(\nu^c\nu^c\varphi_S)
\end{equation}
Here and in the following, we denote an invariant under $A_4$ by a
parenthesis (...). After the electroweak and flavor symmetry
breaking, we have
\begin{equation}
\label{18}w_{\nu}=y_{\nu}(\nu^{c}_1\nu_1+\nu^{c}_2\nu_3+\nu^{c}_3\nu_2)v_u+x_AV_{\xi}(\nu^{c\,2}_1+2\nu^{c}_1\nu^{c}_2)+2x_BV_S(\nu^{c\,2}_1+\nu^{c\,2}_2+\nu^{c\,2}_3-\nu^c_1\nu^c_2-\nu^c_1\nu^c_3-\nu^c_2\nu^c_3)
\end{equation}
where $v_u=\langle h_u\rangle$ is the VEV of the up type Higgs, and
one can always set $y_{\nu}$ to be real by a global phase
transformation of the lepton doublet field. The first term in
Eq.(\ref{18}) contributes to the neutrino Dirac mass matrix
\begin{equation}
\label{19}M_D=y_{\nu}v_u\left(\begin{array}{ccc} 1&0&0\\
0&0&1\\
0&1&0
\end{array}\right)
\end{equation}
The remaining terms lead to neutrino Majorana mass matrix
\begin{equation}
\label{20}M_M=\left(\begin{array}{ccc}A+\frac{2}{3}B&-\frac{1}{3}B&-\frac{1}{3}B\\
-\frac{1}{3}B&\frac{2}{3}B&A-\frac{1}{3}B\\
-\frac{1}{3}B&A-\frac{1}{3}B&\frac{2}{3}B
\end{array}\right)
\end{equation}
where $A=2x_AV_{\xi}$ and $B=6x_BV_S$. The right handed neutrino
mass matrix $M_M$ can be diagonalized by a unitary matrix $U$ as
usual
\begin{equation}
\label{21}U^TM_MU={\rm diag}(|A+B|,|A|,|-A+B|)
\end{equation}
The unitary matrix $U$ is given by
\begin{equation}
\label{22}U=U_{TB}U_{\phi}
\end{equation}
where $U_{TB}$ exactly is the well-known TB mixing matrix
\begin{equation}
\label{23}U_{TB}=\left(\begin{array}{ccc}
\sqrt{\frac{2}{3}}&\frac{1}{\sqrt{3}}&0\\
-\frac{1}{\sqrt{6}}&\frac{1}{\sqrt{3}}&\frac{1}{\sqrt{2}}\\
-\frac{1}{\sqrt{6}}&\frac{1}{\sqrt{3}}&-\frac{1}{\sqrt{2}}
\end{array}\right)
\end{equation}
and $U_{\phi}={\rm
diag}(e^{-i\phi_1/2},e^{-i\phi_2/2},e^{-i\phi_3/2})$ is a matrix of
phase with $\phi_1={\rm arg}(A+B)$, $\phi_2={\rm arg}(A)$ and
$\phi_3={\rm arg}(-A+B)$. The light neutrino mass matrix is given by
the type-I Seesaw formula
\begin{equation}
\label{24}m_{\nu}=-M^{T}_DM^{-1}_MM_D
\end{equation}
It is diagonalized by the unitary transformation $U_{\nu}$, i.e.,
\begin{equation}
\label{25}U^T_{\nu}m_{\nu}U_{\nu}={\rm diag}(m_1,m_2,m_3)
\end{equation}
where $U_{\nu}=iU_{TB}U^{*}_{\phi}$, and $m_{1,2,3}$ are the light
neutrino masses
\begin{equation}
\label{26}m_1=\frac{y^2_{\nu}v^2_{u}}{|A+B|},~~~~m_2=\frac{y^2_{\nu}v^2_u}{|A|},~~~~m_3=\frac{y^2_{\nu}v^2_u}{|-A+B|}
\end{equation}
It is obvious that the AF model is strongly constrained, the
neutrino part of the Lagrangian depends on only three parameters:
$y_{\nu}$, $A$ and $B$, the latter two parameters are in general
complex numbers. Concerning the light neutrino at low energy, only
two parameters $A/y^2_{\nu}$ and $B/y^2_{\nu}$ are involved.
Imposing the constraints of $\Delta m^2_{21}$ and $\Delta m^2_{31}$
measured from the neutrino oscillation experiments, only one real
parameter is left, and it is usually chose to be the lightest
neutrino mass. Detailed studies showed that the flavon VEVs $V_T$,
$V_{\xi}$ and $V_S$ should be approximately of the same order ${\cal
O}(\lambda^2_c\Lambda)$ \cite{Altarelli:2005yx,Bertuzzo:2009im},
where $\lambda_c$ is the Cabibbo angle, and $\Lambda$ is the cutoff
scale of the theory. The neutrino mass spectrum can be either normal
hierarchy (NH) or inverted hierarchy (IH), and the lightest neutrino
mass is tightly constrained as follows
\cite{Bertuzzo:2009im,Hagedorn:2009jy}
\begin{equation}
\label{27}\begin{array}{cr}
0.0044 {\rm \,eV}\leq m_1\leq0.0060 {\rm \,eV},&~~~~~{\rm NH}\\
0.017 {\rm eV}\leq m_3,&~~~~~{\rm IH}
\end{array}
\end{equation}
The above LO predictions for the lepton masses and mixing parameters
could be corrected by both the higher dimensional operators in the
superpotential $w_{\nu}$ compatible with the symmetry of the model
and the shift of the vacuum alignment. Including the NLO operators,
the charged lepton masses are described by the following
superpotential,
\begin{eqnarray}
\nonumber
w_{\ell}&&=y_e\frac{1}{\Lambda^3}e^{c}(\ell\varphi_T)\theta^2h_d+y_{\mu}\frac{1}{\Lambda^2}\mu^{c}(\ell\varphi_T)'\theta
h_d+y_{\tau}\frac{1}{\Lambda}\tau^{c}(\ell\varphi_T)''h_d\\
\nonumber&&+y_e\frac{1}{\Lambda^3}e^{c}(\ell\delta\varphi_T)\theta^2h_d+y_{\mu}\frac{1}{\Lambda^2}\mu^{c}(\ell\delta\varphi_T)'\theta
h_d+y_{\tau}\frac{1}{\Lambda}\tau^{c}(\ell\delta\varphi_T)''h_d\\
\label{32}&&+y_{e1}\frac{1}{\Lambda^4}e^{c}(\ell\varphi_T\varphi_T)\theta^2h_d+y_{\mu1}\frac{1}{\Lambda^3}\mu^{c}(\ell\varphi_T\varphi_T)'\theta
h_d+y_{\tau1}\frac{1}{\Lambda^2}\tau^{c}(\ell\varphi_T\varphi_T)''h_d
\end{eqnarray}
where $\delta\varphi_T$ denotes the shifted vacuum of the flavon
$\varphi_T$. The first line represents the LO contributions, which
leads to a diagonal charged lepton mass matrix, and the latter two
lines are the NLO corrections. Since
$\langle(\varphi_T\varphi_T)_{3_S}\rangle=(V^2_T,0,0)$ which has the
same alignment directions as $\langle\varphi_T\rangle$, consequently
the contributions of $y_{e1}$, $y_{\mu1}$ and $y_{\tau1}$ terms can
be absorbed into the LO results. Taking into account
$\langle\delta\varphi_T\rangle\simeq(1,1,1)\delta V_T$
\cite{Altarelli:2005yx}, the charged lepton mass matrix at NLO is
given by
\begin{equation}
\label{33}m_{\ell}=\left(\begin{array}{ccc}
y_e\frac{V^2_{\theta}}{\Lambda^2}\frac{V_T}{\Lambda}&y_e\frac{V^2_{\theta}}{\Lambda^2}\frac{\delta
V_T}{\Lambda}&y_e\frac{V^2_{\theta}}{\Lambda^2}\frac{\delta
V_T}{\Lambda}\\
y_{\mu}\frac{V_{\theta}}{\Lambda}\frac{\delta
V_{T}}{\Lambda}&y_{\mu}\frac{V_{\theta}}{\Lambda}\frac{V_T}{\Lambda}&y_{\mu}\frac{V_{\theta}}{\Lambda}\frac{\delta
V_T}{\Lambda}\\
y_{\tau}\frac{\delta V_T}{\Lambda}&y_{\tau}\frac{\delta
V_{\tau}}{\Lambda}&y_{\tau}\frac{V_T}{\Lambda}
\end{array}\right)v_d
\end{equation}
where we have redefined $V_T+\delta V_T\rightarrow V_T$. The charged
lepton mass matrix $m_{\ell}$ can be diagonalized by the bi-unitary
transformation, i.e., $U^{\dagger}_{e^{c}}m_{\ell}U_{\ell}={\rm
diag}(m_e,m_{\mu},m_{\tau})$, and the unitary matrix $U_{\ell}$
approximately is
\begin{equation}
\label{34}U_{\ell}\simeq\left(\begin{array}{ccc} 1&(\frac{\delta
V_T}{V_T})^*&(\frac{\delta V_T}{V_T})^*\\
-\frac{\delta V_T}{V_T}&1&(\frac{\delta V_T}{V_T})^*\\
-\frac{\delta V_T}{V_T}&-\frac{\delta V_T}{V_T}&1
\end{array}\right)
\end{equation}
The NLO corrections to the neutrino Dirac mass terms are as follows,
\begin{equation}
\label{28}\frac{y_A}{\Lambda}((\nu^c\ell)_{3_S}\varphi_T)h_u+\frac{y_B}{\Lambda}((\nu^c\ell)_{3_S}\varphi_T)h_u
\end{equation}
where (...$)_{3_{S(A)}}$ denotes the $3_{S(A)}$ product of the two
triplet representations, which can be read directly from the
multiplication rule Eq.(\ref{15}). Substituting the VEVs in
Eq.(\ref{16}), we obtain the NLO correction $\delta M_D$
\begin{equation}
\label{29}\delta M_D=\left(\begin{array}{ccc} 2y_A&0&0\\
0&0&-y_A+y_B\\
0&-y_A-y_B&0
\end{array}\right)\frac{V_{T}}{\Lambda}v_u
\end{equation}
The Majorana mass matrix of the right handed neutrino are corrected
by the following terms at NLO
\begin{eqnarray}
\nonumber&&(x_A\delta\xi+\tilde{x}_{A}\delta\tilde{\xi})(\nu^c\nu^c)+x_B(\nu^{c}\nu^{c}\delta\varphi_S)+\frac{x'_A}{\Lambda}(\nu^{c}\nu^{c}\varphi_T)\xi++\frac{x'_B}{\Lambda}(\nu^{c}\nu^{c}\varphi_T)\tilde{\xi}
+\frac{x'_C}{\Lambda}(\nu^{c}\nu^{c})(\varphi_T\varphi_S)\\
\nonumber&&+\frac{x'_D}{\Lambda}(\nu^{c}\nu^{c})'(\varphi_T\varphi_S)''+\frac{x'_E}{\Lambda}(\nu^{c}\nu^{c})''(\varphi_T\varphi_S)'+\frac{x'_F}{\Lambda}((\nu^c\nu^{c})_{3_S}(\varphi_T\varphi_S)_{3_S})\\
\label{30}&&+\frac{x'_G}{\Lambda}((\nu^c\nu^{c})_{3_S}(\varphi_T\varphi_S)_{3_A})
\end{eqnarray}
where $\delta\xi$, $\delta\tilde{\xi}$ and $\delta\varphi_{S}$
denote the shifted vacuum of $\xi$, $\tilde{\xi}$ and $\varphi_{S}$
respectively. Absorbing the above corrections partly into the LO
results, then $\delta M_M$ can be parameterized by four independent
parameters as follows
\begin{equation}
\label{31}\delta M_M=\left(\begin{array}{ccc}
2\tilde{x}_D&\tilde{x}_B-\tilde{x}_E&\tilde{x}_C\\
\tilde{x}_B-\tilde{x}_E&\tilde{x}_C&-\tilde{x}_D\\
\tilde{x}_C&-\tilde{x}_D&\tilde{x}_B+2\tilde{x}_E
\end{array}\right)\frac{V_T}{\Lambda}B
\end{equation}
Therefore at NLO the light neutrino mass matrix is given by
$m_{\nu}=-(M^{T}_D+\delta M^{T}_D)(M_M+\delta M_M)^{-1}(M_D+\delta
M_D)$. Performing the standard perturbation diagonalization,
$m_{\nu}$ can be diagonalized by the unitary transformation
$U'_{\nu}$ as $U^{'T}_{\nu}m_{\nu}U'_{\nu}={\rm
diag}(m'_1,m'_2,m'_3)$. To first order in the expansion parameter
$V_T/\Lambda\equiv\varepsilon$, the light neutrino masses
$m'_{1,2,3}$ are
\begin{eqnarray}
\nonumber&&m'_1=-\frac{y_{\nu}^2v^2_u}{A+B}-\frac{y_{\nu}v^2_u}{2(A+B)^2}[4y_AA+4y_AB+(\tilde{x}_B+\tilde{x}_C-2\tilde{x}_D-2\tilde{x}_E)y_{\nu}B]\varepsilon\\
\nonumber&&m'_2=-\frac{y_{\nu}^2v^2_u}{A}+\frac{y^2_{\nu}v^2_uB}{A^2}(\tilde{x}_B+\tilde{x}_C)\varepsilon\\
\label{32}&&m'_{3}=\frac{y^2_{\nu}v^2_u}{A-B}+\frac{y_{\nu}v^2_u}{2(A-B)^2}[-4y_AA+4y_AB+(\tilde{x}_B+\tilde{x}_C+2\tilde{x}_D+2\tilde{x}_E)y_{\nu}B]\varepsilon
\end{eqnarray}
The mixing matrix $U'_{\nu}$ is given by
\begin{equation}
\label{33}U_{\nu}=\left(\begin{array}{ccc}
\sqrt{\frac{2}{3}}-\frac{1}{\sqrt{3}}s^{*}_{12}&\frac{1}{\sqrt{3}}+\sqrt{\frac{2}{3}}s_{12}&\sqrt{\frac{2}{3}}s_{13}+\frac{1}{\sqrt{3}}s_{23}\\
-\frac{1}{\sqrt{6}}-\frac{1}{\sqrt{3}}s^{*}_{12}-\frac{1}{\sqrt{2}}s^{*}_{13}&\frac{1}{\sqrt{3}}-\frac{1}{\sqrt{6}}s_{12}-\frac{1}{\sqrt{2}}s^{*}_{23}&\frac{1}{\sqrt{2}}-\frac{1}{\sqrt{6}}s_{13}+\frac{1}{\sqrt{3}}s_{23}\\
-\frac{1}{\sqrt{6}}-\frac{1}{\sqrt{3}}s^{*}_{12}+\frac{1}{\sqrt{2}}s^{*}_{13}&\frac{1}{\sqrt{3}}-\frac{1}{\sqrt{6}}s_{12}+\frac{1}{\sqrt{2}}s^{*}_{23}&-\frac{1}{\sqrt{2}}-\frac{1}{\sqrt{6}}s_{13}+\frac{1}{\sqrt{3}}s_{23}
\end{array}\right)
\end{equation}
The small parameters $s_{12}$, $s_{13}$ and $s_{23}$ are
\begin{eqnarray}
\nonumber&&
s_{12}=\frac{1}{\sqrt{2}(|B|^2+AB^{*}+A^{*}B)y_{\nu}}\Big\{A^{*}[4y_AA+2y_BB+(-2\tilde{x}_D+\tilde{x}_E)y_{\nu}B]\varepsilon+(A+B)[4y^{*}_AA^{*}\\
\nonumber&&~~+2y^{*}_BB^{*}+(-2\tilde{x}^{*}_D+\tilde{x}^{*}_E)y_{\nu}B^{*}]\varepsilon^{*}\Big\}\\
\nonumber&&s_{13}=\frac{1}{4\sqrt{3}(AB^{*}+A^{*}B)y_{\nu}}\Big\{-B(A^{*}-B^{*})[4y_B+3(\tilde{x}_B-\tilde{x}_C)y_{\nu}]\varepsilon+B^{*}(A+B)[4y^{*}_B+3(\tilde{x}^{*}_B-\tilde{x}^{*}_C)y_{\nu}]\varepsilon^{*}\Big\}\\
\label{34}&&s_{23}=\frac{1}{\sqrt{6}(AB^{*}+A^{*}B-|B|^2)y_{\nu}}\Big\{B(A^{*}-B^{*})(2y_{B}+3\tilde{x}_Ey_{\nu})\varepsilon-B^{*}A(2y^{*}_B+3\tilde{x}^{*}_Ey_{\nu})\varepsilon^{*}\Big\}
\end{eqnarray}
Taking into account the NLO corrections to the charged lepton mass
matrix, the leptonic PMNS matrix becomes
$U_{PMNS}=U^{\dagger}U_{\nu}$, consequently the parameters of the
lepton mixing matrix are modified as
\begin{eqnarray}
\nonumber&&|U_{e3}|=\bigg|\frac{1}{6\sqrt{2}(AB^{*}+A^{*}B)y_{\nu}}\Big\{-B(A^{*}-B^{*})[4y_B+3(\tilde{x}_{B}-\tilde{x}_C)y_{\nu}]\varepsilon+B^{*}(A+B)[4y^{*}_B+3(\tilde{x}^{*}_B\\
\nonumber&&-\tilde{x}^{*}_C)y_{\nu}]\varepsilon^{*}\Big\}+\frac{1}{3\sqrt{2}(AB^{*}+A^{*}B-|B|^2)y_{\nu}}\Big\{B(A^{*}-B^{*})(2y_B+3\tilde{x}_Ey_{\nu})\varepsilon-B^{*}A(2y^{*}_B+3\tilde{x}^{*}_Ey_{\nu})\varepsilon^{*}\Big\}\bigg|\\
\nonumber&&\sin^2\theta_{21}=\frac{1}{3}-\frac{2}{3}\Big[\frac{\delta
V_T}{V_T}+(\frac{\delta
V_T}{V_T})^{*}\Big]+\frac{1}{3(|B|^2+AB^{*}+A^{*}B)y_{\nu}}\Big\{(2A^{*}+B^{*})[4y_AA+2y_BB+(-2\tilde{x}_D\\
\nonumber&&+\tilde{x}_E)y_{\nu}B]\varepsilon+(2A+B)[4y^{*}_AA^{*}+2y^{*}_BB^{*}+(-2\tilde{x}^{*}_D+\tilde{x}^{*}_E)y_{\nu}B^{*}]\varepsilon^{*}\Big\}\\
\nonumber&&\sin^2\theta_{23}=\frac{1}{2}+\frac{1}{2}[\frac{\delta
V_T}{V_T}+(\frac{\delta
V_T}{V_T})^{*}]-\frac{|B|^2}{12(AB^{*}+A^{*}B)y_{\nu}}\Big\{[4y_B+3(\tilde{x}_B-\tilde{x}_C)y_{\nu}]\varepsilon+[4y^{*}_B+3(\tilde{x}^{*}_B-\\
\label{35}&&-\tilde{x}^{*}_C)y_{\nu}]\varepsilon^{*}\Big\}-\frac{|B|^2}{6(AB^{*}+A^{*}B-|B|^2)y_{\nu}}\Big[(2y_B+3\tilde{x}_Ey_{\nu})\varepsilon+(2y^{*}_B+3\tilde{x}^{*}_Ey_{\nu})\varepsilon^{*}\Big]
\end{eqnarray}
It is obvious that all the three mixing angle receive corrections of
the same order of magnitude about $\lambda^2_c$, which is allowed by
the current neutrino oscillation data in Eq.(\ref{1}). In
particular, the reactor mixing angle $\theta_{13}$ becomes of order
$\lambda^2_c$ after including the NLO corrections. In order to see
clearly the behavior of the mixing parameters after including the
subleading contributions, we perform a numerical analysis. The LO
and NLO parameters $y_{\nu}$, $y_A$, $y_B$, $\tilde{x}_B$,
$\tilde{x}_C$, $\tilde{x}_D$ and $\tilde{x}_E$ are treated as random
complex number with absolute value between 0 and 2, $A$ and $B$ in
the right handed neutrino mass matrix Eq.(\ref{20}) are random
complex number with absolute value in the range of $10^{12}-10^{16}$
GeV, and the parameters $\delta V_T/V_T$ and $V_T/\Lambda$ have been
fixed at the indicative value of 0.04. Furthermore, we require the
oscillation parameters to lie in their allowed $3\sigma$ range given
in Eq.(\ref{1}). The results are presented in Fig. \ref{fig:AF_NLO},
where the Jarlskog invariant $J$ of CP violation in neutrino
oscillation is defined as
\begin{equation}
\label{add3}J=Im[(U_{PMNS})_{e1}(U_{PMNS})_{\mu2}(U_{PMNS})^{*}_{e2}(U_{PMNS})^{*}_{\mu1}]
\end{equation}
As we can see from this figure, in AF model $\theta_{13}$ should be
much smaller than the present upper bound, and it prefer to lie in
the $1\sigma$ range. Our analytical estimates are confirmed.

\subsection{$S_4$ model of Ding\label{sec:S4_model}}

The total flavor symmetry of this model is $S_4\times Z_3\times Z_4$
\cite{Ding:2009iy}, where the auxiliary symmetry $Z_3\times Z_4$ is
crucial to eliminate unwanted couplings and to insure the needed
vacuum alignment. It is remarkable that the realistic pattern of
fermion masses and flavor mixing in both the lepton and quark sector
have been reproduced in this model, and the mass hierarchies are
controlled by the spontaneous breaking of the flavor symmetry
instead of the FN mechanism. The matter fields and the flavons of
the model and their classification under the flavor symmetry are
shown in Table \ref{tab:S4trans}, where the quark fields have been
omitted. In this section, the convention for the group theory of
$S_4$ is the same as that in Ref.\cite{Ding:2009iy}.

\begin{figure}[h]
\begin{center}
\begin{tabular}{c}
\includegraphics[scale=1,width=5.0cm]{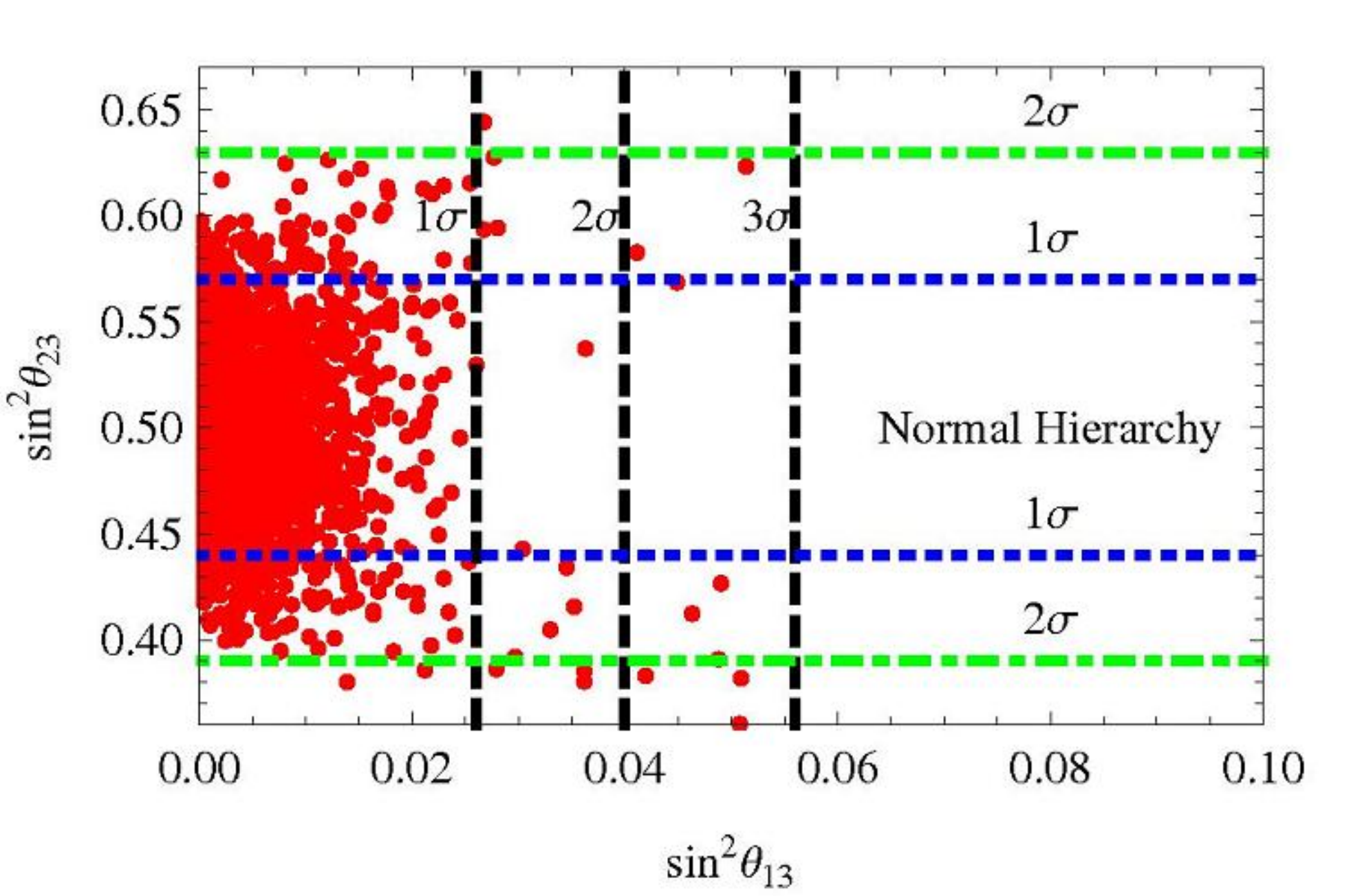}
\hspace{1cm}
\includegraphics[scale=1,width=5.0cm]{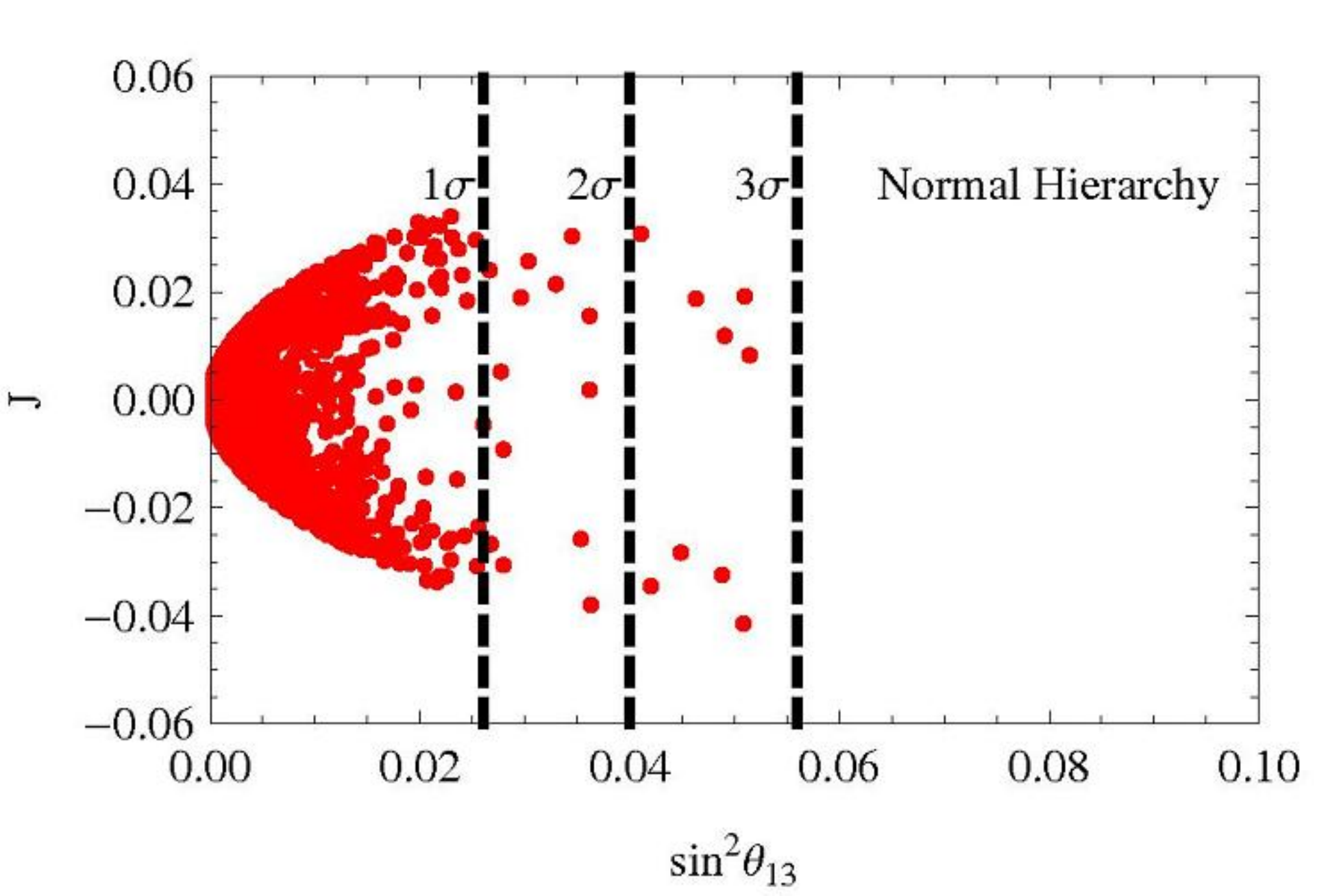}\\\\
\includegraphics[scale=1,width=5.0cm]{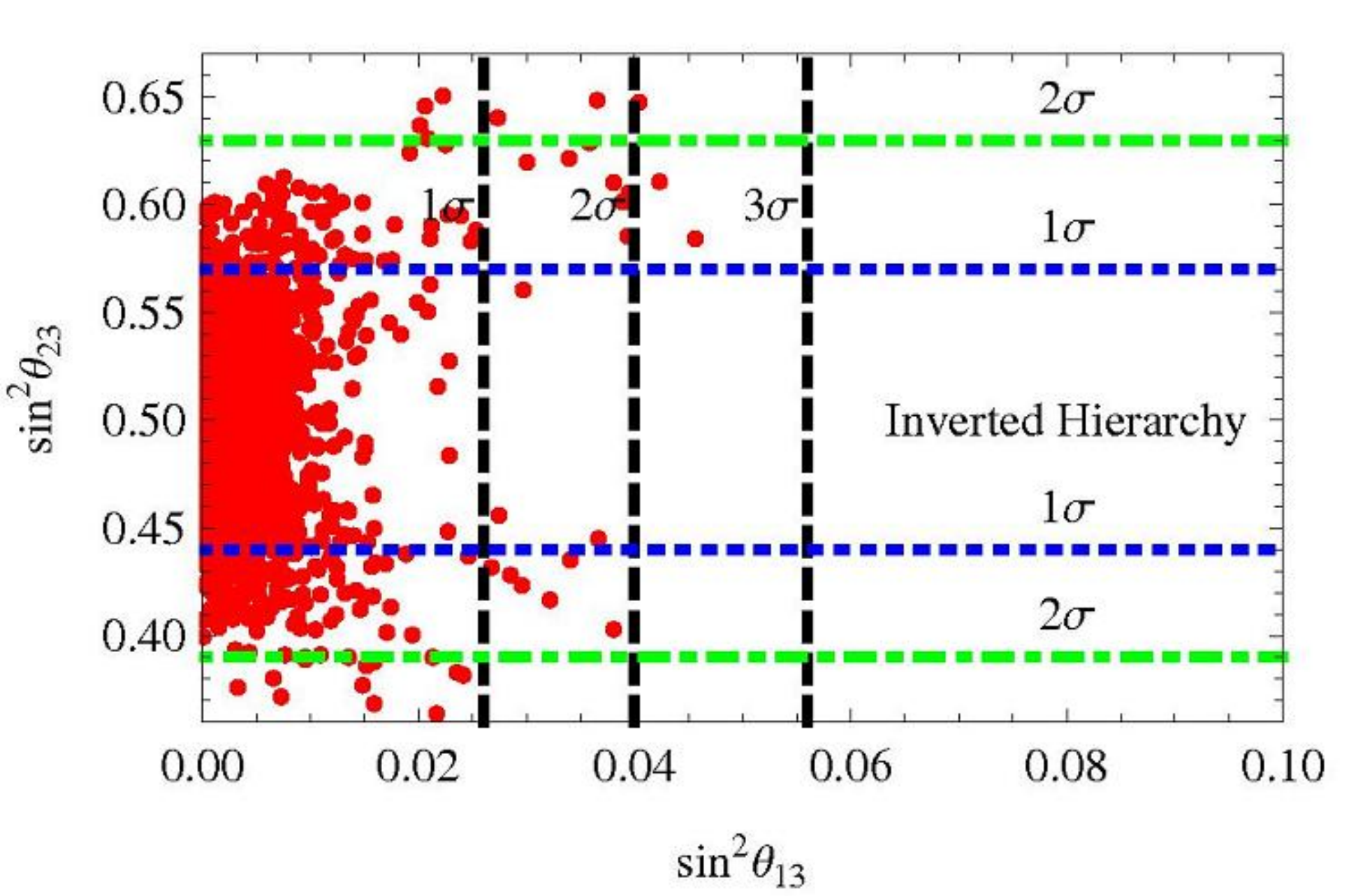}
\hspace{1cm}
\includegraphics[scale=1,width=5.0cm]{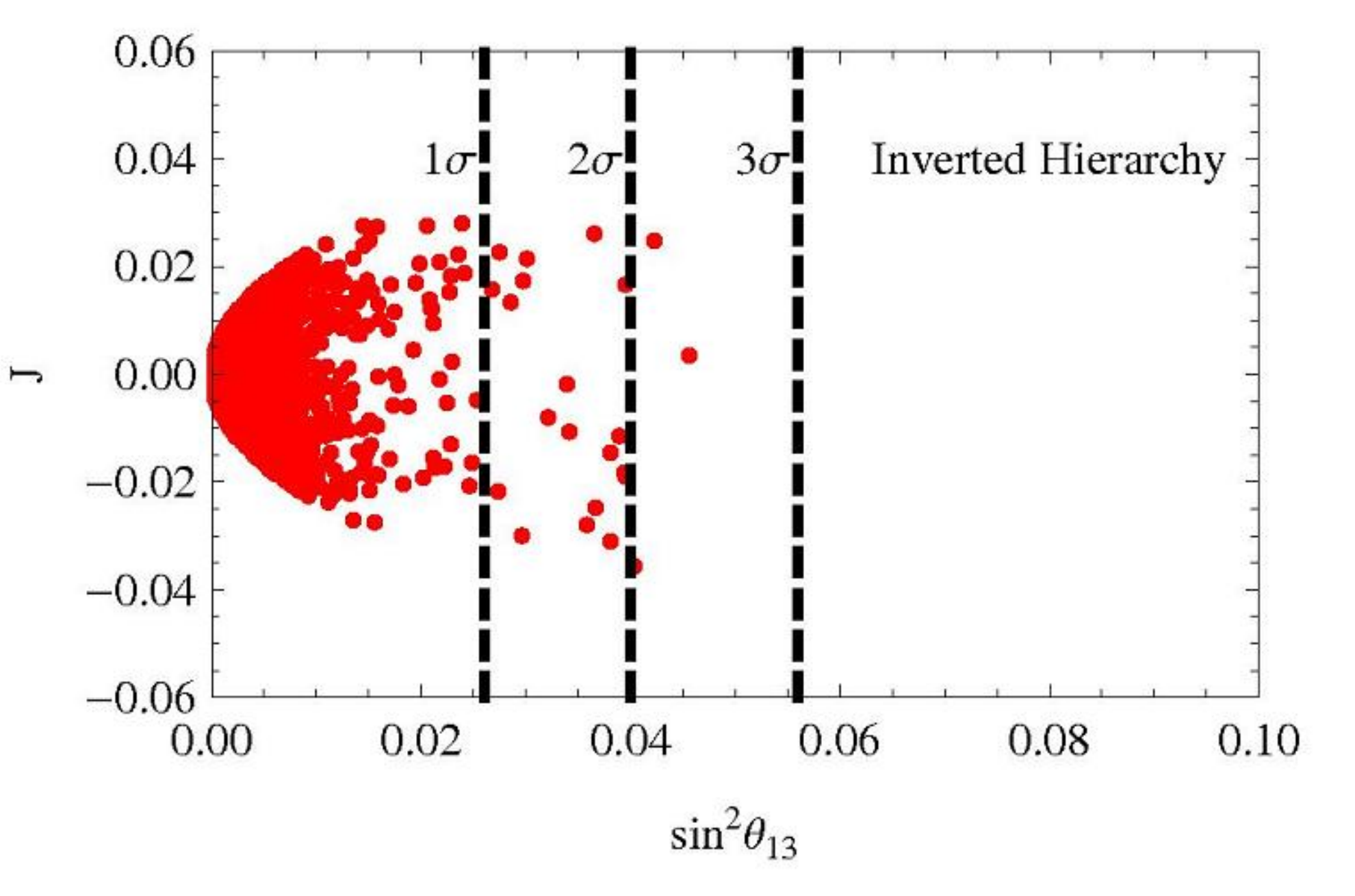}
\end{tabular}
\caption{\label{fig:Ding_NLO}Scatter plot of $\sin^2\theta_{23}$ and
the Jarlskog invariant $J$ against $\sin^2\theta_{13}$ for both
normal hierarchy and inverted hierarchy neutrino mass spectrum in
the $S_4$ model of Ref.\cite{Ding:2009iy}.}
\end{center}
\end{figure}

\begin{table}[hptb]
\begin{center}
\begin{tabular}{|c|c|c|c|c|c|c|c|c|c|c|c|c|}\hline\hline
& $\ell$  & $e^{c}$ & $\mu^{c}$ &  $\tau^{c}$ &$\nu^c$ &
$h_{u,d}$&$\varphi$ & $\chi$ & $\zeta$ & $\eta$ & $\phi$ & $\Delta$
\\\hline

$\rm{S_4}$& $3_1$& $1_1$ & $1_2$&$1_1$ &  $3_1$  &  $1_1$&$3_1$ &
$3_2$& $1_2$ & 2   & $3_1$   & $1_2$  \\\hline

$\rm{Z_{3}}$& $\omega$ & $\omega^2$& $\omega^2$&  $\omega^2$  & $1$
&1 & 1 &1 &1 &  $\omega^2$  & $\omega^2$   & $\omega^2$
\\\hline

$\rm{Z_{4}}$& 1 &i & -1&  -i  & 1  & 1& i &i & 1&  1  & 1  & -1
\\\hline\hline
\end{tabular}
\end{center}
\caption{\label{tab:S4trans}The transformation rules of the matter
fields and the flavons under the symmetry groups $S_4$, $Z_3$ and
$Z_4$ in the $S_4$ model of Ref. \cite{Ding:2009iy}. $\omega$ is the
third root of unity, i.e.
$\omega=e^{i\frac{2\pi}{3}}=(-1+i\sqrt{3})/2$.}
\end{table}
Explicit calculation demonstrated that the LO vacuum alignment is as
follows \cite{Ding:2009iy},
\begin{eqnarray}
\nonumber&&\langle\varphi\rangle=(0,V_{\varphi},0),~~~\langle\chi\rangle=(0,V_{\chi},0)~~~\langle\zeta\rangle=V_{\zeta}\\
\label{36}&&\langle\eta\rangle=(V_{\eta},V_{\eta}),~~~\langle\phi\rangle=(V_{\phi},V_{\phi},V_{\phi}),~~~\langle\Delta\rangle=V_{\Delta}
\end{eqnarray}
In this model, the charged lepton masses are described by the
following superpotential
\begin{eqnarray}
\nonumber&&w_{\ell}=\frac{y_{e1}}{\Lambda^3}\;e^{c}(\ell\varphi)_{1_1}(\varphi\varphi)_{1_1}h_d+\frac{y_{e2}}{\Lambda^3}\;e^{c}((\ell\varphi)_2(\varphi\varphi)_2)_{1_1}h_d+\frac{y_{e3}}{\Lambda^3}\;e^{c}((\ell\varphi)_{3_1}(\varphi\varphi)_{3_1})_{1_1}h_d\\
\nonumber&&~~+\frac{y_{e7}}{\Lambda^3}\;e^{c}((\ell\varphi)_2(\chi\chi)_2)_{1_1}h_d+\frac{y_{e8}}{\Lambda^3}\;e^{c}((\ell\varphi)_{3_1}(\chi\chi)_{3_1})_{1_1}h_d+\frac{y_{e9}}{\Lambda^3}\;e^{c}((\ell\chi)_2(\varphi\varphi)_2)_{1_1}h_d\\
\label{add1}&&~~+\frac{y_{e10}}{\Lambda^3}\;e^{c}((\ell\chi)_{3_1}(\varphi\varphi)_{3_1})_{1_1}h_d+\frac{y_{\mu}}{2\Lambda^2}\mu^{c}(\ell(\varphi\chi)_{3_2})_{1_2}h_d+\frac{y_{\tau}}{\Lambda}\tau^{c}(\ell\varphi)_{1_1}h_d+...
\end{eqnarray}
Taking into account the vacuum alignment in Eq.(\ref{36}), we find
that the charged lepton mass matrix is diagonal at LO, and the
charged lepton masses are given by
\begin{equation}
\label{add2}m_e=\Big|y_e\frac{V^3_{\varphi}}{\Lambda^3}v_d\Big|,~~~m_{\mu}=\Big|y_{\mu}\frac{V_{\varphi}V_{\chi}}{\Lambda^2}v_d\Big|,~~~m_{\tau}=\Big|y_{\tau}\frac{V_{\varphi}}{\Lambda}v_d\Big|
\end{equation}
where $y_e$ is the linear combination of $y_{ei}$ ($i=1-10$). At LO
the superpotential contributing to the neutrino mass is as follows
\begin{eqnarray}
\label{37}w_{\nu}=\frac{y_{\nu1}}{\Lambda}((\nu^{c}\ell)_2\eta)_{1_1}h_u+\frac{y_{\nu2}}{\Lambda}((\nu^{c}\ell)_{3_1}\phi)_{1_1}h_u+\frac{1}{2}M(\nu^c\nu^c)_{1_1}
\end{eqnarray}
The neutrino Dirac and Majorana mass matrices can be
straightforwardly read out as
\begin{equation}
\label{38}M_{D}=\left(\begin{array}{ccc}2b&a-b&a-b\\
a-b&a+2b&-b\\
a-b&-b&a+2b\end{array}\right)v_{u},~~~~~M_M=\left(\begin{array}{ccc}
M&0&0\\
0&0&M\\
0&M&0\end{array}\right)
\end{equation}
where we denote $a=y_{\nu1}{V_{\eta}}/{\Lambda}$ and
$b=y_{\nu2}{V_{\phi}}/{\Lambda}$, both of them should be of order
$\lambda^2_c$. The heavy right handed neutrino mass matrix $M_M$ can
be diagonalized as follows
\begin{equation}
\label{39}U^{T}M_MU={\rm diag}(|M|,|M|,|M|)
\end{equation}
It is obvious that the right handed neutrinos are exactly
degenerate, and the unitary matrix $U$ is
\begin{equation}
\label{40}U=\left(\begin{array}{ccc}1&0&0\\
0&\frac{1}{\sqrt{2}}&\frac{1}{\sqrt{2}}\\
0&\frac{1}{\sqrt{2}}&-\frac{1}{\sqrt{2}}
\end{array}\right)U_{\alpha}
\end{equation}
Here $U_{\alpha}={\rm diag}
(e^{-i\alpha/2},e^{-i\alpha/2},ie^{-i\alpha/2})$, and $\alpha={\rm
arg}(M)$ is the complex phase of $M$. The light neutrino mass matrix
is given by the Seesaw relation $m_{\nu}=-M^{T}_DM^{-1}_MM_D$, which
is exactly diagonalized by the TB mixing matrix
\begin{equation}
\label{41}U^{T}_{\nu}m_{\nu}U_{\nu}={\rm diag}(m_1,m_2,m_3)
\end{equation}
The unitary matrix $U_{\nu}$ is written as
\begin{equation}
\label{42}U_{\nu}=U_{TB}\,{\rm
diag}(e^{-i\alpha_1/2},e^{-i\alpha_2/2},e^{-i\alpha_3/2})
\end{equation}
where the phase $\alpha_{1,2,3}$ and the light neutrino masses
$m_{1,2,3}$ are given by
\begin{eqnarray}
\nonumber&&\alpha_1={\rm arg}(-(a-3b)^2/M),~~~\alpha_2={\rm
arg}(-4a^2/M),~~~\alpha_3={\rm arg}((a+3b)^2/M)\\
\label{43}&&m_1=|(a-3b)^2|\frac{v^2_u}{|M|},~~~~~m_2=4|a^2|\frac{v^2_u}{|M|},~~~~~m_3=|(a+3b)^2|\frac{v^2_u}{|M|}
\end{eqnarray}
Similar to the AF model, this model is very predictive, there are
only three independent parameter $a$, $b$ and $M$ in the neutrino
sector. The neutrino mass spectrum can be normal hierarchy or
inverted hierarchy as well. Taking into account the mass square
difference $\Delta m^2_{21}$ and $\Delta m^2_{31}$ measured in the
neutrino oscillation experiments, we obtain the following limit for
the lightest neutrino mass
\begin{eqnarray}
\nonumber&&m_1\geq0.011\;{\rm eV},~~~~~{\rm NH}\\
\label{44}&&m_3>0.0\;{\rm eV},~~~~~~~~{\rm IH}
\end{eqnarray}
In the following, we briefly discuss the NLO corrections, please
read Ref.\cite{Ding:2009iy} for details. After including the higher
dimensional operators allowed by the symmetry at NLO, the
off-diagonal entries of the charged lepton mass matrix become
non-zero and of the order of the diagonal term in each row
multiplied by $\epsilon$, which parameterizes the ratio
$VEV/\Lambda$ with order ${\cal O}(\lambda^2_c)$. Therefore the
charged lepton mass matrix can generally be written as
\begin{equation}
\label{45}m_{\ell}=\left(\begin{array}{ccc}
m^{\ell}_{11}\epsilon^2&m^{\ell}_{12}\epsilon^3&m^{\ell}_{13}\epsilon^3\\
m^{\ell}_{21}\epsilon^2&m^{\ell}_{22}\epsilon&m^{\ell}_{23}\epsilon^2\\
m^{\ell}_{31}\epsilon&m^{\ell}_{32}\epsilon&m^{\ell}_{33}
\end{array}\right)\epsilon v_d
\end{equation}
where the coefficients $m^{\ell}_{ij}$($i,j=$1, 2, 3) are order one
unspecified constants. The matrix $m^{\dagger}_{\ell}m_{\ell}$ is
diagonalized by setting $\ell$ to $U_{\ell}\ell$, where $U_{\ell}$
approximately is given by
\begin{equation}
\label{46}U_{\ell}\simeq\left(\begin{array}{ccc}
1&(\frac{m^{\ell}_{21}}{m^{\ell}_{22}}\varepsilon)^{*}&(\frac{m^{\ell}_{31}}{m^{\ell}_{33}}\varepsilon)^{*}\\
-\frac{m^{\ell}_{21}}{m^{\ell}_{22}}\varepsilon&1&(\frac{m^{\ell}_{32}}{m^{\ell}_{33}}\varepsilon)^{*}\\
-\frac{m^{\ell}_{31}}{m^{\ell}_{33}}\varepsilon&-\frac{m^{\ell}_{32}}{m^{\ell}_{33}}\varepsilon&1
\end{array}\right)
\end{equation}
The NLO correction to the Majorana masses of the right handed
neutrino arises at order $1/\Lambda$, the corresponding higher
dimensional operator is $(\nu^{c}\nu^{c})_{1_1}\zeta^2/\Lambda$,
whose contribution can be completely absorbed into the redefinition
of the mass parameter $M$. Consequently the right handed neutrinos
are highly degenerate, and we only need to consider the NLO
corrections to the neutrino Dirac couplings as follows
\begin{eqnarray}
\label{47}\frac{y_{\nu1}}{\Lambda}(\nu^{c}\ell\delta\eta)_{1_1}h_u+\frac{y_{\nu2}}{\Lambda}(\nu^{c}\ell\delta\phi)_{1_1}h_u+\frac{x_{\nu1}}{\Lambda^2}(\nu^{c}\ell\eta)_{1_2}\zeta
h_u+\frac{x_{\nu2}}{\Lambda^2}(\nu^{c}\ell\phi)_{1_2}\zeta h_u
\end{eqnarray}
A part of the above corrections can be absorbed into the LO results,
then the NLO corrections to $M_D$ can be parameterized as
\begin{equation}
\label{48}\delta M_D=\left(\begin{array}{ccc}
0&\tilde{b}\epsilon&\tilde{a}\epsilon-\tilde{b}\epsilon\\
-\tilde{b}\epsilon&\tilde{a}\epsilon&\tilde{b}\epsilon\\
\tilde{a}\epsilon+\tilde{b}\epsilon&-\tilde{b}\epsilon&0
\end{array}\right)v_u
\end{equation}
where the magnitudes of $\tilde{a}$ and $\tilde{b}$ are expected to
be of the same order as those of $a$ and $b$. In first order of
$\epsilon$, the parameters of the lepton mixing matrix are modified
as
\begin{eqnarray}
\nonumber&&|U_{e3}|=\frac{1}{\sqrt{2}}\Big|\frac{1}{6(|a|^2+9|b|^2)(ab^{*}+a^{*}b)}[(a+3b)^2(a^{*}\tilde{a}^{*}
+6b^{*}\tilde{b}^{*})\epsilon^{*}-(a^{*}-3b^{*})^2(a\tilde{a}+6b\tilde{b})\epsilon]\\
\nonumber&&~~~~+\Big(\frac{m^{\ell}_{21}}{m^{\ell}_{22}}\epsilon\Big)^{*}-\Big(\frac{m^{\ell}_{31}}{m^{\ell}_{33}}\epsilon\Big)^{*}\Big|\\
\nonumber&&\sin^2\theta_{12}=\frac{1}{3}\Big[1-\frac{m^{\ell}_{21}}{m^{\ell}_{22}}\epsilon-\frac{m^{\ell}_{31}}{m^{\ell}_{33}}\epsilon-\Big(\frac{m^{\ell}_{21}}{m^{\ell}_{22}}\epsilon\Big)^{*}-\Big(\frac{m^{\ell}_{31}}{m^{\ell}_{33}}\epsilon\Big)^{*}\Big]\\
\nonumber&&\sin^2\theta_{23}=\frac{1}{2}+\frac{1}{2(|a|^{2}+9|b|^2)(ab^{*}+a^{*}b)}\Big[ab(a^{*}\tilde{a}^{*}+6b^{*}\tilde{b}^{*})\epsilon^{*}+a^{*}b^{*}(a\tilde{a}+6b\tilde{b})\epsilon\Big]\\
\label{49}&&~~~~+\frac{1}{2}\Big[\frac{m^{\ell}_{32}}{m^{\ell}_{33}}\epsilon+\Big(\frac{m^{\ell}_{32}}{m^{\ell}_{33}}\epsilon\Big)^{*}\Big]
\end{eqnarray}
As we can see, NLO contributions introduce corrections of order
$\lambda^2_c$ to all mixing angles, and the reactor angle
$\theta_{13}$ becomes non-zero. Then we turn to a numerical
analysis. Since $a$, $b$, $\tilde{a}$ and $\tilde{b}$ are expected
to be of order $\lambda^2_c$, they are treated as complex numbers
with absolute value between 0.01 and 0.1, the parameters
$m^{\ell}_{21}/m^{\ell}_{22}$, $m^{\ell}_{31}/m^{\ell}_{33}$ and
$m^{\ell}_{32}/m^{\ell}_{33}$ in the charged lepton mixing matrix
$U_{\ell}$ are taken to be complex numbers with absolute value in
the range of 0-2, the heavy neutrino mass $|M|$ is allowed to vary
from $10^{11}$ GeV to $10^{14}$ GeV, and the expansion parameter
$\epsilon$ is set to the indicative value 0.04. The correlation for
$\sin^2\theta_{23}-\sin^2\theta_{13}$ and $J-\sin^2\theta_{13}$ is
showed in Fig.\ref{fig:Ding_NLO}, It is obvious that rather small
$\theta_{13}$ is favored, which is consistent with our theoretical
analysis.

\section{Predictions for LFV in AF model\label{sec:LFVA4}}
We first study the branch ratios of the rare LFV processes at LO of
the model, Since the number of independent parameters is rather
small in this case, these branching ratios are closely related to
the neutrino oscillation parameters. As has been shown in section
\ref{sec:A4_model}, The charged lepton mass matrix is diagonal at
LO, and the right handed neutrino mass matrix $M_M$ is diagonalized
by the unitary matrix $U=U_{TB}U_{\phi}$, consequently in the base
where both the right handed neutrino and charged lepton mass
matrices are real and diagonal, the neutrino Yukawa coupling matrix
is given by
\begin{equation}
\label{50}\hat{\mathbf{Y}}_{\nu}=U^{T}{\mathbf
Y}_{\nu}=y_{\nu}\left(\begin{array}{ccc}\sqrt{\frac{2}{3}}e^{-i\phi_1/2}&-\frac{1}{\sqrt{6}}e^{-i\phi_1/2}&-\frac{1}{\sqrt{6}}e^{-i\phi_1/2}\\
\frac{1}{\sqrt{3}}e^{-i\phi_2/2}&\frac{1}{\sqrt{3}}e^{-i\phi_2/2}&\frac{1}{\sqrt{3}}e^{-i\phi_2/2}\\
0&-\frac{1}{\sqrt{2}}e^{-i\phi_3/2}&\frac{1}{\sqrt{2}}e^{-i\phi_3/2}\end{array}\right)
\end{equation}
As a result we can straightforwardly obtain the off-diagonal
elements of $\hat{\mathbf{Y}}^{\dagger}_{\nu}{\mathbf
L}\hat{\mathbf{Y}}_{\nu}$, which is directly related to the LFV
branching ratios via Eqs.(\ref{8}), (\ref{10}), (\ref{12}) and
(\ref{13})
\begin{eqnarray}
\nonumber&&(\hat{\mathbf{Y}}^{\dagger}_{\nu}{\mathbf
L}\hat{\mathbf{Y}}_{\nu})_{12}=(\hat{\mathbf{Y}}^{\dagger}_{\nu}{\mathbf
L}\hat{\mathbf{Y}}_{\nu})_{21}=\frac{1}{3}y^2_{\nu}\ln\frac{m_2}{m_1}\\
\nonumber&&(\hat{\mathbf{Y}}^{\dagger}_{\nu}{\mathbf
L}\hat{\mathbf{Y}}_{\nu})_{13}=(\hat{\mathbf{Y}}^{\dagger}_{\nu}{\mathbf
L}\hat{\mathbf{Y}}_{\nu})_{31}=\frac{1}{3}y^2_{\nu}\ln\frac{m_2}{m_1}\\
\label{51}&&(\hat{\mathbf{Y}}^{\dagger}_{\nu}{\mathbf
L}\hat{\mathbf{Y}}_{\nu})_{23}=(\hat{\mathbf{Y}}^{\dagger}_{\nu}{\mathbf
L}\hat{\mathbf{Y}}_{\nu})_{32}=\frac{1}{6}y^2_{\nu}\ln\frac{m_1m^2_2}{m^3_3}
\end{eqnarray}
Obviously the LFV processes are tightly related to the light
neutrino mass, and the branching ratio is proportional to
$y^4_{\nu}$. We notice that the branching ratio is independent of
the grand unification scale $M_G$, and it tends to zero if the
neutrino mass spectrum is degenerate. Whereas the branching ratio
could become considerable large if the neutrino mass spectrum is
strongly normal hierarchy or strongly inverted hierarchy. It is
remarkable that we have $(\hat{\mathbf{Y}}^{\dagger}_{\nu}{\mathbf
L}\hat{\mathbf{Y}}_{\nu})_{12}=(\hat{\mathbf{Y}}^{\dagger}_{\nu}{\mathbf
L}\hat{\mathbf{Y}}_{\nu})_{13}$, which is related to the $\mu-\tau$
symmetry of the light neutrino mass matrix. As a result, the ratio
of the branching ratios for $\tau\rightarrow e\gamma$ and
$\mu\rightarrow e\gamma$ approximately is
\begin{equation}
\label{52}\frac{Br(\tau\rightarrow e\gamma)}{Br(\mu\rightarrow
e\gamma)}\simeq Br(\tau\rightarrow
e\bar{\nu}_{e}\nu_{\tau})\simeq17.84\%
\end{equation}
This ratio is almost independent of the SUSY breaking parameters.
Given the present experimental bound $Br(\mu\rightarrow
e\gamma)<1.2\times10^{-11}$ \cite{Amsler:2008zzb}, this result
implies that $\tau\rightarrow e\gamma$ has rates much below the
present and expected future sensitivity \cite{Amsler:2008zzb}. We
note that this ratio would be corrected drastically by the NLO
contributions. For the IH neutrino mass spectrum, we have
\begin{eqnarray}
\nonumber&&m_1=\sqrt{m^2_3+|\Delta m^2_{31}|},
~~m_2=\sqrt{m^2_3+|\Delta m^2_{31}|+\Delta m^2_{21}}, \\
\label{53}&&\ln\frac{m_2}{m_1}=\frac{1}{2}\ln(\frac{m^2_3+|\Delta
m^2_{31}|+\Delta m^2_{21}}{m^2_3+|\Delta
m^2_{31}|})\simeq\frac{\Delta m^2_{21}}{2(m^2_3+|\Delta m^2_{31}|)}
\end{eqnarray}
As a result, in the case of IH spectrum, the LFV processes
$\mu\rightarrow e\gamma$, $\tau\rightarrow e\gamma$,
$\mu\rightarrow3e$, $\tau\rightarrow3e$ and $\mu-e$ conversion in
nuclein generally have rather small branching ratios at LO, which
should be smaller than those of the NH case. While for
$\tau\rightarrow\mu\gamma$ and $\tau\rightarrow3\mu$, the branching
ratios for NH and IH spectrum should be approximately of the same
order.

The analytical results in Eq.({\ref{51}) allow us to estimate the
branching ratios of the LFV processes via the formula
Eq.(\ref{10})-Eq.(\ref{13}). For definiteness, we will present our
results only for the well-known mSUGRA point SPS3
\cite{Allanach:2002nj}, which is taken as the reference example. The
SPS3 point is in the co-annihilation region for the SUSY dark
matter, and the values of the universal SSB parameters are as
follows
\begin{equation}
\label{54}m_0=90 \,{\rm GeV},~~M_{1/2}=400\, {\rm GeV},~~A_0=0\,
{\rm GeV},~~~\tan\beta=10
\end{equation}
Since no suppression is expected for parameter that is unrelated to
the breaking of the flavor symmetry, the coupling $y_{\nu}$ should
be of order 1. We set $y_{\nu}$ to be equal to 0.5 in the following
numerical analysis, the corresponding predictions for
$Br(\ell_i\rightarrow\ell_j\gamma)$, $Br(\ell_i\rightarrow3\ell_j)$,
$CR(\mu-e,Al)$ and $CR(\mu-e,Ti)$ are shown in
Fig.\ref{fig:AF_NH_LO} and Fig.\ref{fig:AF_IH_LO} for NH and IH
spectrum respectively. The uncertainties of both $\Delta m^2_{21}$
and $\Delta m^2_{31}$ are taken into account, they are allowed to
vary within their $3\sigma$ allowed range in Eq.(\ref{1}). We can
clearly see that the LFV processes $\tau\rightarrow e\gamma$,
$\tau\rightarrow\mu\gamma$, $\tau\rightarrow 3e$ and
$\tau\rightarrow 3\mu$ are below the present and future experimental
precision at $B$ factory for both the NH and IH spectrum. As has
been stressed above, the predictions of $Br(\mu\rightarrow
e\gamma)$, $Br(\tau\rightarrow e\gamma)$, $Br(\mu\rightarrow3e)$,
$Br(\tau\rightarrow3e)$, $CR(\mu-e,Al)$ and $CR(\mu-e,Ti)$ for IH
are really much smaller than those of NH case, the rates of
$\tau\rightarrow\mu\gamma$ and $\tau\rightarrow3\mu$ for NH and IH
roughly have the same order.

\begin{figure}[hptb]
\begin{center}
\begin{tabular}{c}
\includegraphics[scale=1,width=3.75cm]{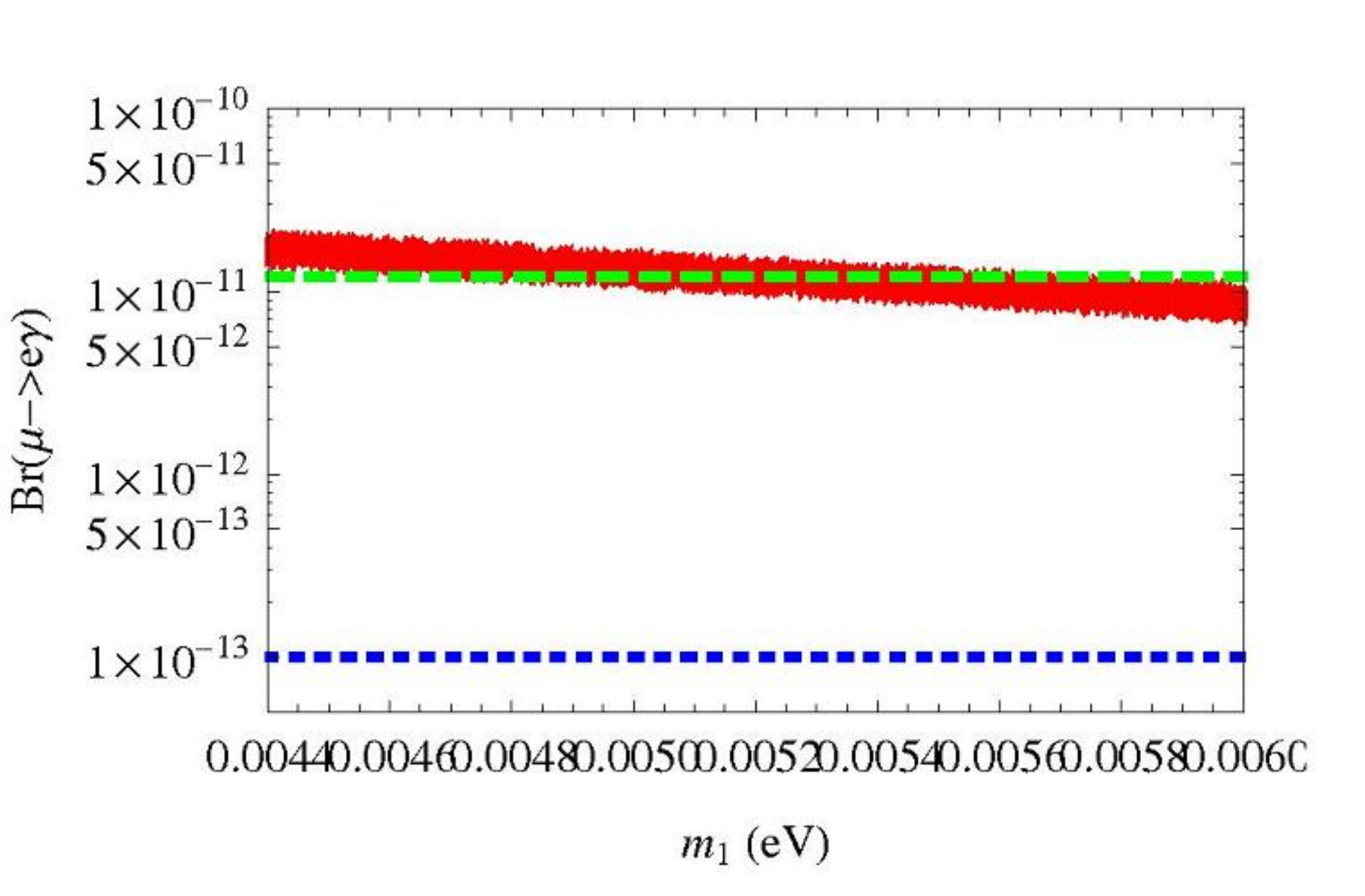}
\includegraphics[scale=1,width=3.75cm]{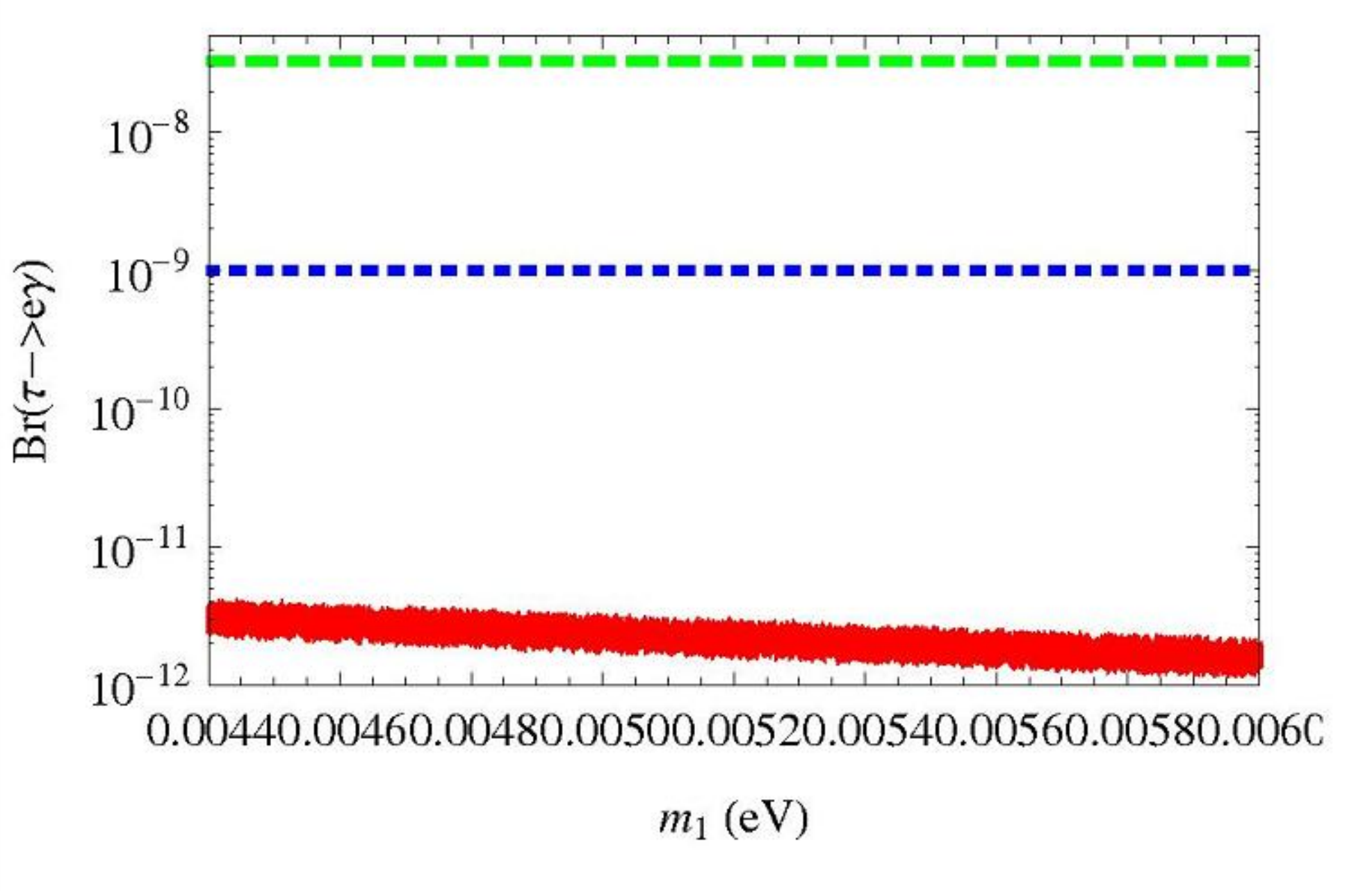}
\includegraphics[scale=1,width=3.75cm]{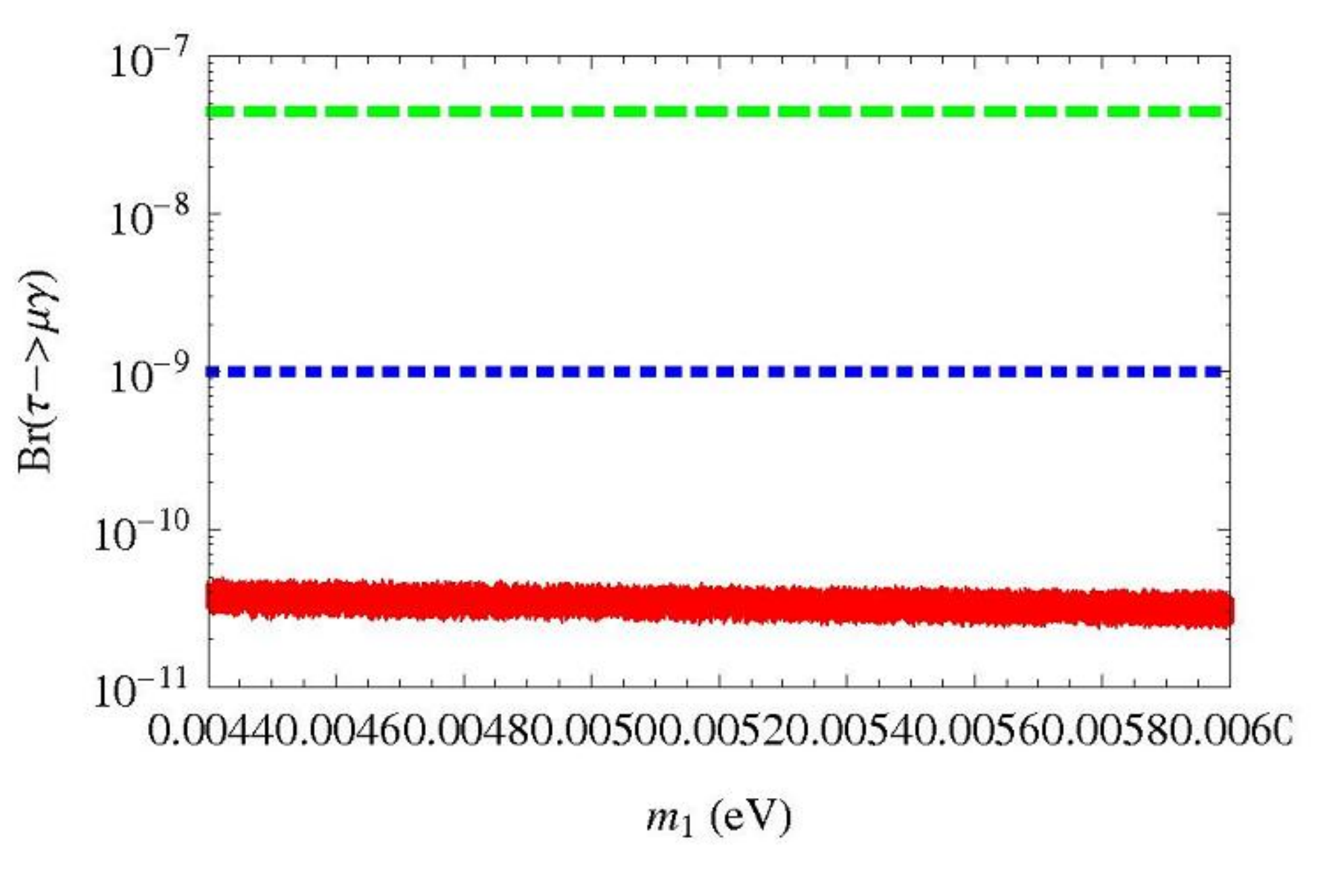}
\includegraphics[scale=1,width=3.75cm]{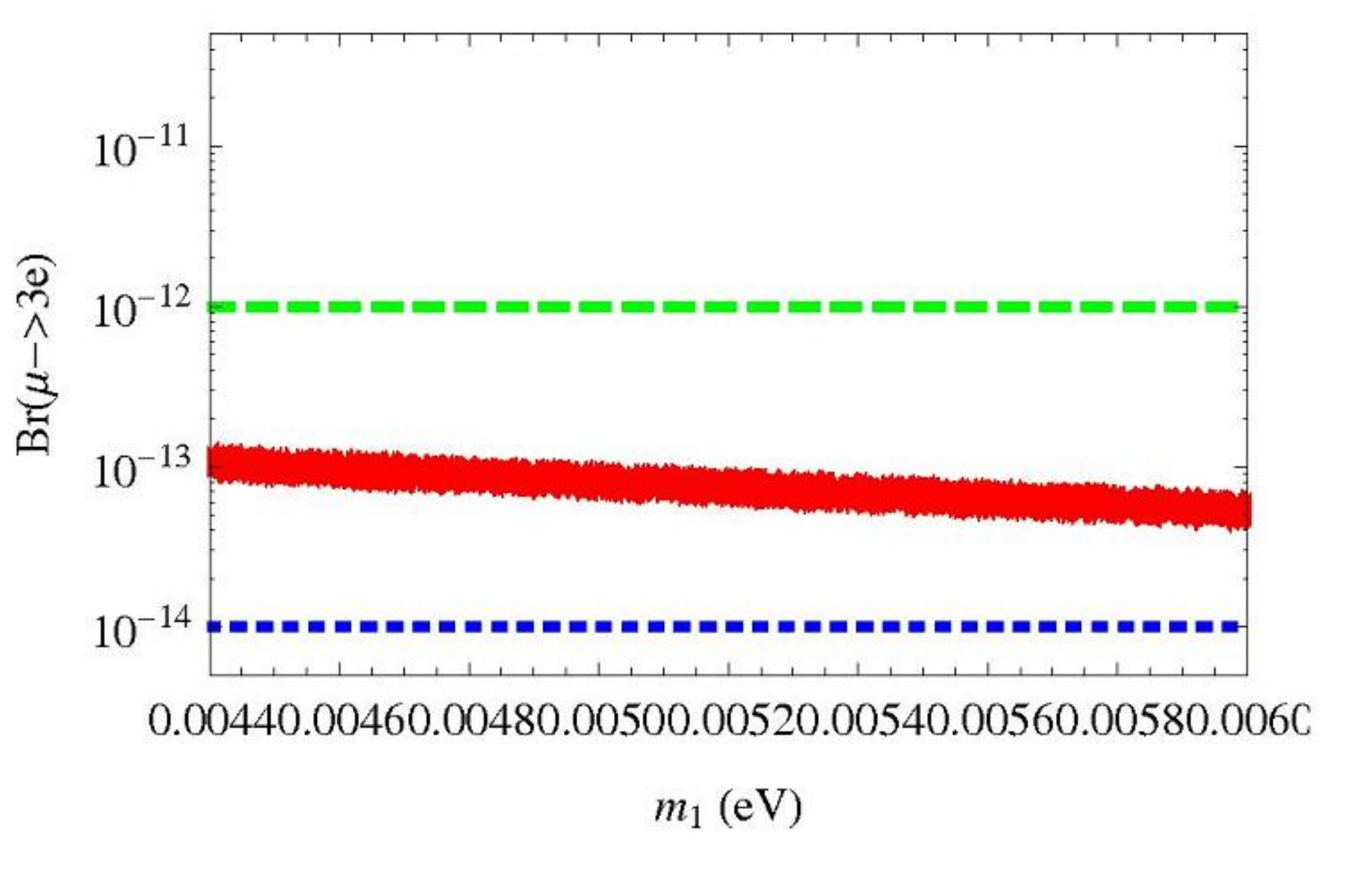}\\\\
\includegraphics[scale=1,width=3.75cm]{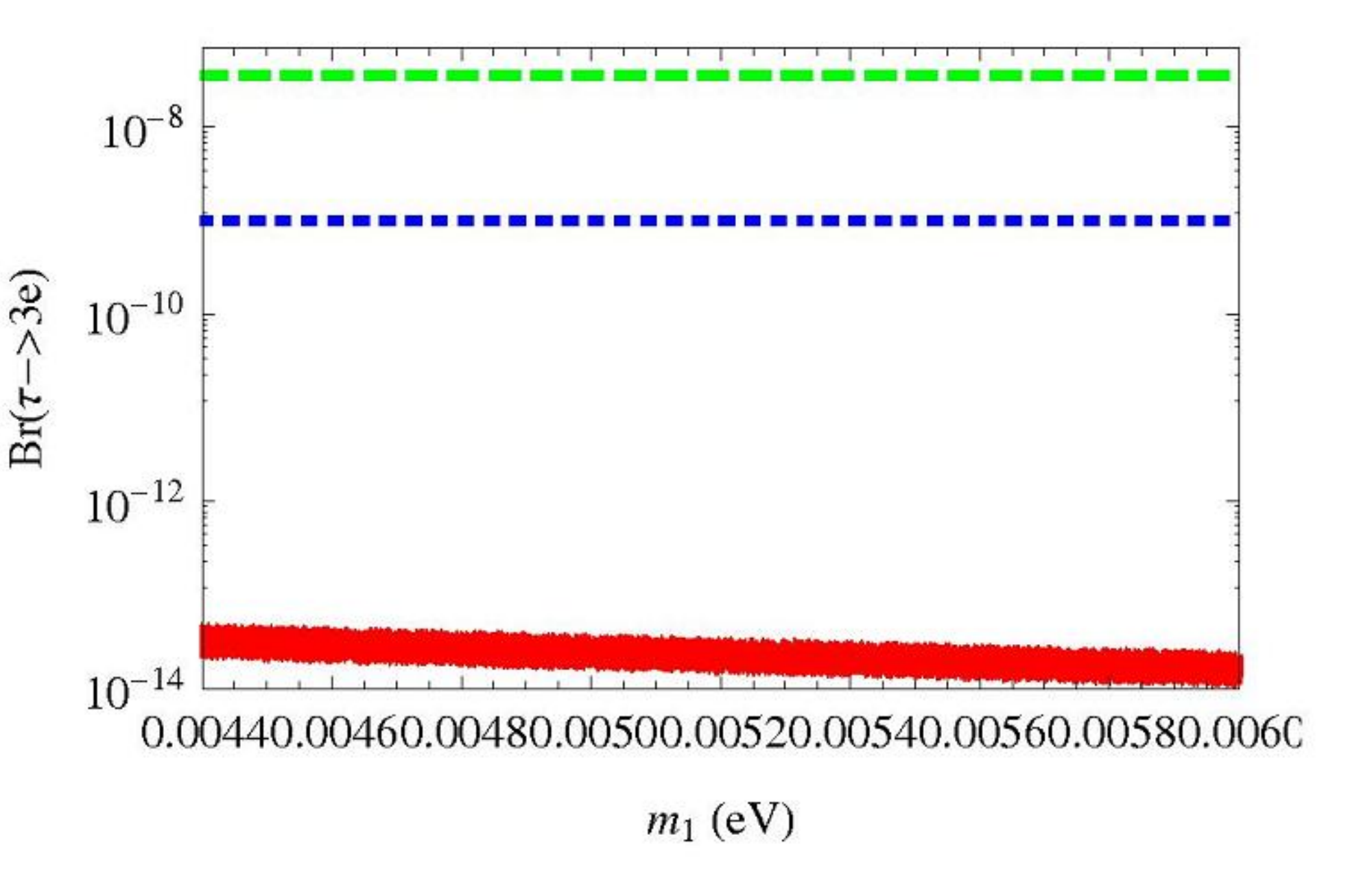}
\includegraphics[scale=1,width=3.75cm]{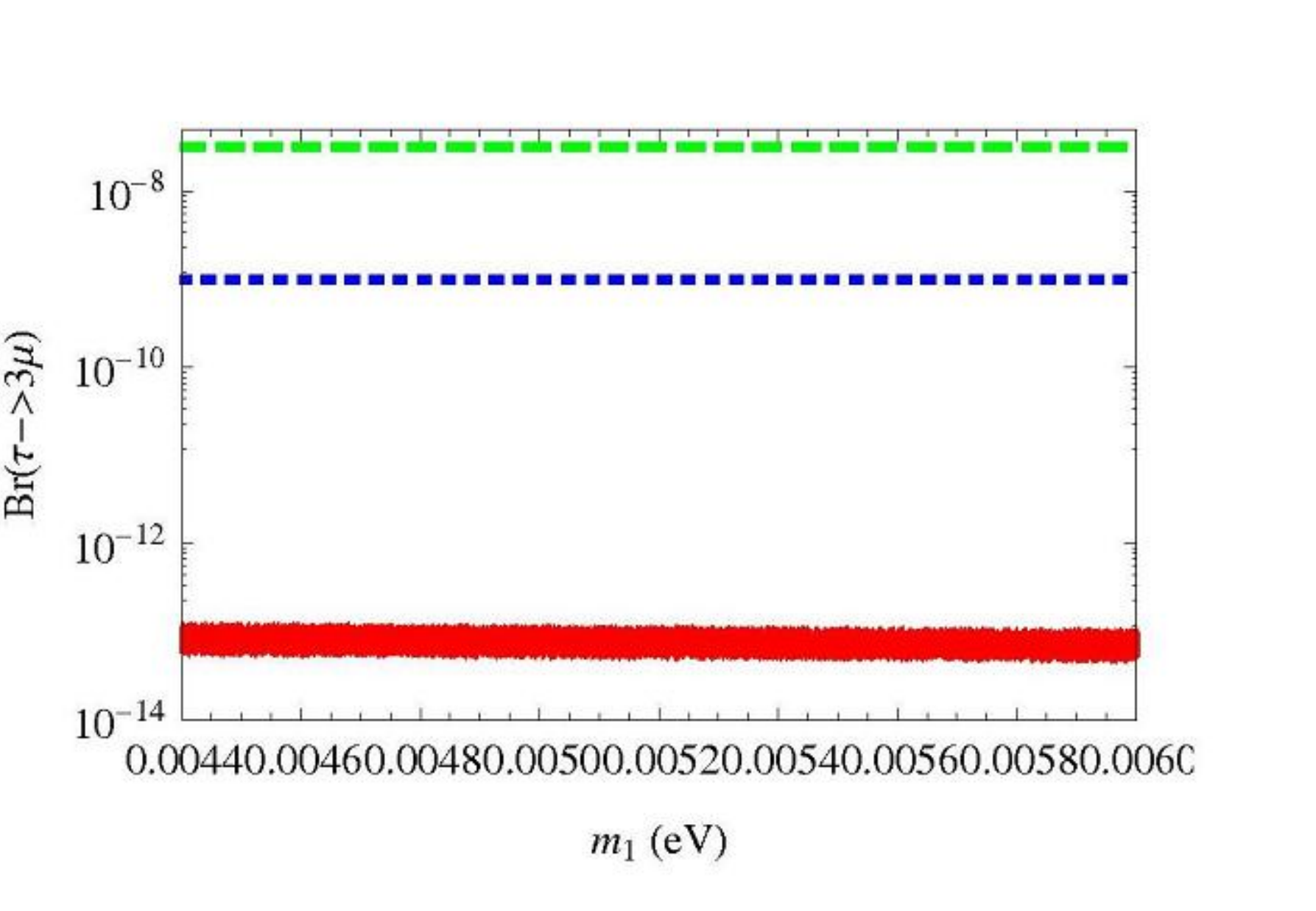}
\includegraphics[scale=1,width=3.75cm]{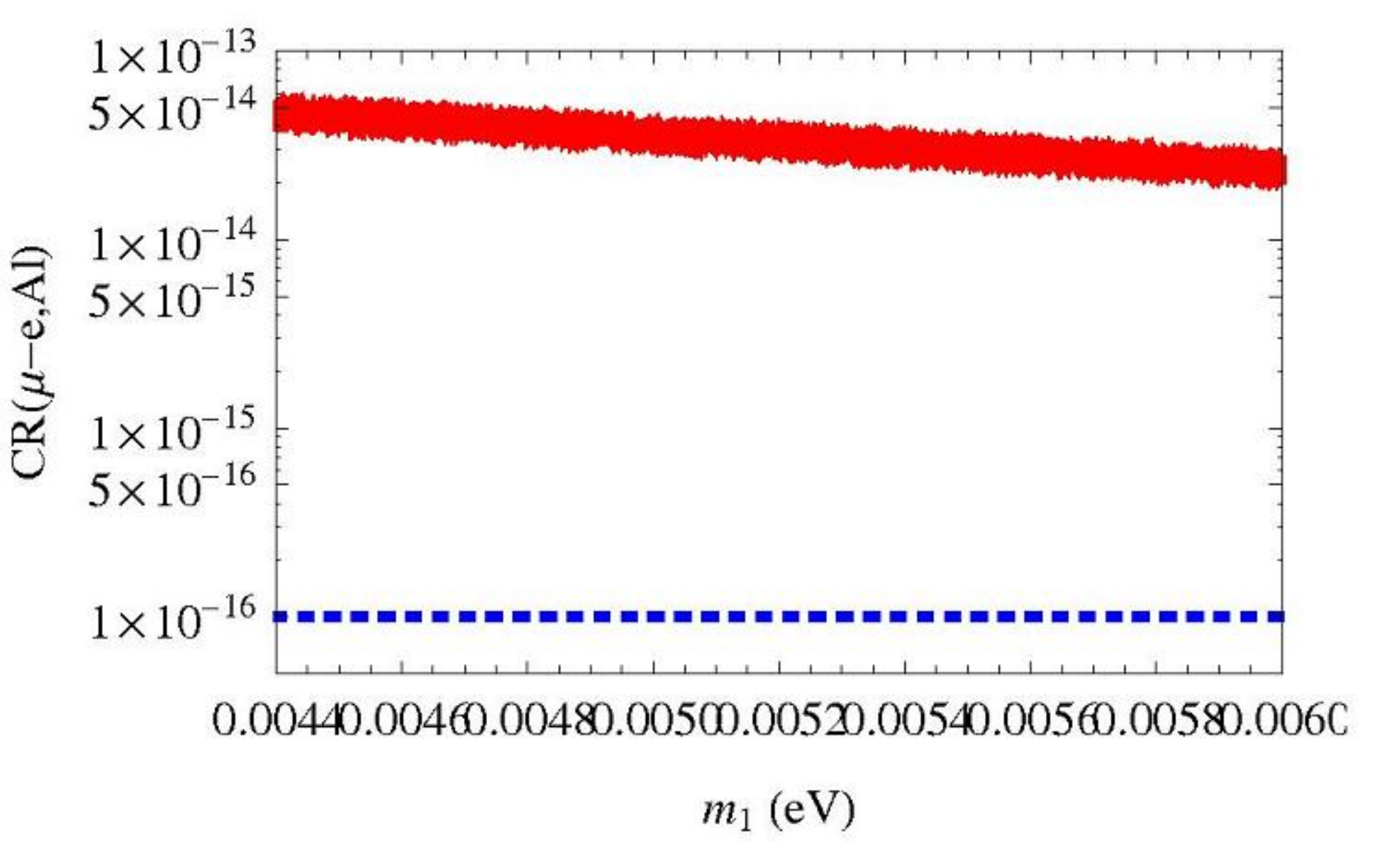}
\includegraphics[scale=1,width=3.75cm]{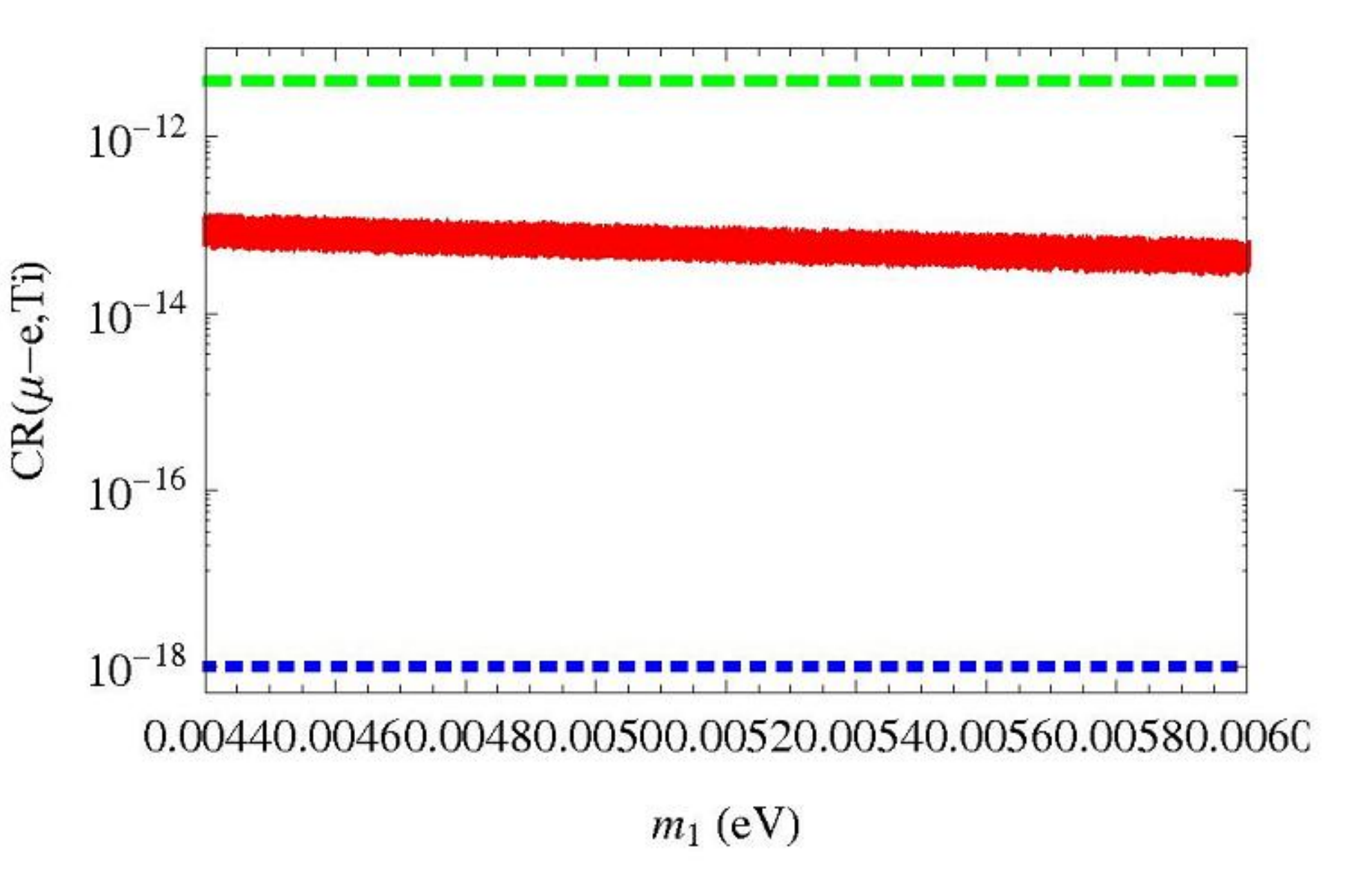}
\end{tabular}
\caption{\label{fig:AF_NH_LO}Dependence of
$Br(\ell_i\rightarrow\ell_j\gamma)$, $Br(\ell_i\rightarrow3\ell_j)$,
$CR(\mu-e,Al)$ and $CR(\mu-e,Ti)$ on the lightest neutrino mass
$m_1$ in AF model for the normal hierarchy spectrum. The bands have
been obtained by varying $\Delta m^2_{21}$ and $\Delta m^2_{31}$ in
their $3\sigma$ experimental range. The dashed and dotted lines
represent the present and future experimental sensitivity
respectively. There is still no upper bound for $CR(\mu-e,Al)$ so
far.}
\end{center}
\end{figure}

\begin{figure}[hptb]
\begin{center}
\begin{tabular}{c}
\includegraphics[scale=1,width=3.75cm]{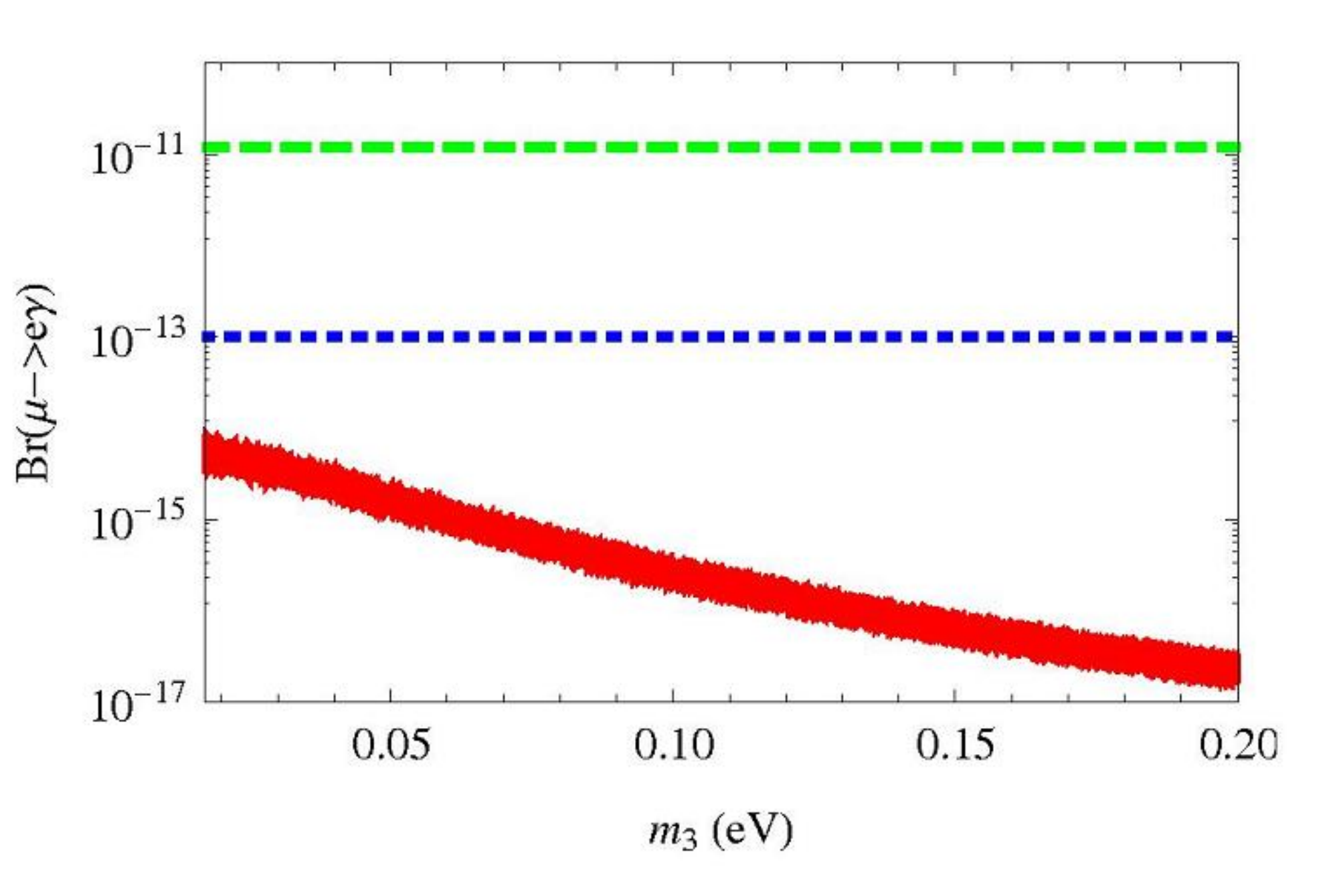}
\includegraphics[scale=1,width=3.75cm]{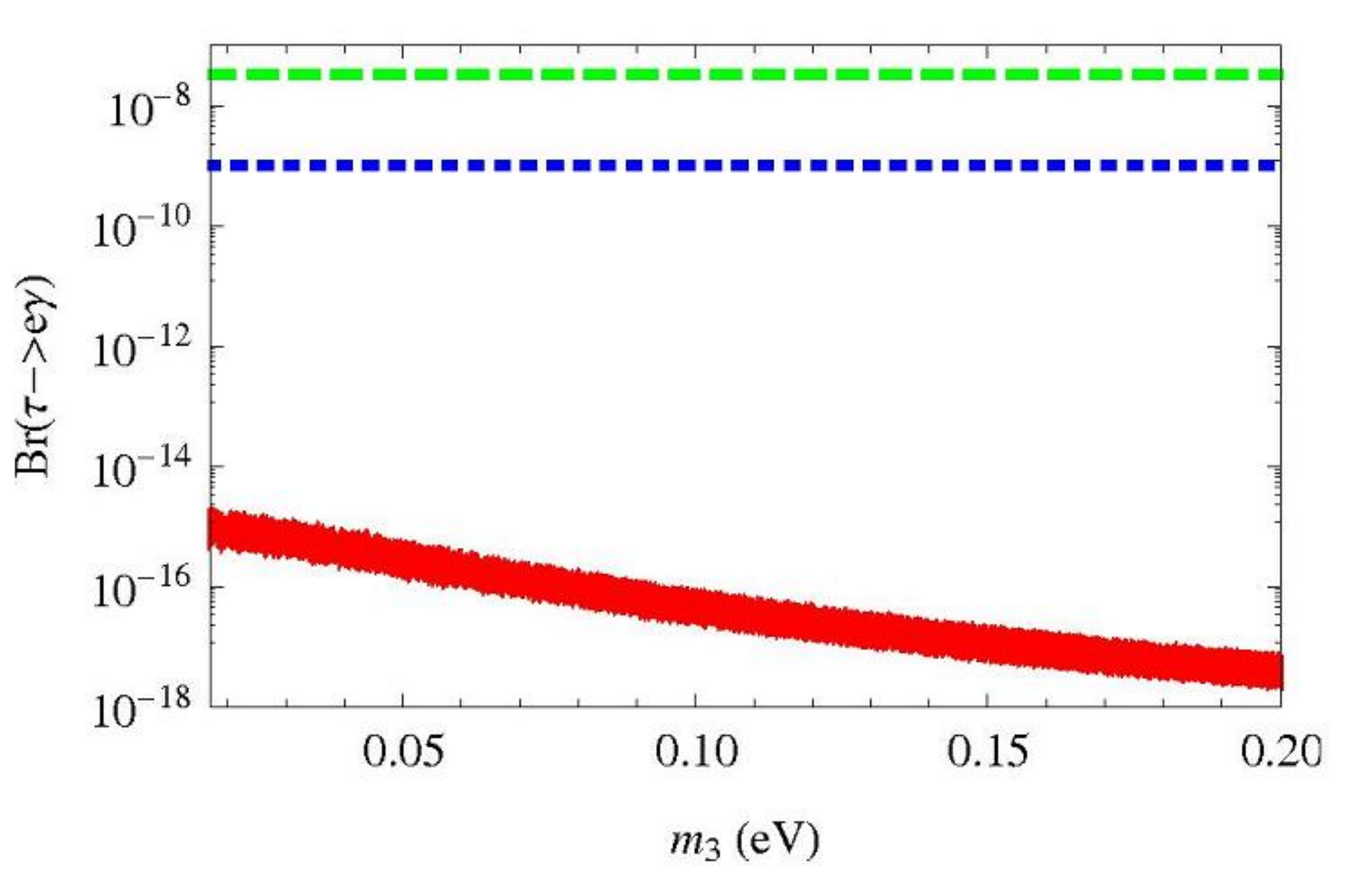}
\includegraphics[scale=1,width=3.75cm]{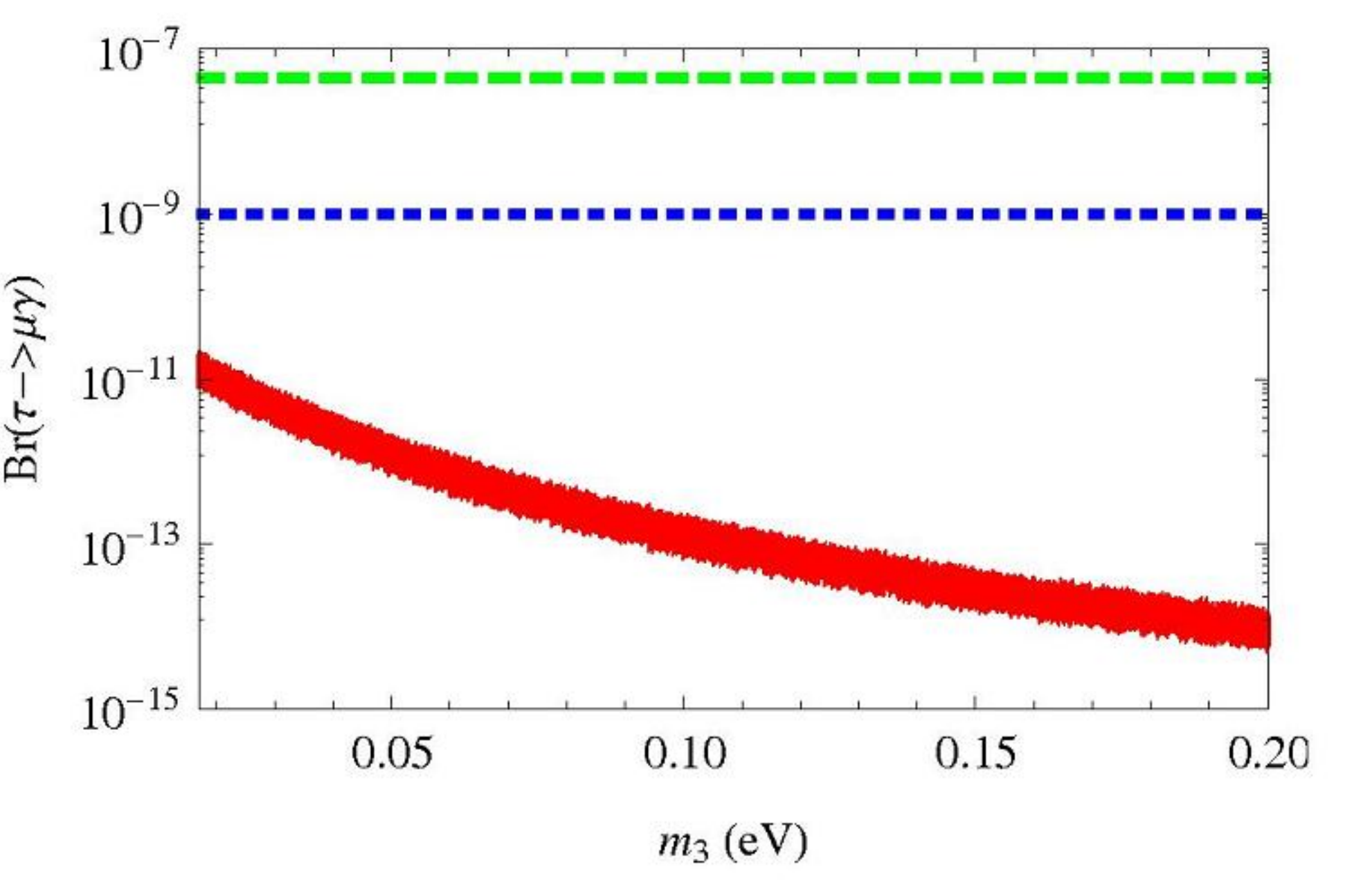}
\includegraphics[scale=1,width=3.75cm]{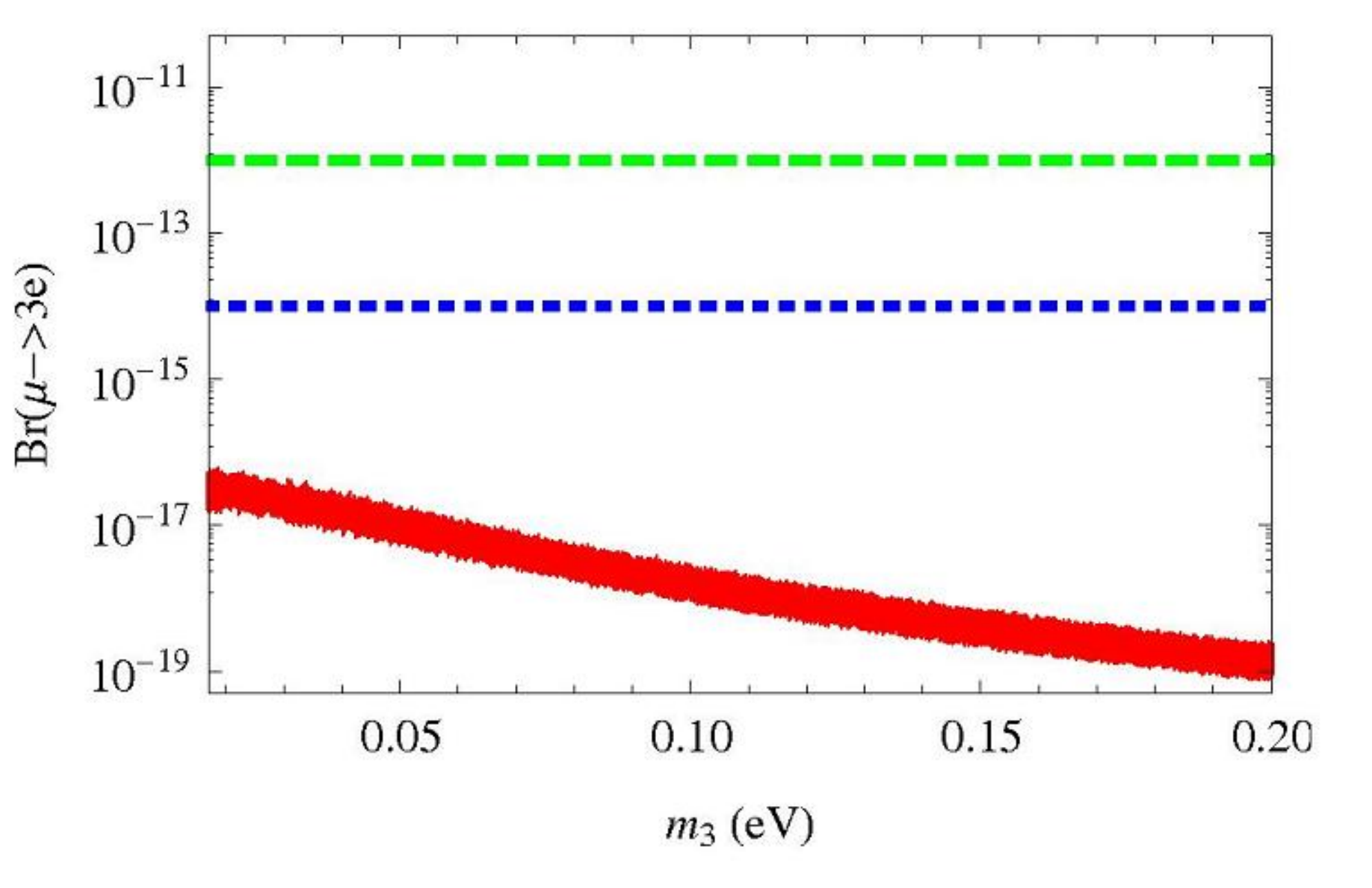}\\\\
\includegraphics[scale=1,width=3.75cm]{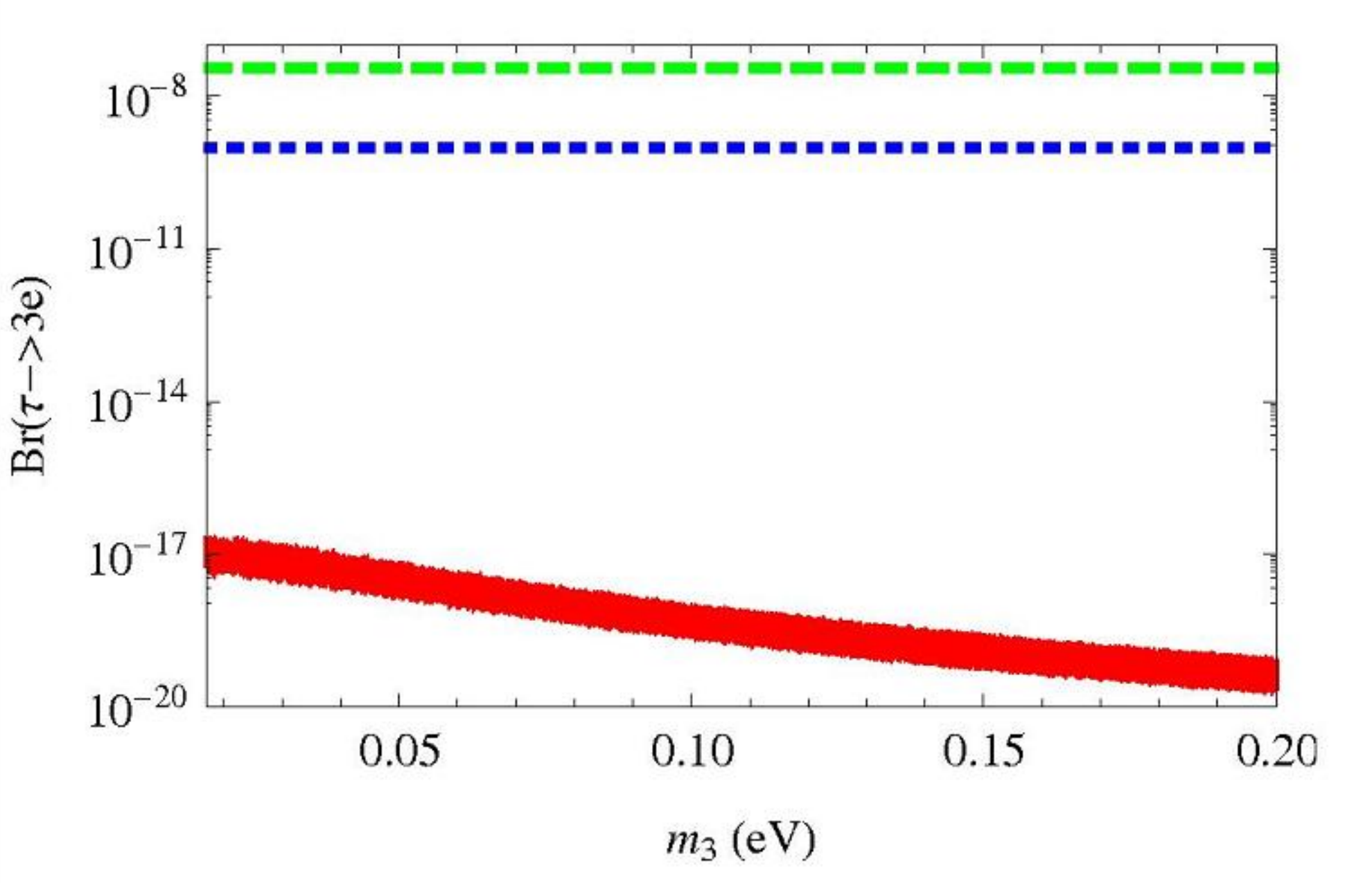}
\includegraphics[scale=1,width=3.75cm]{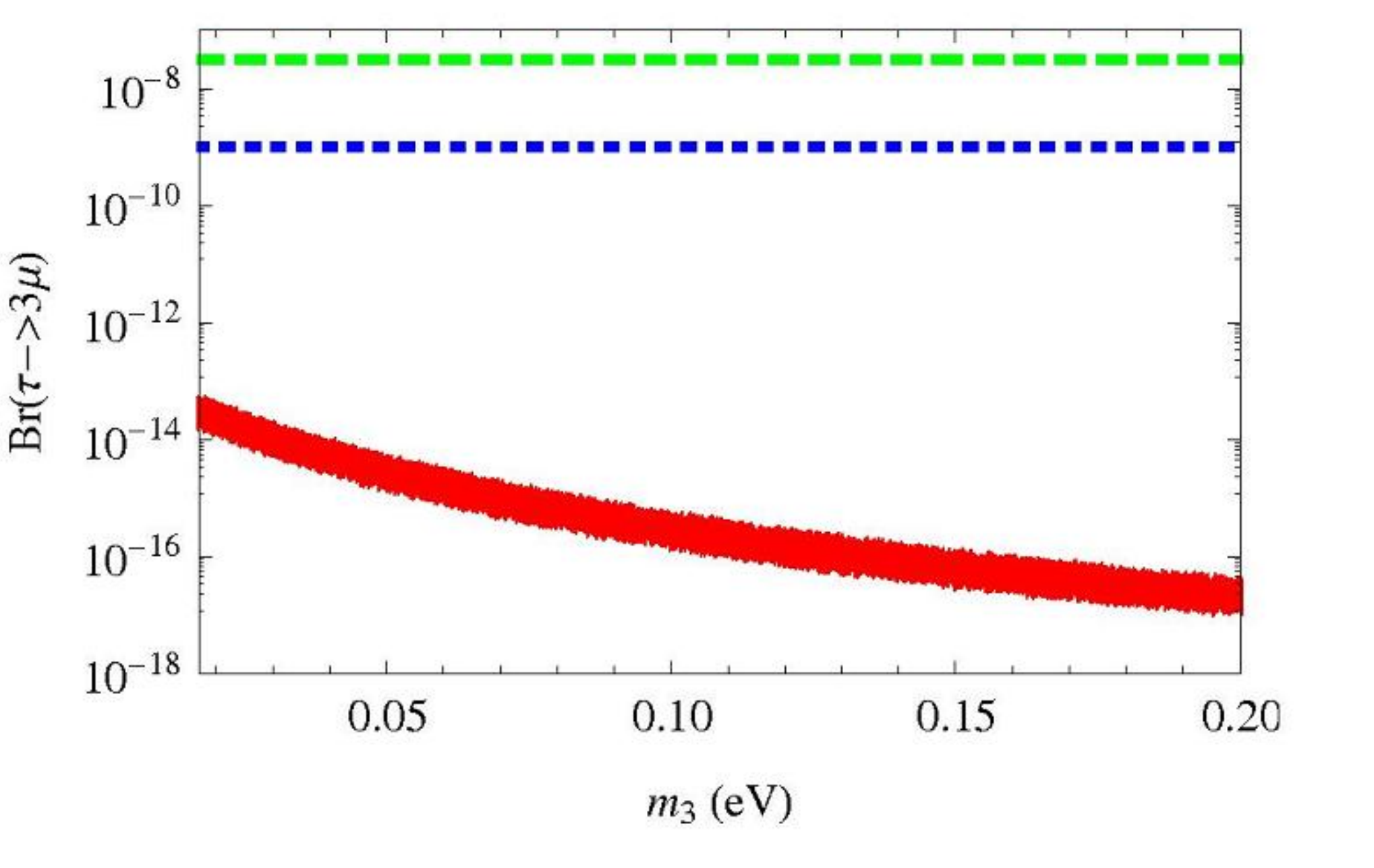}
\includegraphics[scale=1,width=3.75cm]{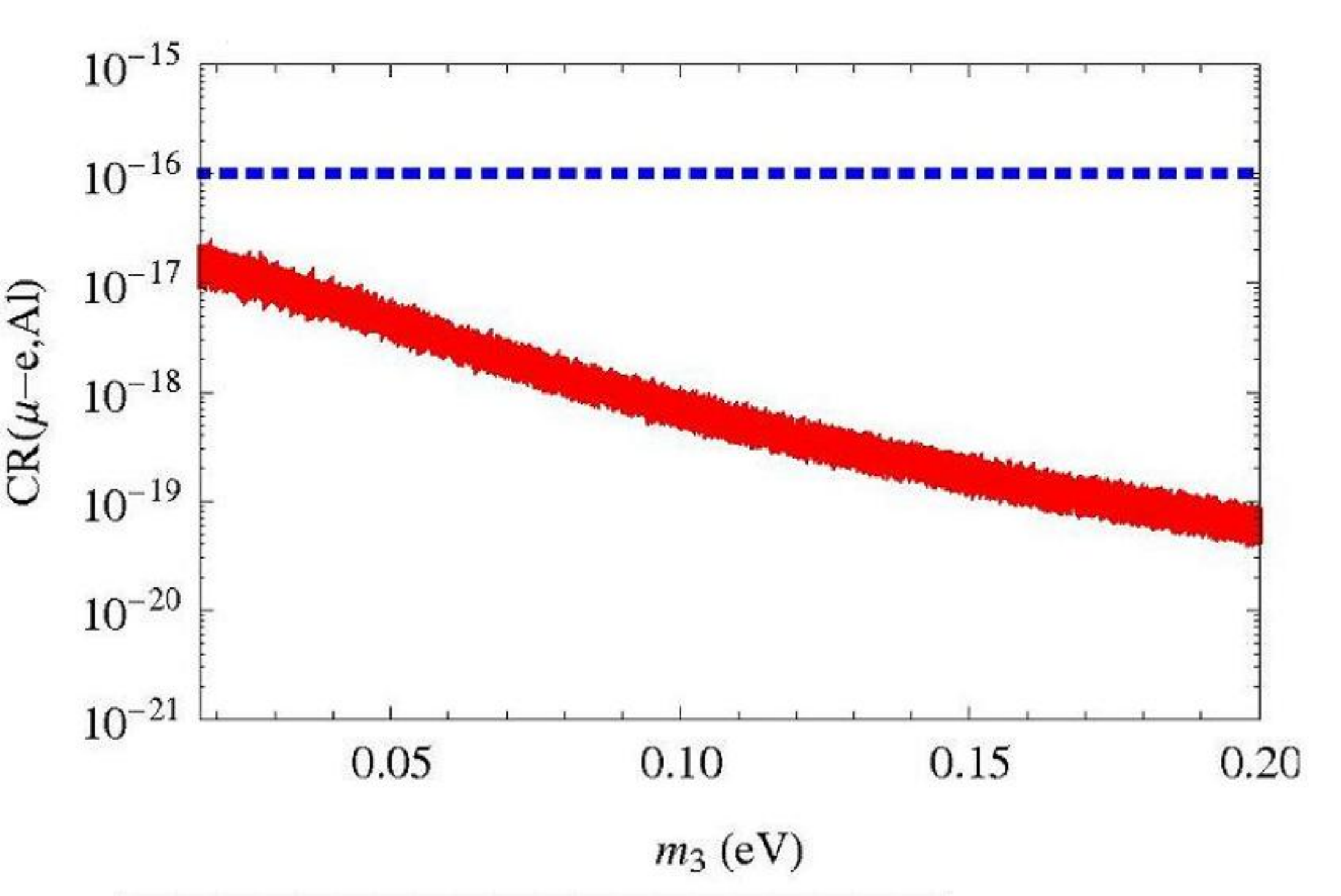}
\includegraphics[scale=1,width=3.75cm]{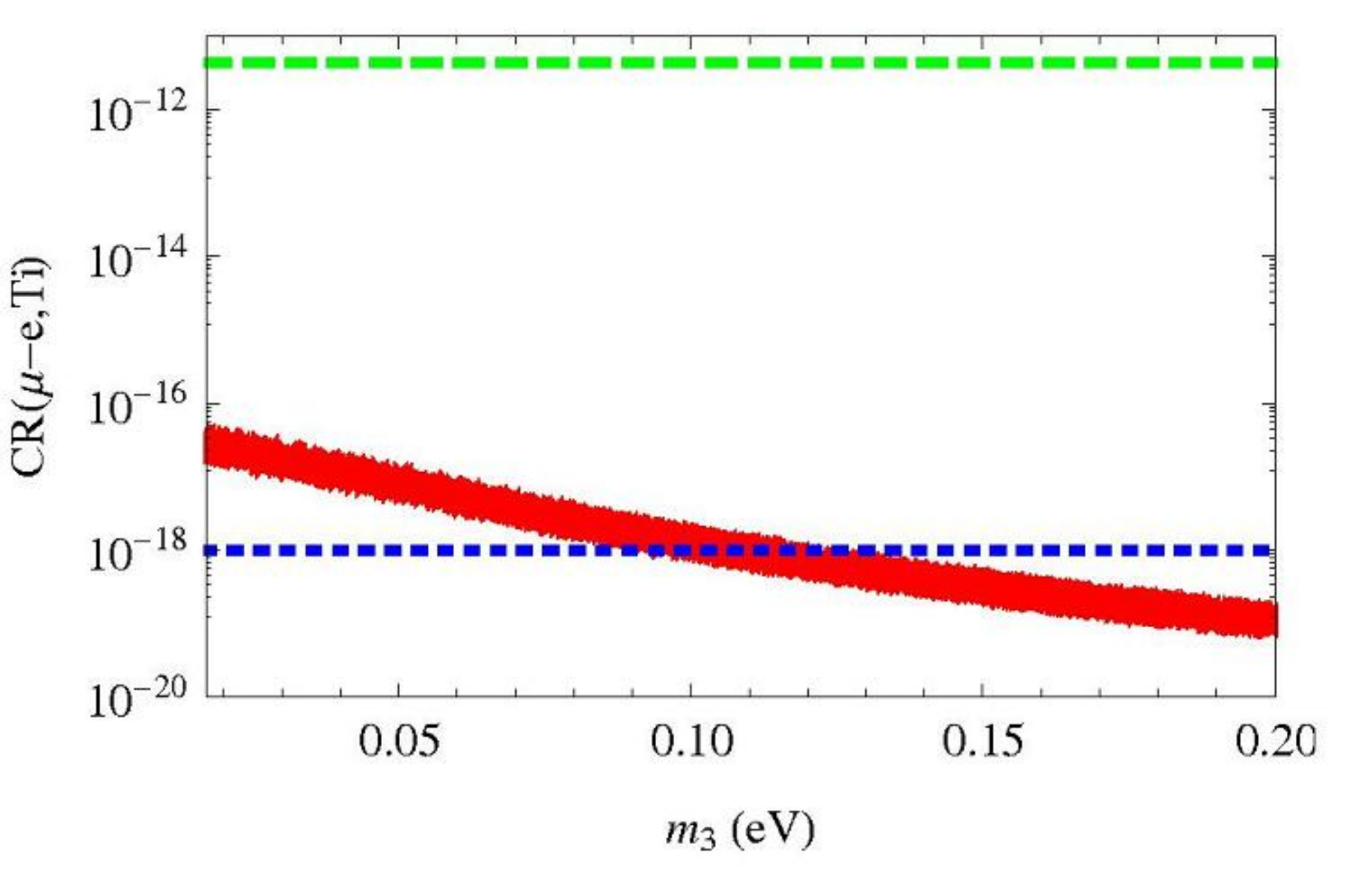}
\end{tabular}
\caption{\label{fig:AF_IH_LO}Dependence of
$Br(\ell_i\rightarrow\ell_j\gamma)$, $Br(\ell_i\rightarrow3\ell_j)$,
$CR(\mu-e,Al)$ and $CR(\mu-e,Ti)$ on the lightest neutrino mass
$m_3$ in AF model for the inverted hierarchy spectrum. The bands
have been obtained by varying $\Delta m^2_{21}$ and $\Delta
m^2_{31}$ in their $3\sigma$ experimental range. The dashed and
dotted lines represent the present and future experimental
sensitivity respectively.}
\end{center}
\end{figure}

For the NH mass spectrum, numerical calculations demonstrate that
the ratio between $Br(\tau\rightarrow e\gamma)$ and
$Br(\mu\rightarrow e\gamma)$ indeed is $17.84\%$, and
$Br(\tau\rightarrow\mu\gamma)$ is about 2-4 times as large as
$Br(\mu\rightarrow e\gamma)$. Therefore we conclude that the
branching ratios of $\mu\rightarrow e\gamma$, $\tau\rightarrow
e\gamma$ and $\tau\rightarrow \mu\gamma$ roughly are of the same
order, the same result has been obtained in Ref.
\cite{Feruglio:2008ht}. It is notable that the current experiment
limit $Br(\mu\rightarrow e\gamma)<1.2\times10^{-11}$ already
constrains the model strongly, the lightest neutrino mass $m_1$ very
close to the lower bound 0.0044 eV is disfavored. The rates of
$\mu\rightarrow3e$, $\mu-e$ conversion in Al and $\mu-e$ conversion
in Ti are below the present upper bound, while they are above the
expected future experimental sensitivity, consequently it is very
promising to detect these three rare processes in near future. If
this turns out to be true, the parameter space of the model would be
strongly constrained. We plot the constraint on the parameters $m_1$
and $y_{\nu}$ imposed by the observation of $\mu\rightarrow
e\gamma$, $\mu\rightarrow3e$ and $\mu-e$ conversion in Al and Ti in
Fig.\ref{fig:AF_NH_CS}. It is obvious that the branching ratios of
these LFV processes are more sensitive to the parameter $y_{\nu}$
than $m_1$. For $y_{\nu}$ of order 1, we should be able to observe
$\mu\rightarrow e\gamma$, $\mu\rightarrow3e$ and $\mu-e$ conversion
in Al and Ti at next generation experiments in the case of NH
spectrum, in particularly the $\mu-e$ conversion processes, which is
an even more robust test to the AF model.

\begin{figure}[hptb]
\begin{center}
\begin{tabular}{c}
\includegraphics[scale=1,width=5.0cm]{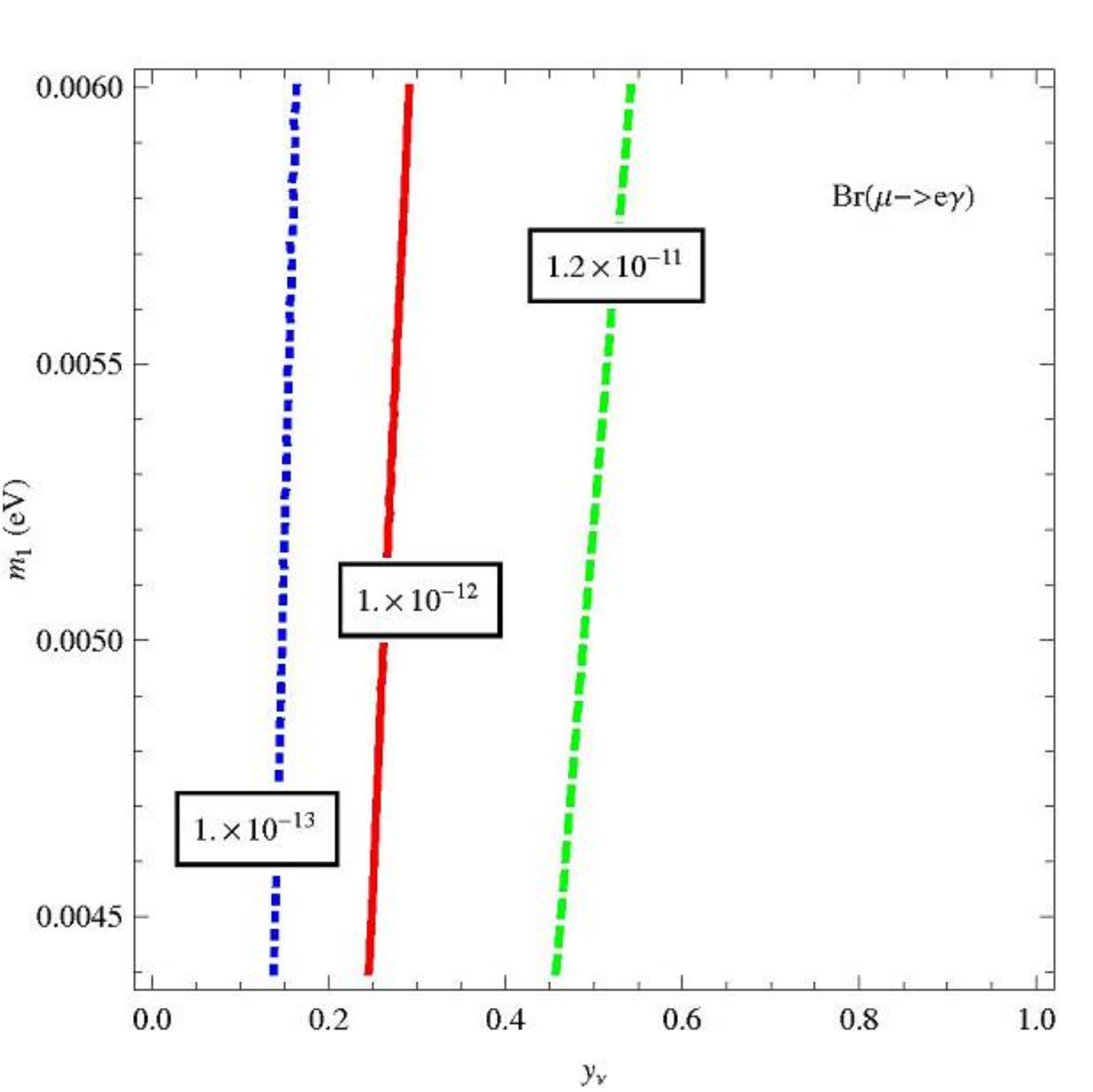}
\hspace{1cm}
\includegraphics[scale=1,width=5.0cm]{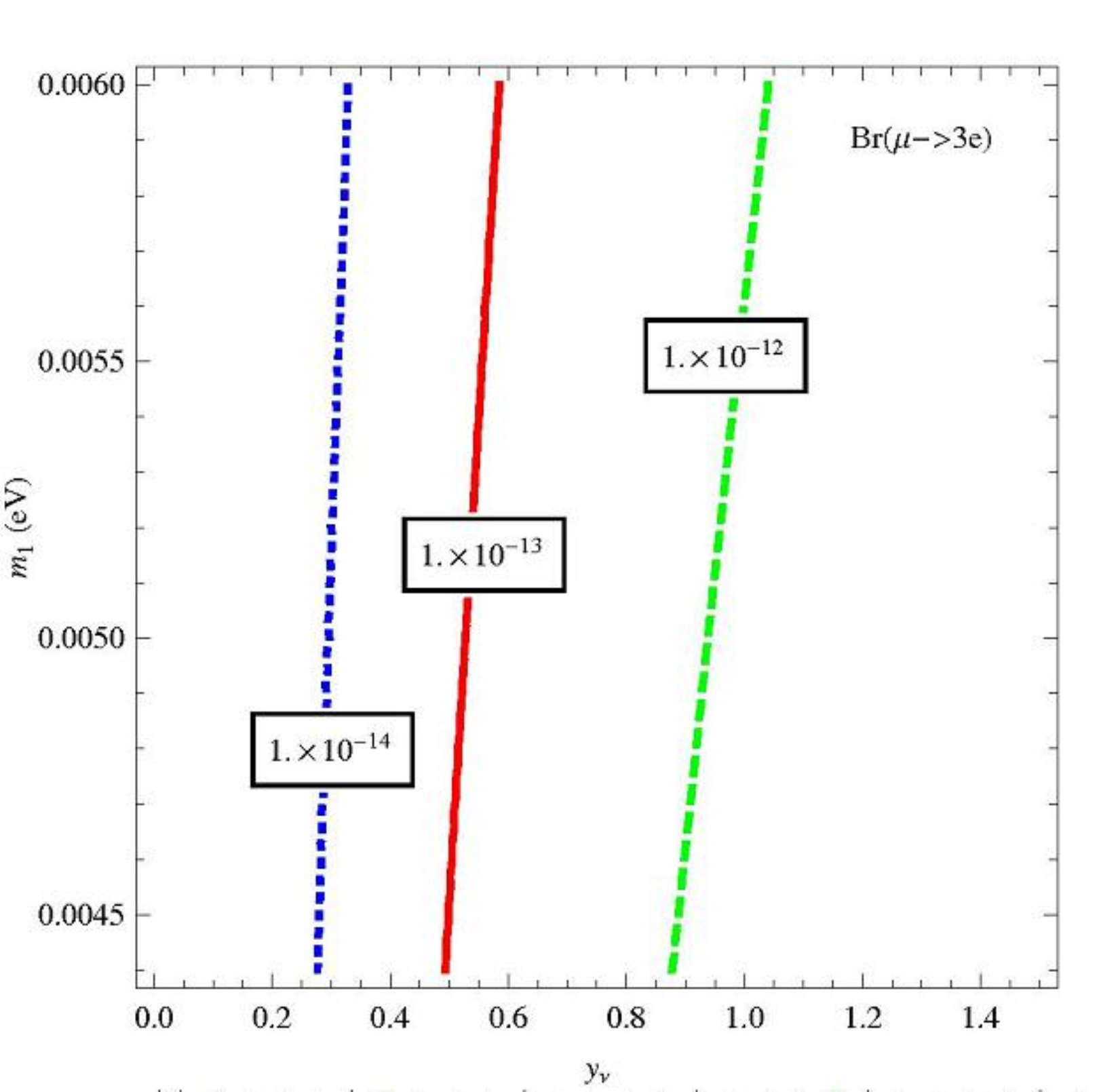}\\\\
\includegraphics[scale=1,width=5.0cm]{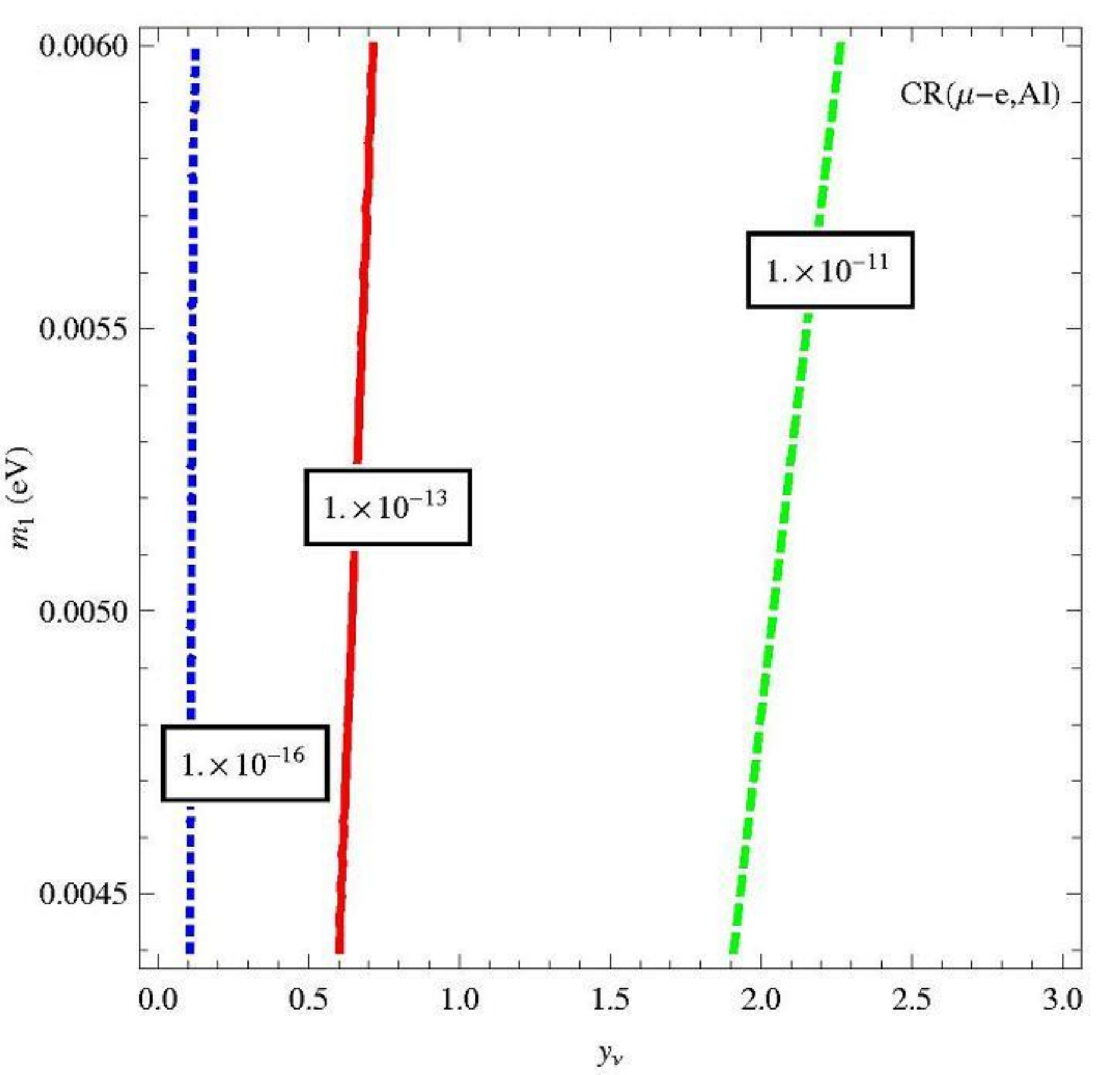}
\hspace{1cm}
\includegraphics[scale=1,width=5.0cm]{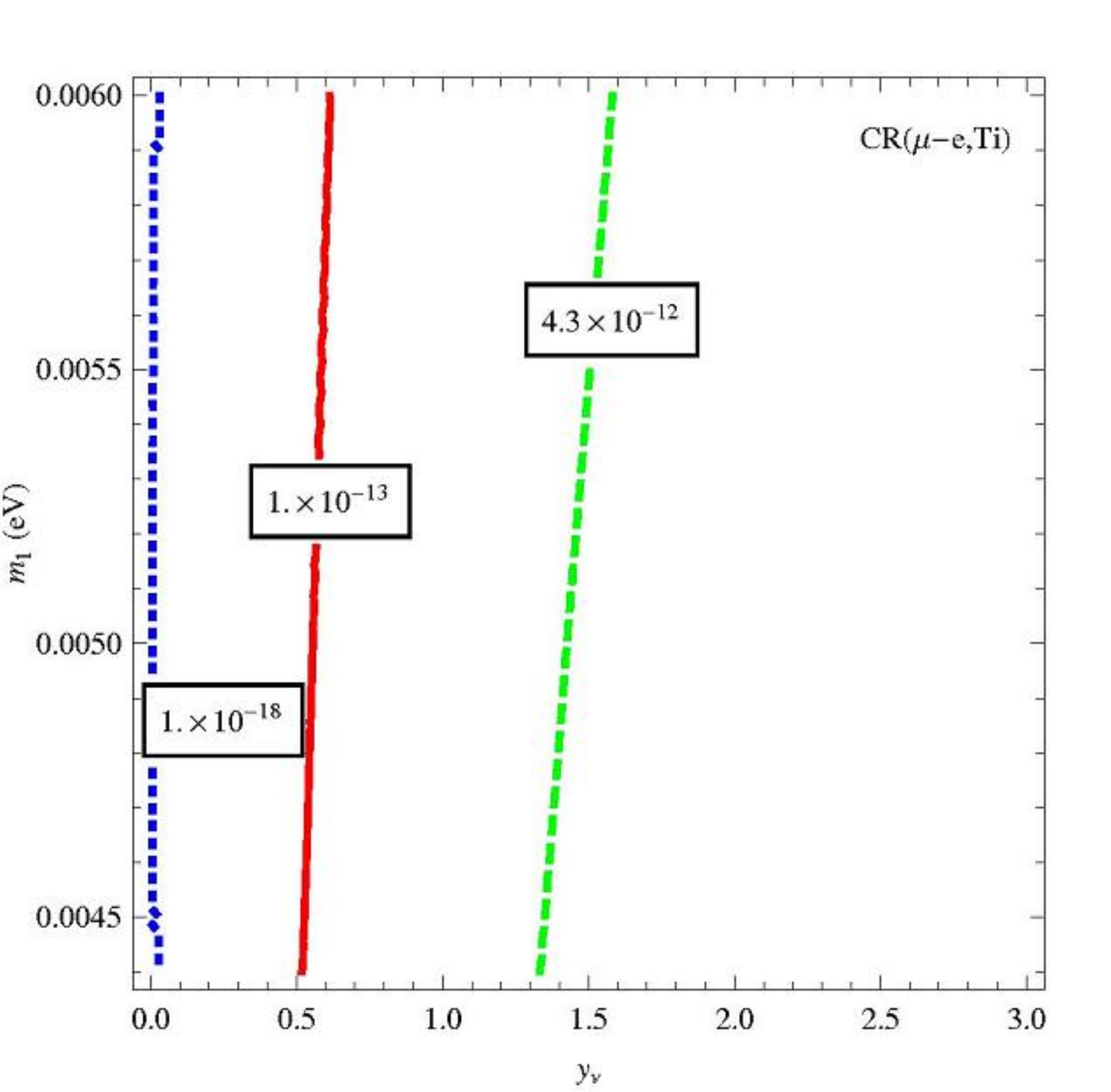}
\end{tabular}
\caption{\label{fig:AF_NH_CS}The contour plots for
$Br(\mu\rightarrow e\gamma)$, $Br(\mu\rightarrow 3e)$,
$CR(\mu-e,Al)$ and $CR(\mu-e,Ti)$ in the parameter space of
$y_{\nu}$ and $m_1$ for the NH spectrum in AF model, where $\Delta
m^2_{21}$ and $\Delta m^2_{31}$ are chose to be the best fit values
$7.65\times10^{-5}{\rm \,eV^2}$ and $2.40\times10^{-3}{\rm \,eV^2}$
respectively.}
\end{center}
\end{figure}

For the IH spectrum, the branching ratios for the $\mu e$ and $\tau
e$ involved LFV processes are indeed smaller than those of NH case,
as has been stressed below Eq.(\ref{53}). We notice that all the LFV
processes are below both the present and future experiment
sensitivity except the $\mu-e$ conversion in Ti. However, the rates
for $\mu\rightarrow e\gamma$, $\tau\rightarrow e\gamma$,
$\mu\rightarrow3e$, $\tau\rightarrow3e$ and $\mu-e$ conversion in Al
and Ti at LO are so small that they may be corrected considerably by
the NLO terms. Therefore it is very necessary to include the NLO
contributions and exploit the physical effects on the LFV
observables.

We perform a numerical analysis by treating all the NLO effective
couplings $y_A$, $y_{B}$, $\tilde{x}_B$, $\tilde{x}_C$,
$\tilde{x}_D$ and $\tilde{x}_E$ in Eq. (\ref{31}) and Eq. (\ref{33})
as random complex numbers with absolute value between 0 and 2. The
LO parameters $A$ and $B$ in the heavy right handed Majorana mass
matrix Eq.(\ref{20}) are taken to be random complex numbers with
absolute value between $10^{12}$ GeV and $10^{16}$ GeV, while
$\frac{\delta V_T}{V_T}$ and $\frac{V_T}{\Lambda}$ has been fixed at
the indicative values of 0.04. In order to compare with the leading
order predictions, the coupling $y_{\nu}$ is set equal to 0.5 as
well in the numerical analysis. The scatter plot of the LFV
branching ratios vs the lightest neutrino mass for NH and IH
spectrum are showed in Fig. \ref{fig:AF_NH_NLO} and Fig.
\ref{fig:AF_IH_NLO} respectively. These plots display only the
points corresponding to choices of the parameters reproducing
$\Delta m^2_{21}$, $\Delta m^2_{31}$ and the mixing angles within
their allowed $3\sigma$ interval given in Eq.(\ref{1}).

\begin{figure}[hptb]
\begin{center}
\begin{tabular}{c}
\includegraphics[scale=1,width=3.75cm]{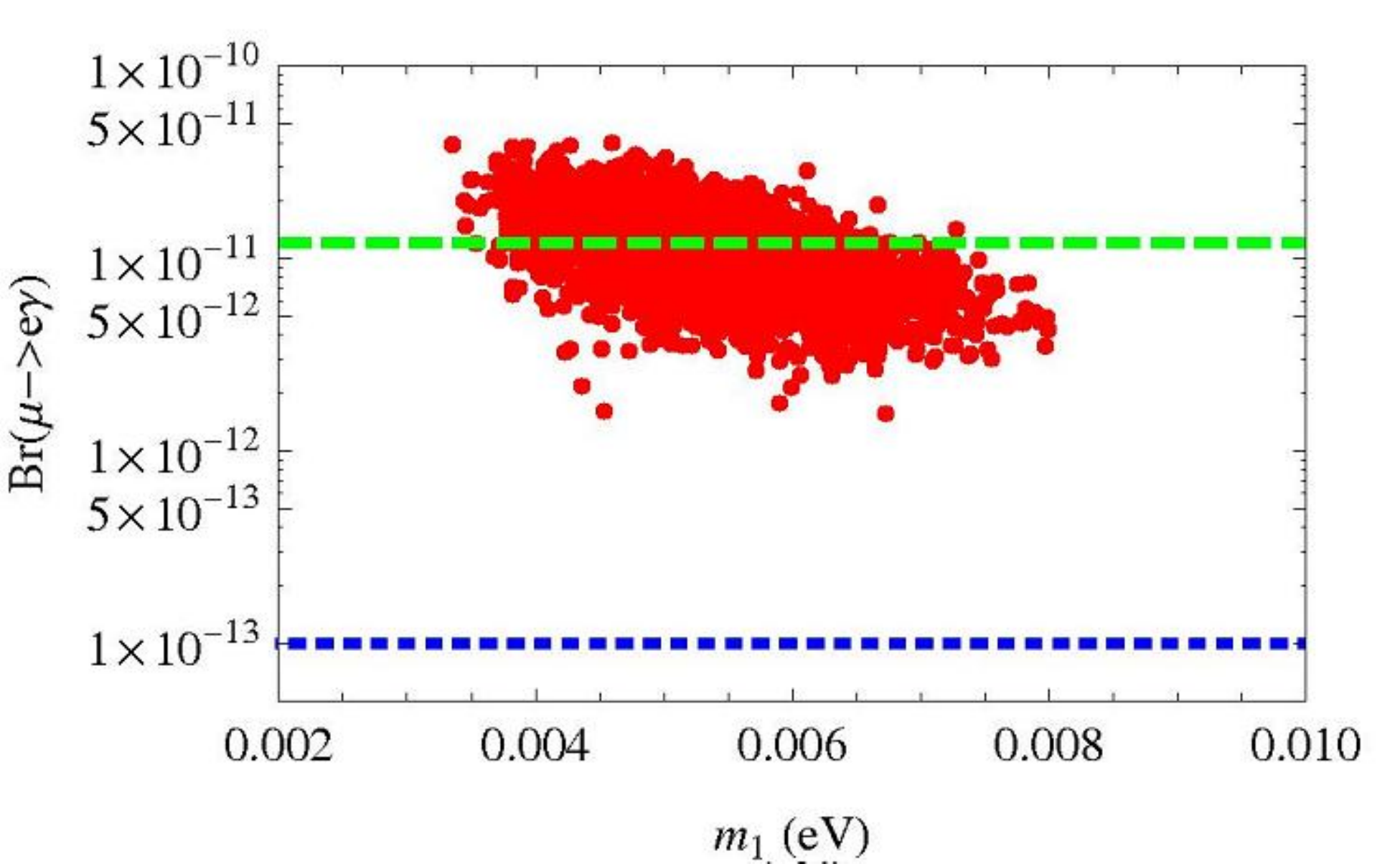}
\includegraphics[scale=1,width=3.75cm]{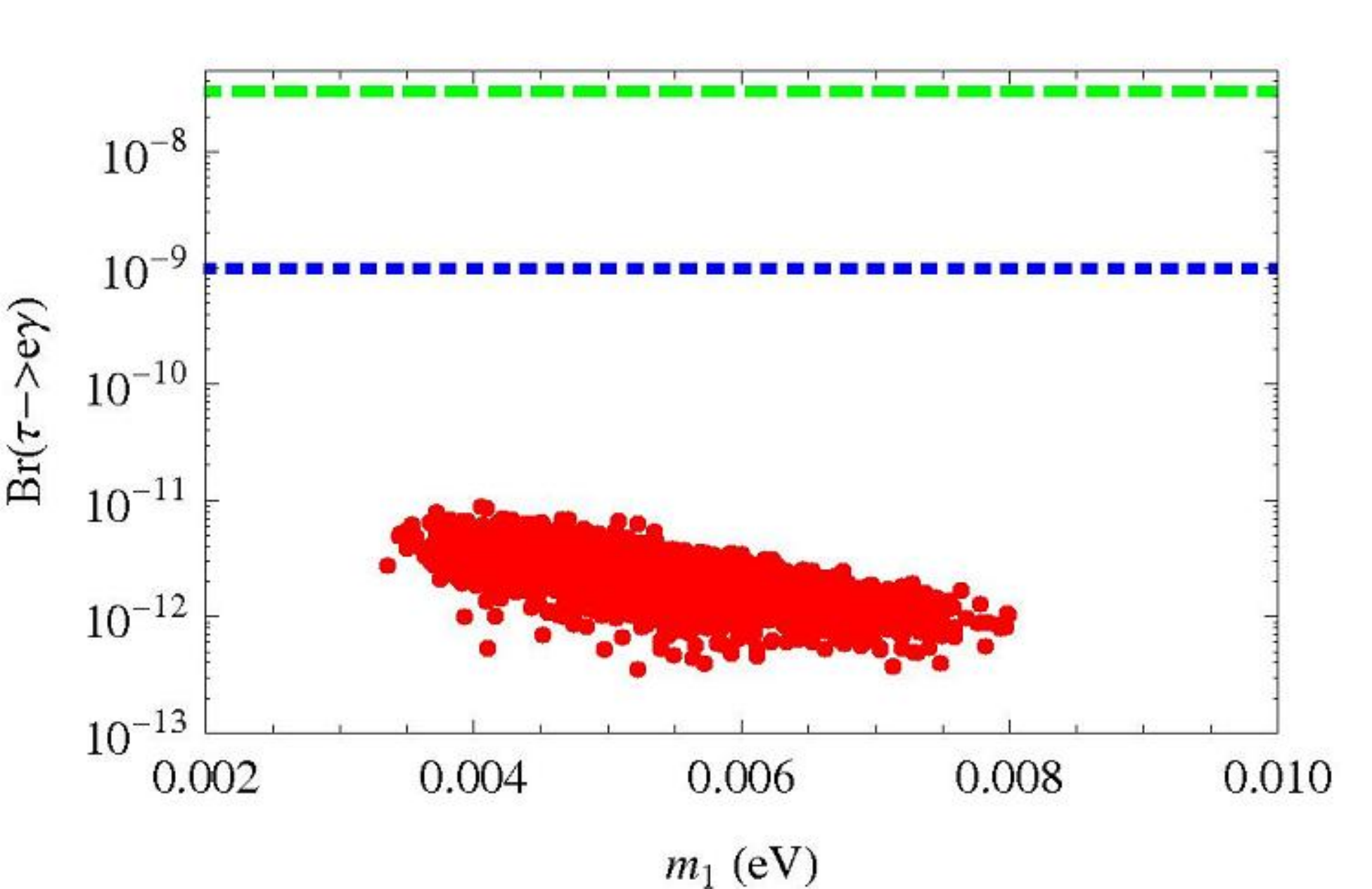}
\includegraphics[scale=1,width=3.75cm]{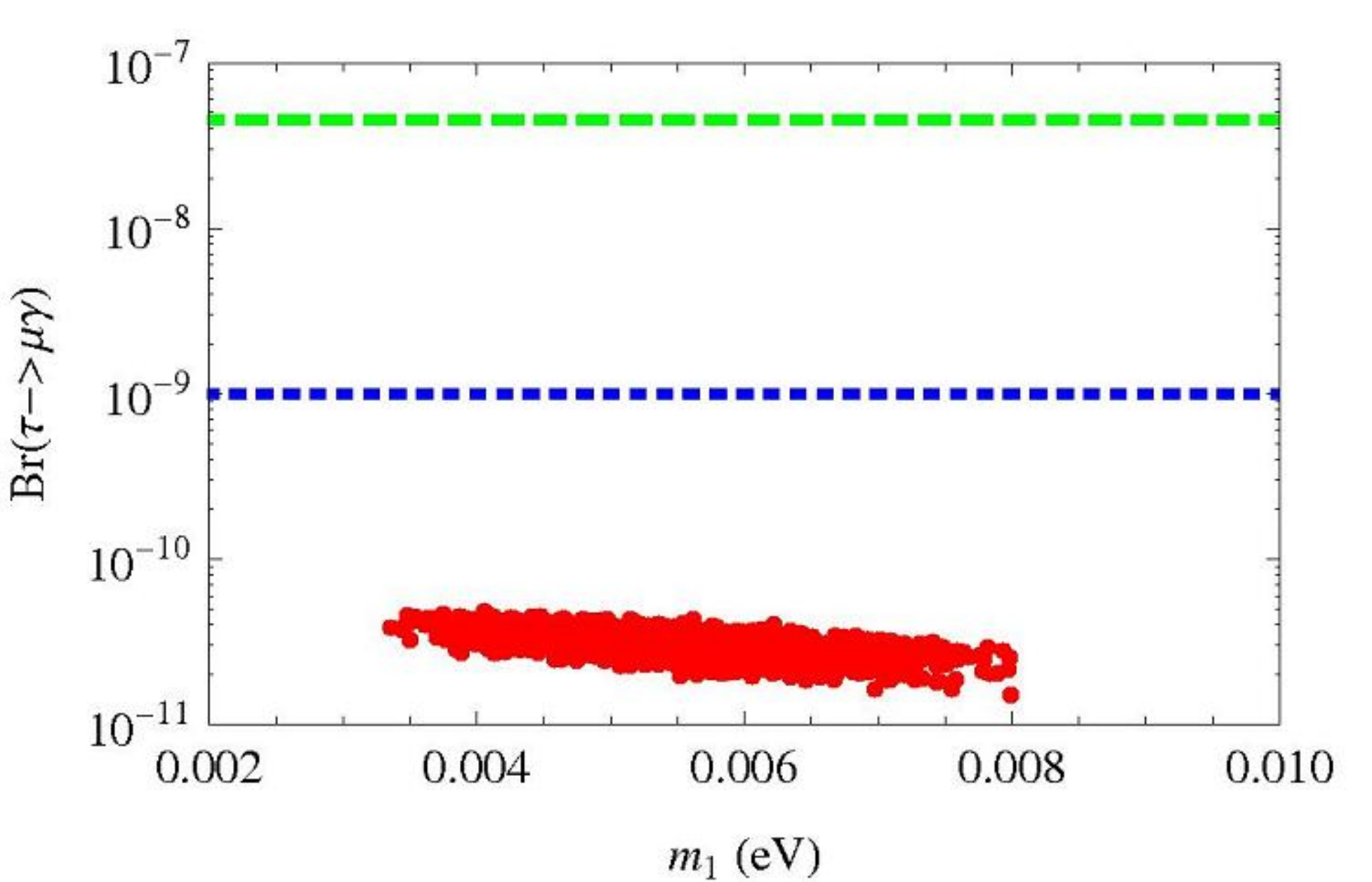}
\includegraphics[scale=1,width=3.75cm]{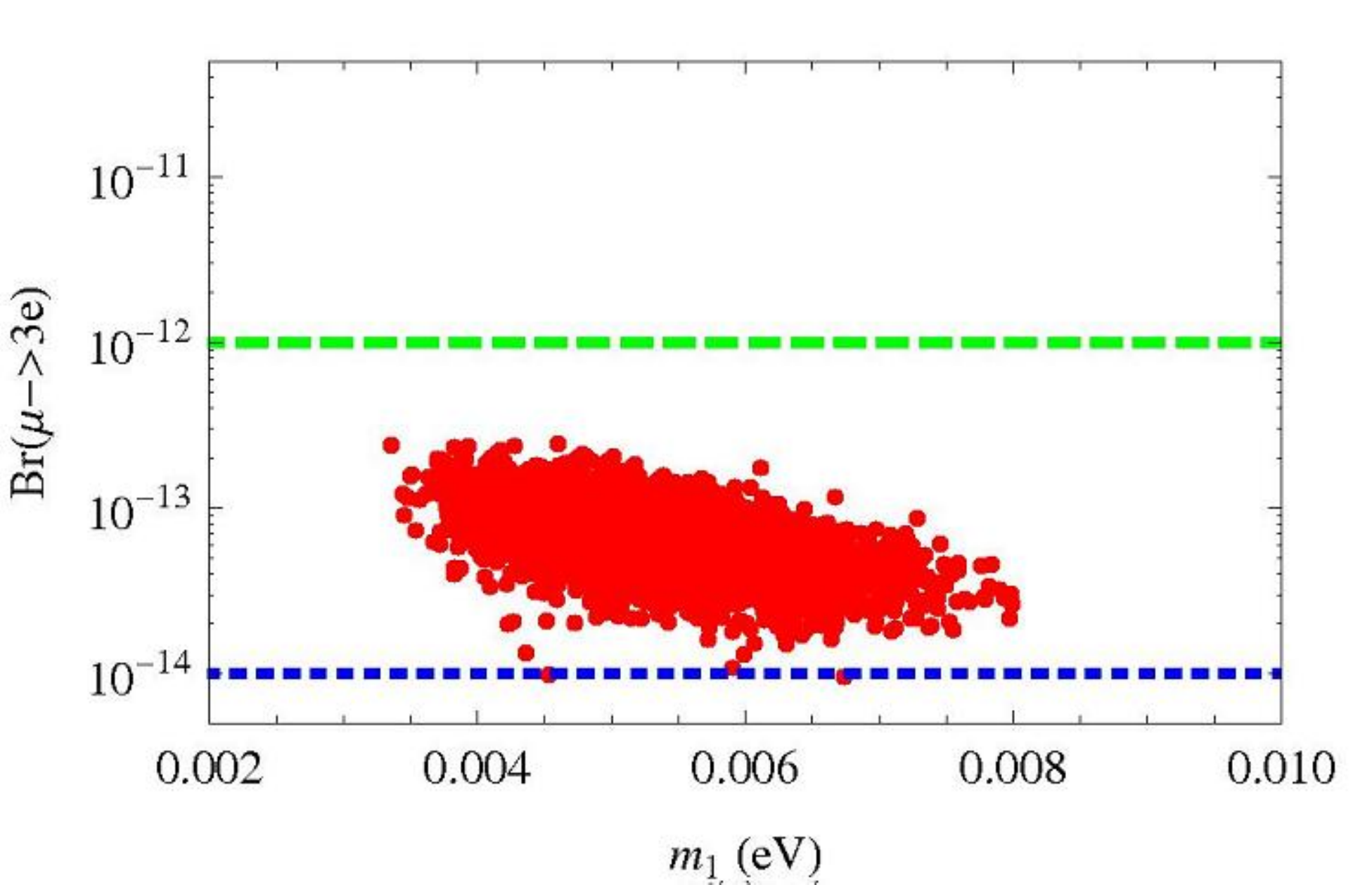}\\\\
\includegraphics[scale=1,width=3.75cm]{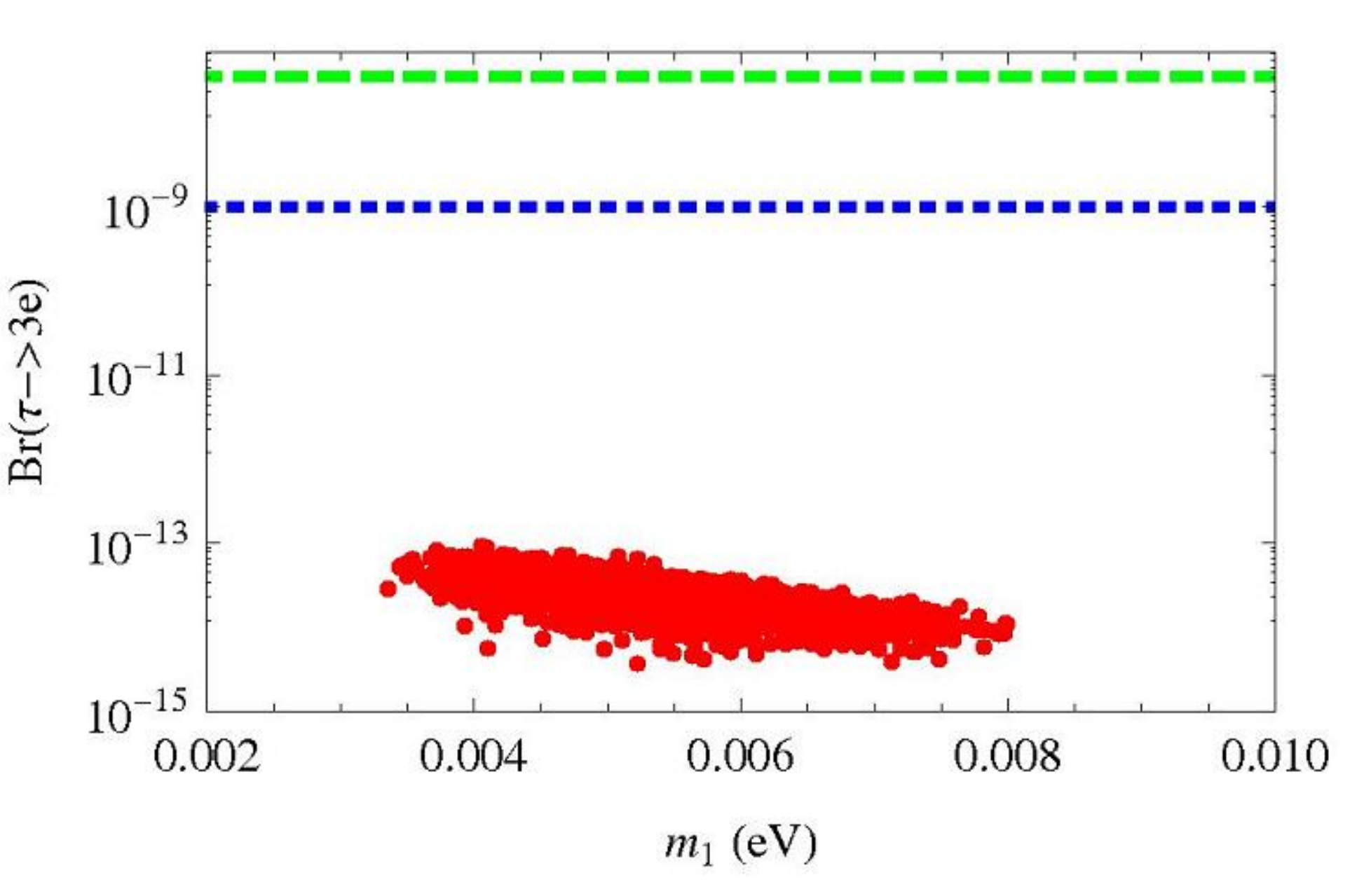}
\includegraphics[scale=1,width=3.75cm]{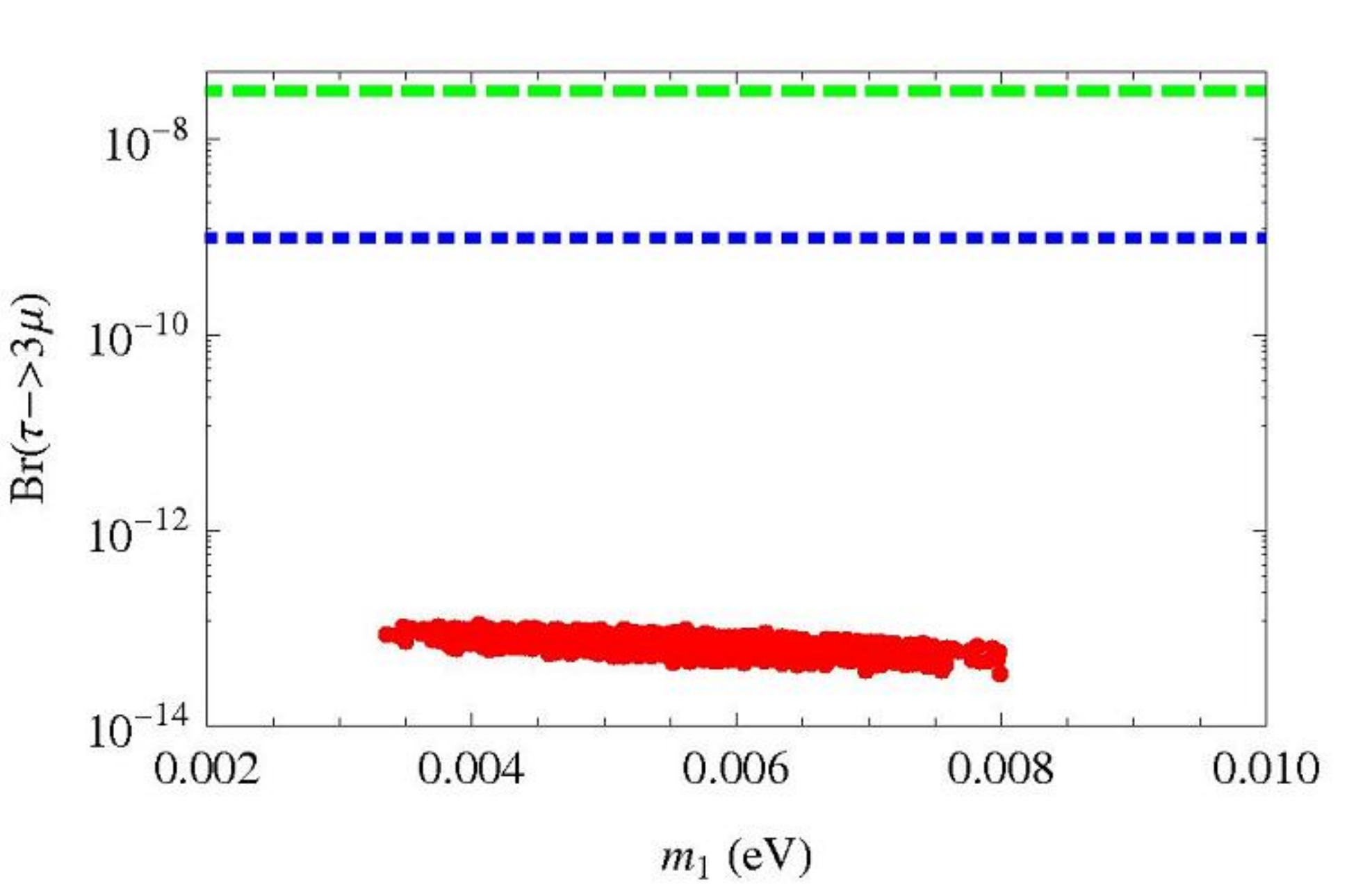}
\includegraphics[scale=1,width=3.75cm]{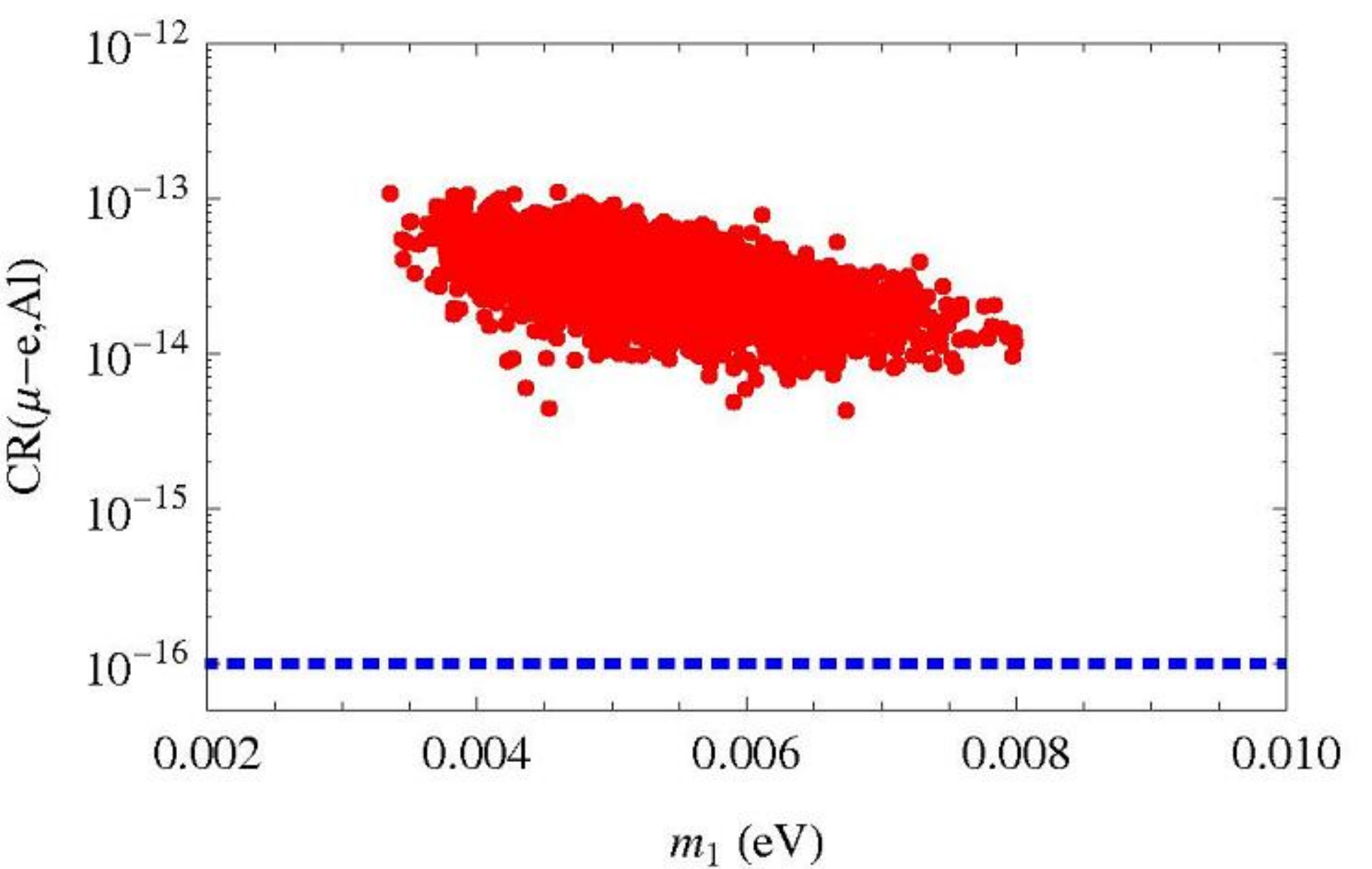}
\includegraphics[scale=1,width=3.75cm]{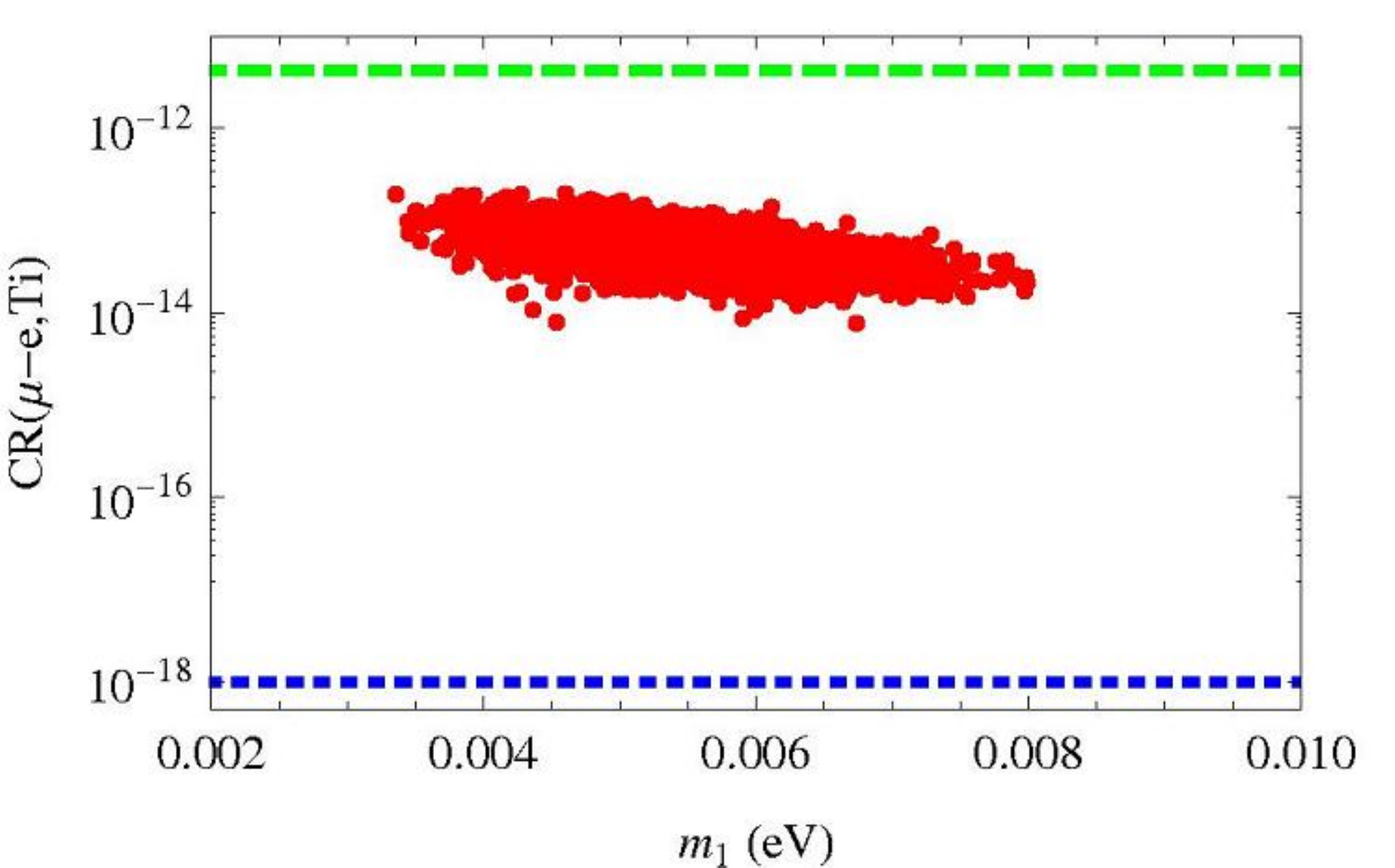}
\end{tabular}
\caption{\label{fig:AF_NH_NLO} Scatter plot of
$Br(\ell_i\rightarrow\ell_j\gamma)$, $Br(\ell_i\rightarrow3\ell_j)$,
$CR(\mu-e,Al)$ and $CR(\mu-e,Ti)$ against the lightest neutrino mass
$m_1$ in AF model for the normal hierarchy spectrum. The dashed and
dotted lines denote the present and future experimental sensitivity
respectively. There is no upper bound for $CR(\mu-e,Al)$ so far.}
\end{center}
\end{figure}

\begin{figure}[hptb]
\begin{center}
\begin{tabular}{c}
\includegraphics[scale=1,width=3.75cm]{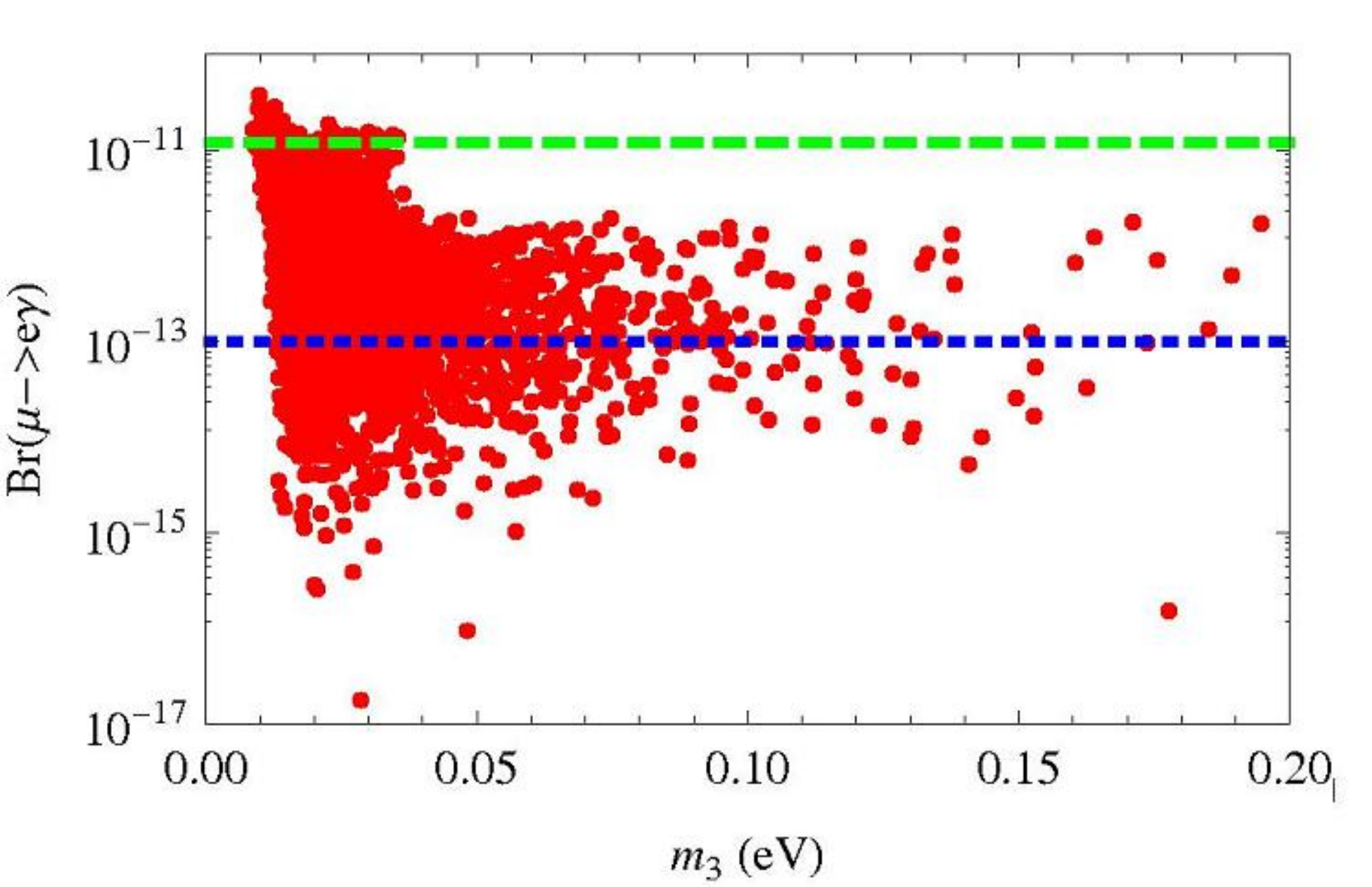}
\includegraphics[scale=1,width=3.75cm]{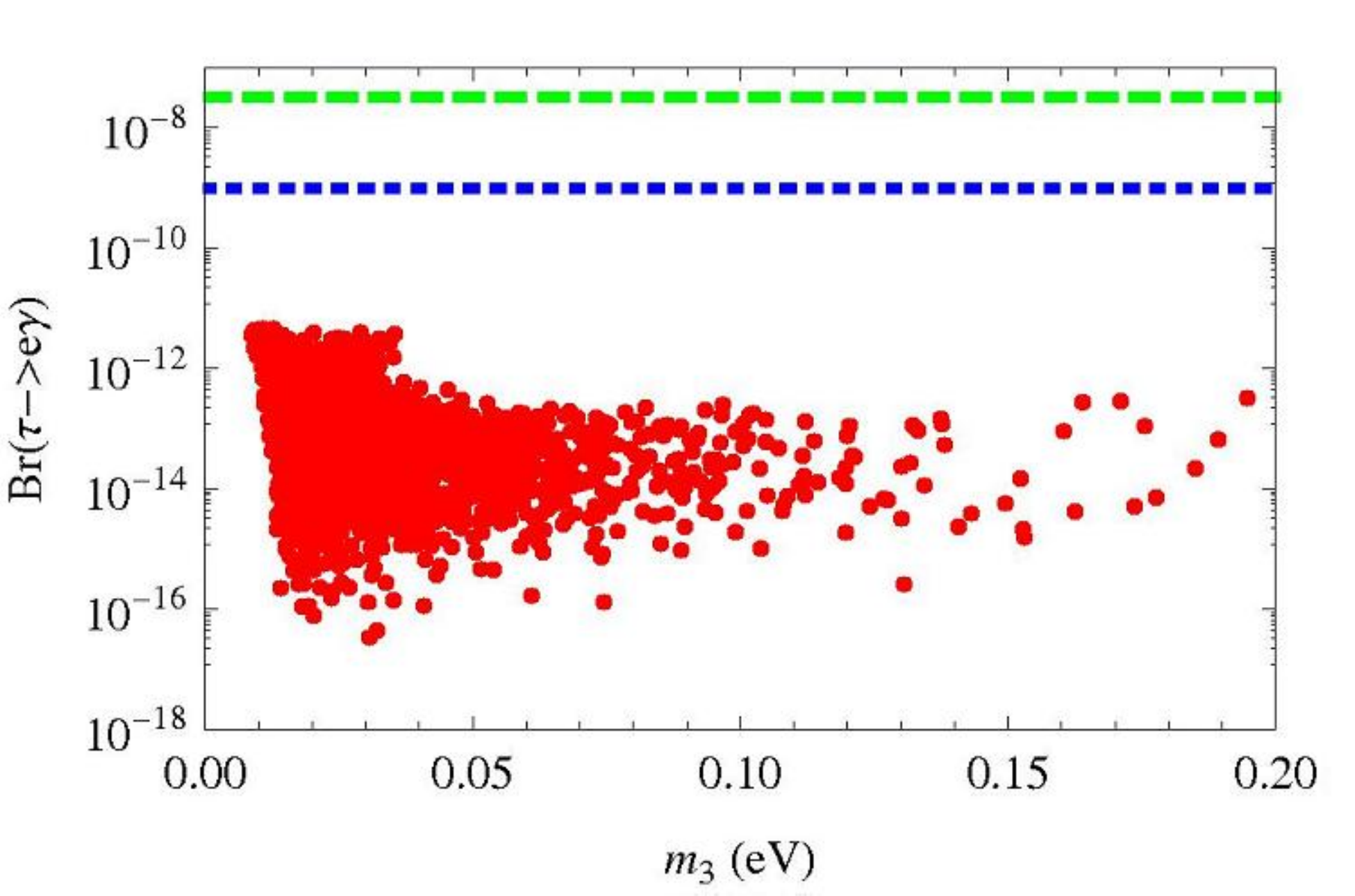}
\includegraphics[scale=1,width=3.75cm]{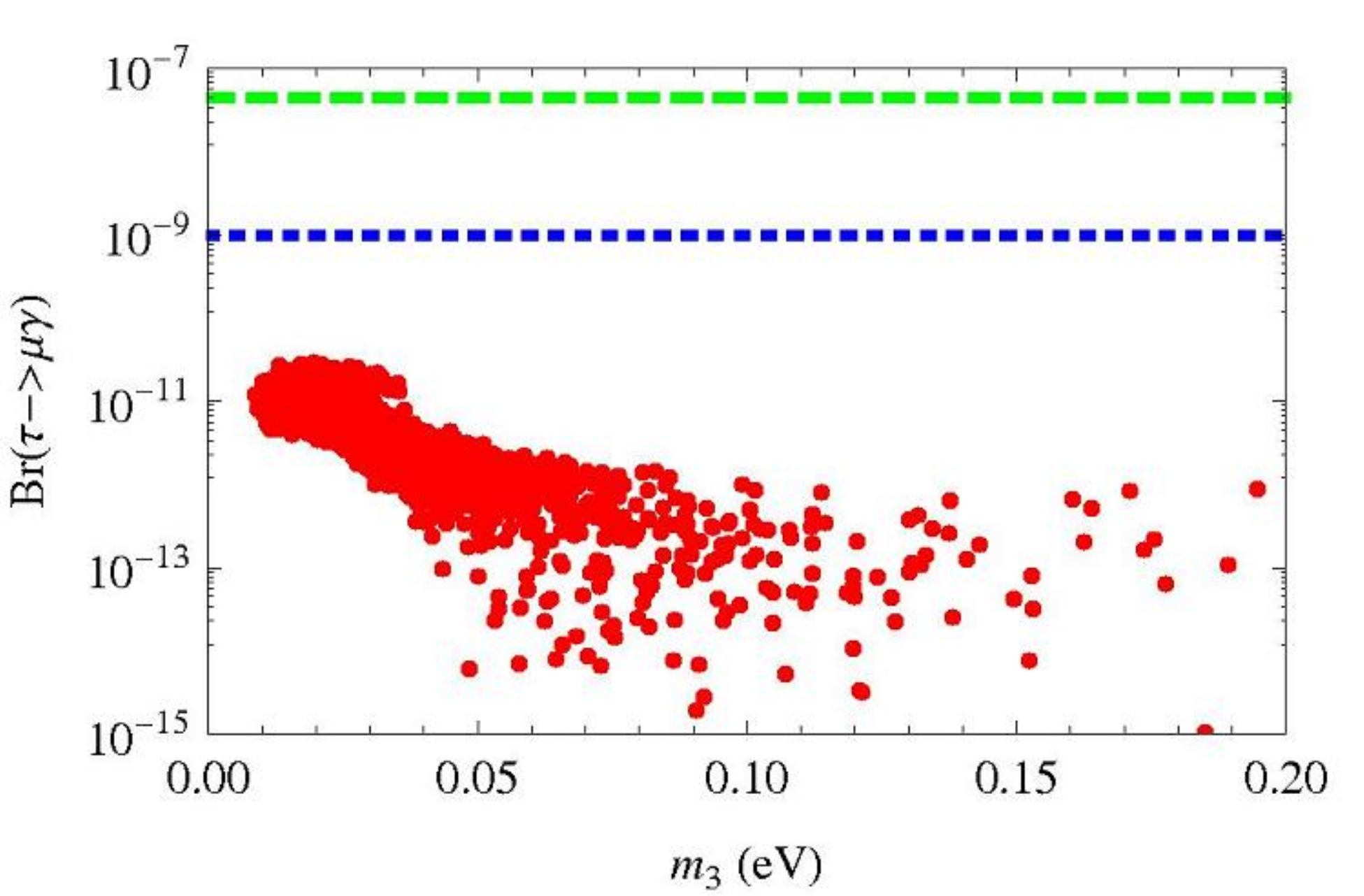}
\includegraphics[scale=1,width=3.75cm]{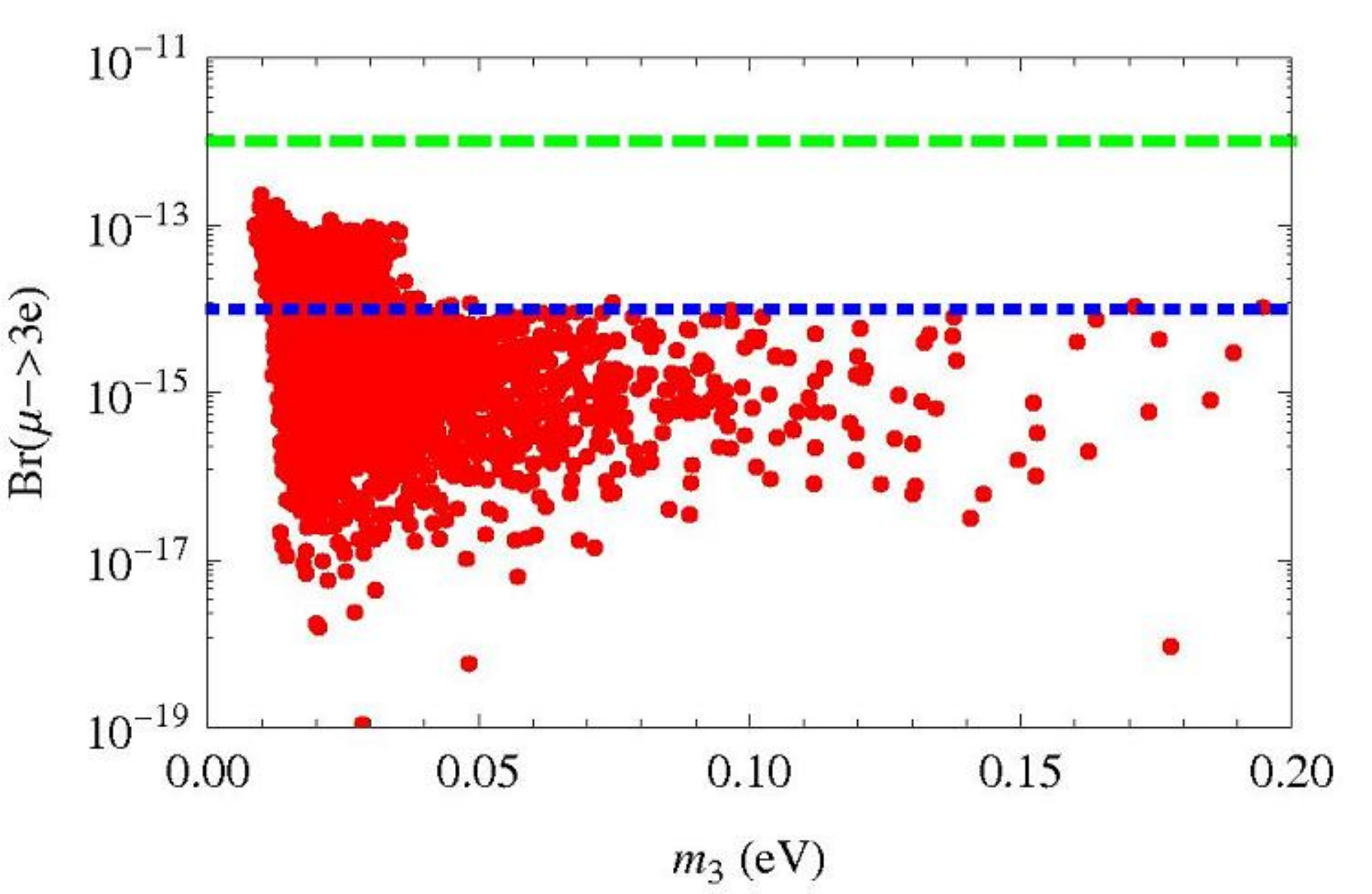}\\\\
\includegraphics[scale=1,width=3.75cm]{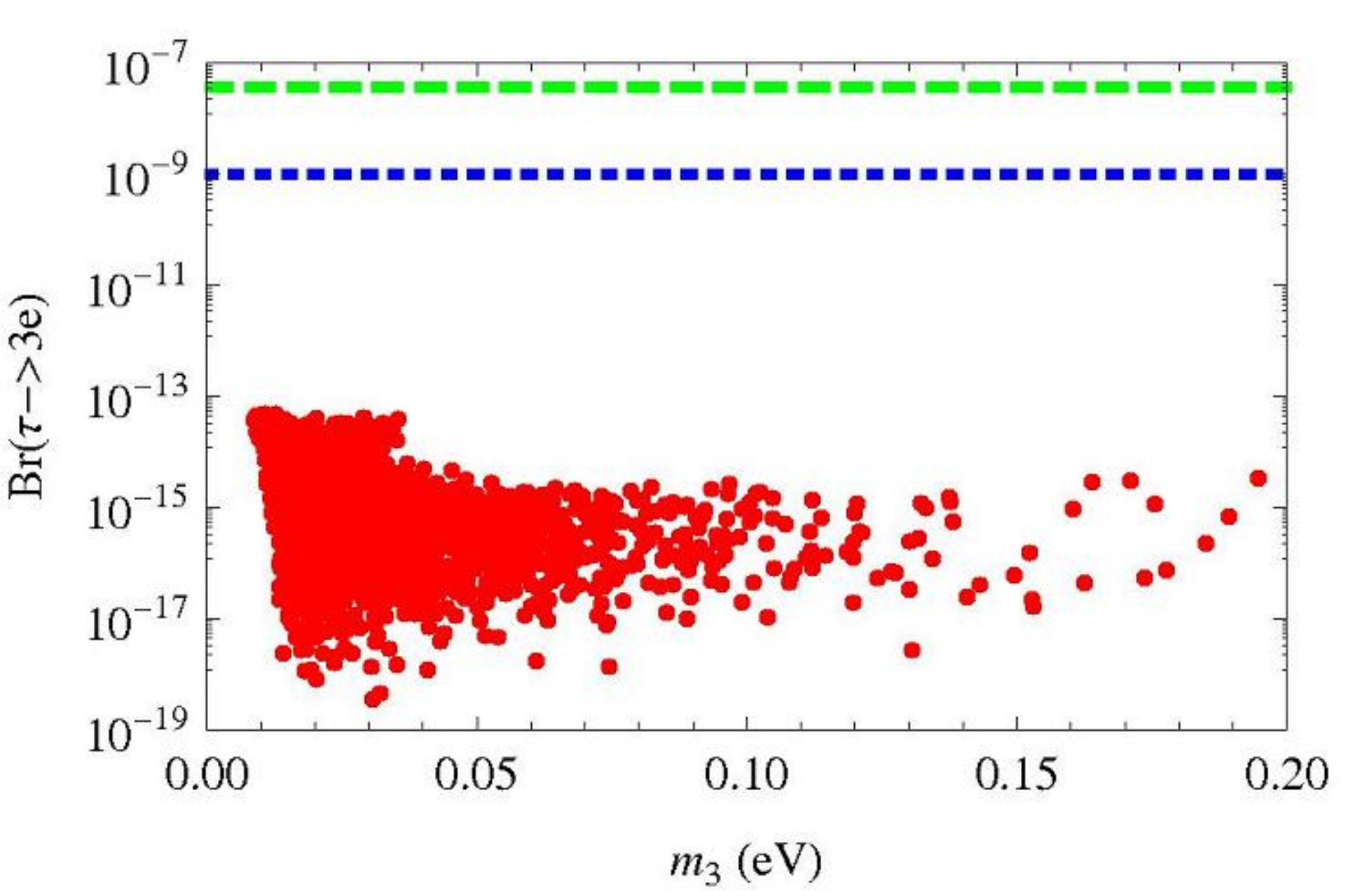}
\includegraphics[scale=1,width=3.75cm]{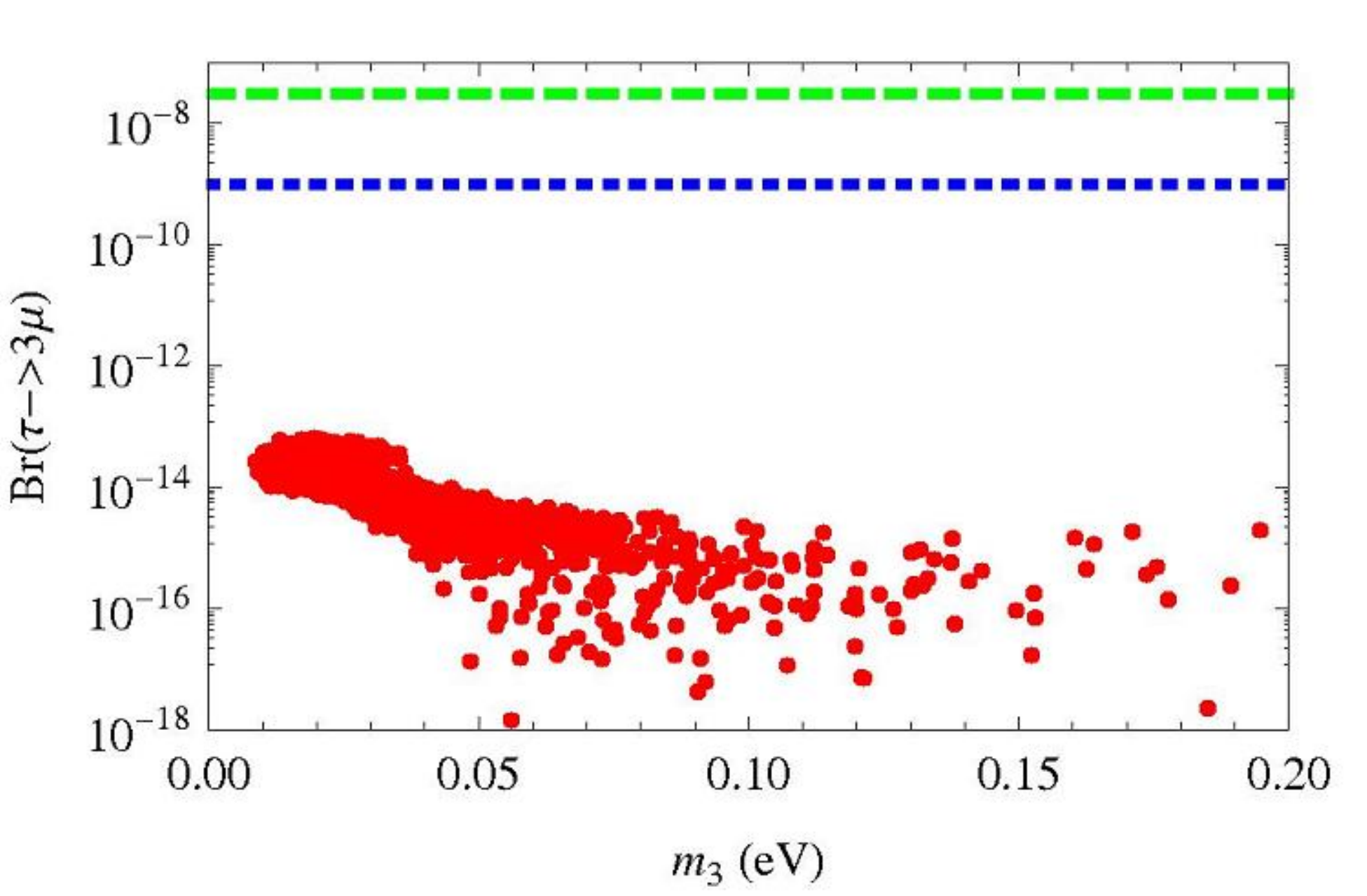}
\includegraphics[scale=1,width=3.75cm]{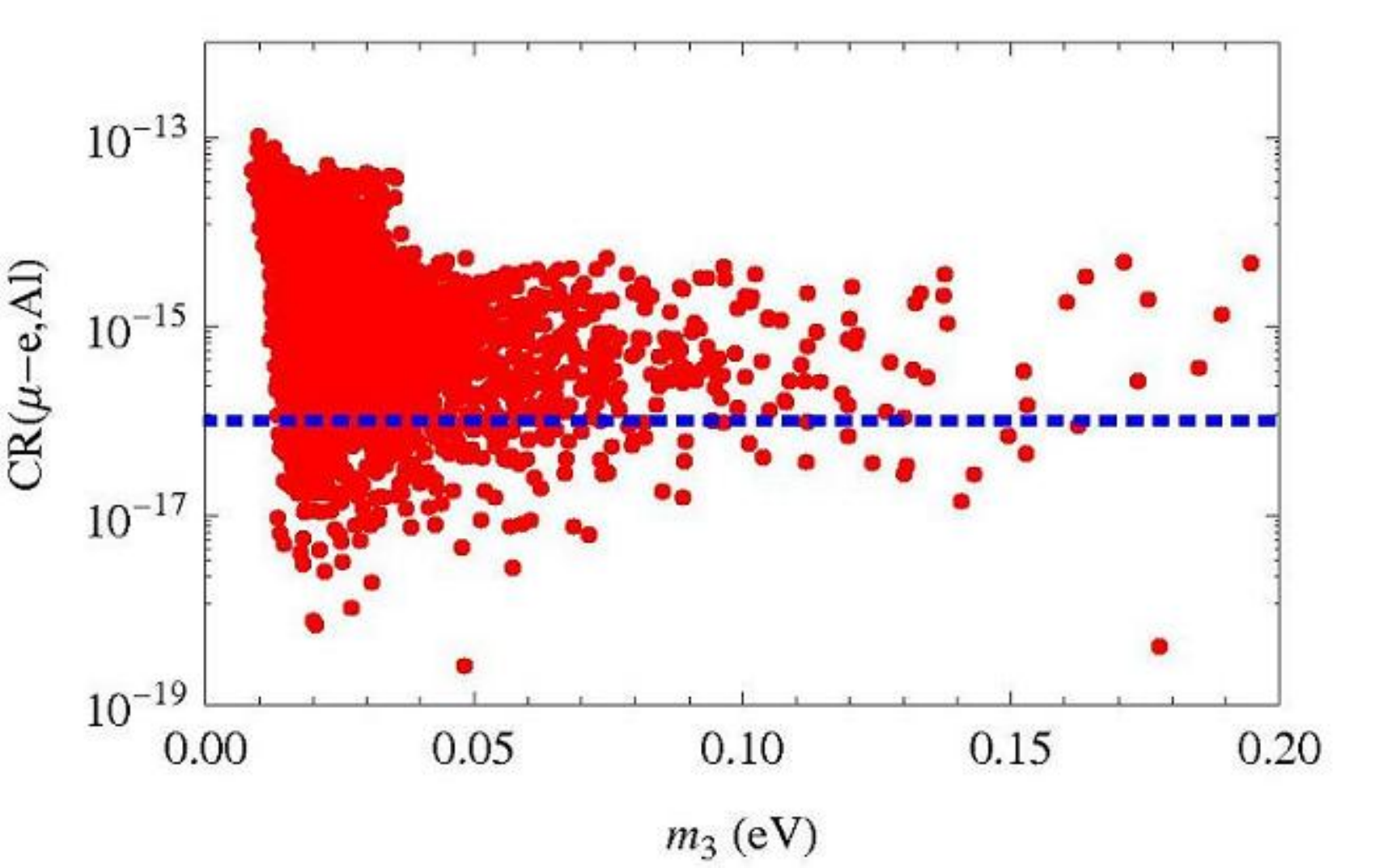}
\includegraphics[scale=1,width=3.75cm]{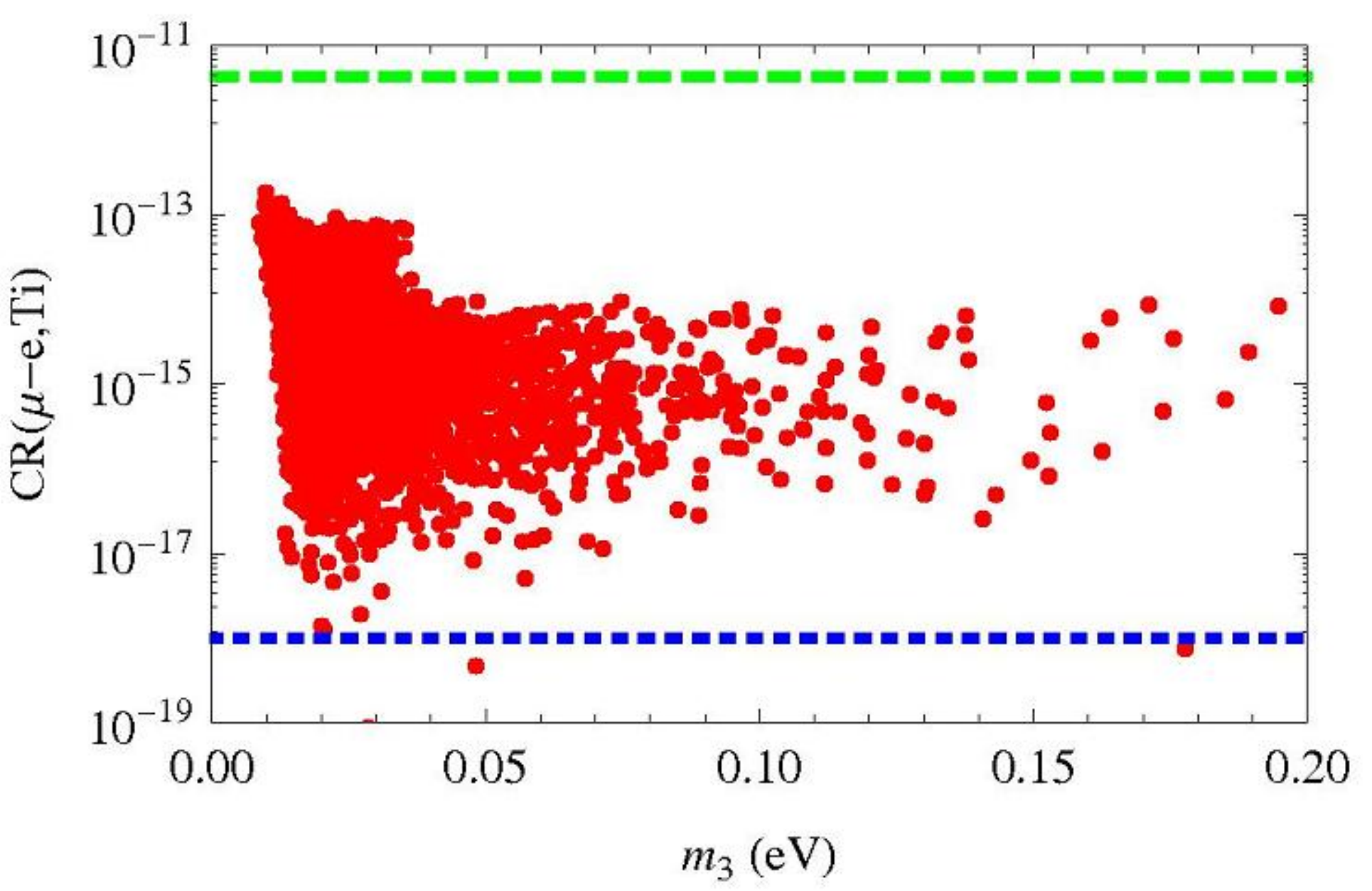}
\end{tabular}
\caption{\label{fig:AF_IH_NLO} Scatter plot of
$Br(\ell_i\rightarrow\ell_j\gamma)$, $Br(\ell_i\rightarrow3\ell_j)$,
$CR(\mu-e,Al)$ and $CR(\mu-e,Ti)$ against the lightest neutrino mass
$m_3$ in AF model for the inverted hierarchy spectrum. The dashed
and dotted lines represent the present and future experimental
sensitivity respectively.}
\end{center}
\end{figure}

We see that the LFV branching ratios for NH are modified slightly by
the NLO contribution, the above discussions for NH case at LO still
apply here, and the next generation experiments of $\mu\rightarrow
e\gamma$, $\mu\rightarrow3e$ and $\mu-e$ conversion in Al and Ti are
crucial tests to the AF model. However, for IH spectrum the rates of
the processes $\mu\rightarrow e\gamma$, $\tau\rightarrow e\gamma$,
$\mu\rightarrow 3e$, $\tau\rightarrow3e$ and $\mu-e$ conversion in
Al and Ti are enhanced considerably. We note that
$Br(\tau\rightarrow\mu\gamma)$ and $Br(\tau\rightarrow3\mu)$ are
qualitatively the same as the LO results, this is because that
$\tau\rightarrow\mu\gamma$ and $\tau\rightarrow3\mu$ are related
with the elements $(\hat{\mathbf{Y}}^{\dagger}_{\nu}{\mathbf
L}\hat{{\mathbf Y}}_{\nu})_{23}$, which are not suppressed at LO and
are much larger than NLO corrections. It is notable that $\mu-e$
conversion in Ti is within the sensitivity of the next generation
experiment for the whole parameter space considered, the signals of
$\mu\rightarrow e\gamma$ and $\mu-e$ conversion in Al should be
detected in near future in a very large part of the parameter space,
and $\mu\rightarrow3e$ could be observed only in a marginal part of
the parameter space. However, $\tau\rightarrow e\gamma$,
$\tau\rightarrow\mu\gamma$, $\tau\rightarrow3e$ and
$\tau\rightarrow3\mu$ are still below the expected future
sensitivity. We note that the LO result $\frac{Br(\tau\rightarrow
e\gamma)}{Br(\mu\rightarrow e\gamma)}\simeq17.84\%$ is destroyed
completely after including the NLO corrections, the variation of
$\frac{Br(\tau\rightarrow e\gamma)}{Br(\mu\rightarrow e\gamma)}$ vs
the lightest neutrino mass is presented in Fig. \ref{fig:AF_ratio}.

In short summary, for both NH and IH spectrum of AF model, $\mu-e$
conversion in Ti can be observed in all the parameter space
considered, $\mu\rightarrow e\gamma$ and $\mu-e$ conversion in Al
should be observed at least in a very significant part of the
parameter space, and $\mu\rightarrow3e$ may be observed on for NH
spectrum. Of all the LFV processes, the $\mu-e$ conversion in Ti
should be an even more robust one, since its sensitivity would be
improved drastically in near future.

\begin{figure}[hptb]
\begin{center}
\begin{tabular}{c}
\includegraphics[scale=1,width=6.0cm]{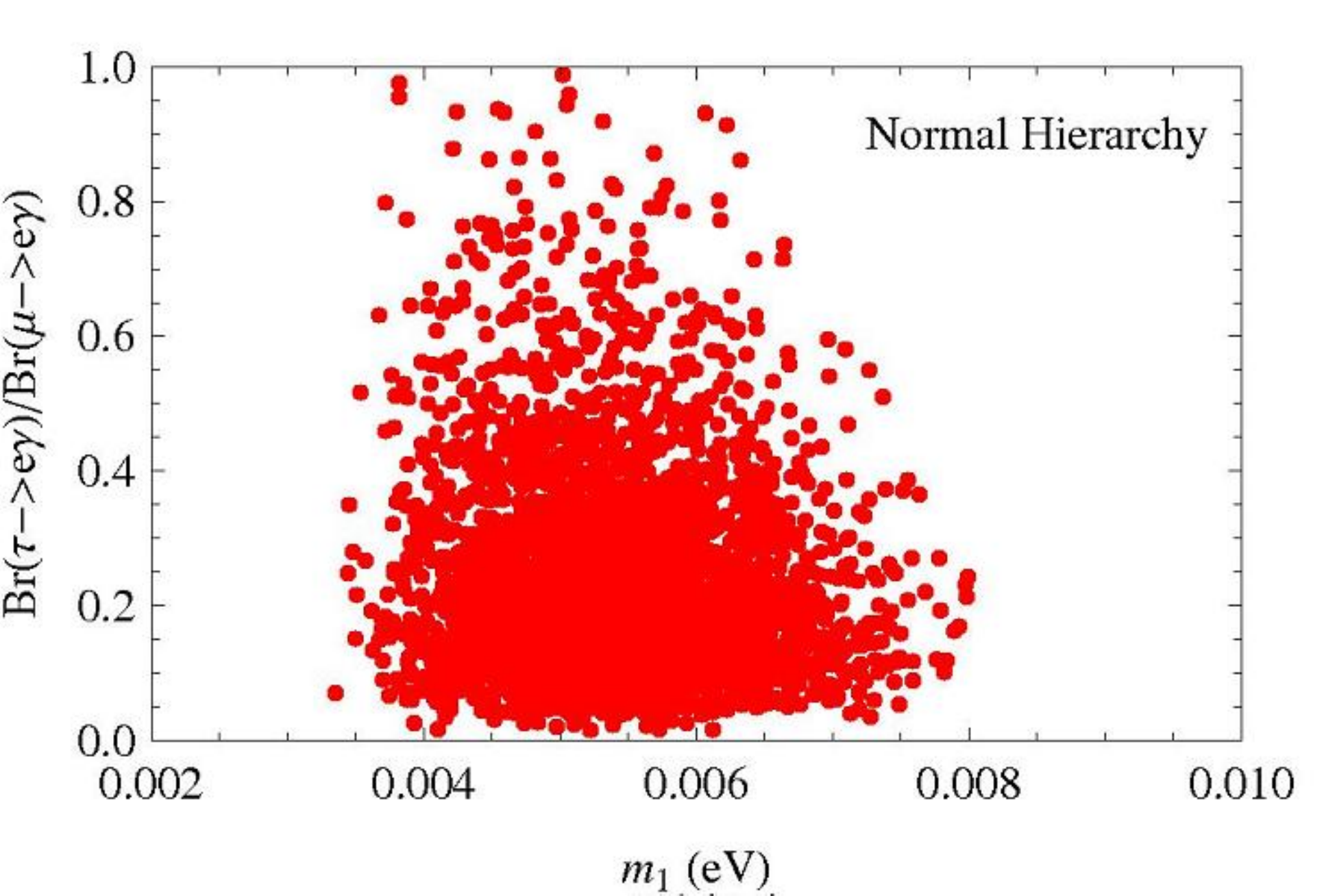}
\hspace{1cm}
\includegraphics[scale=1,width=6.0cm]{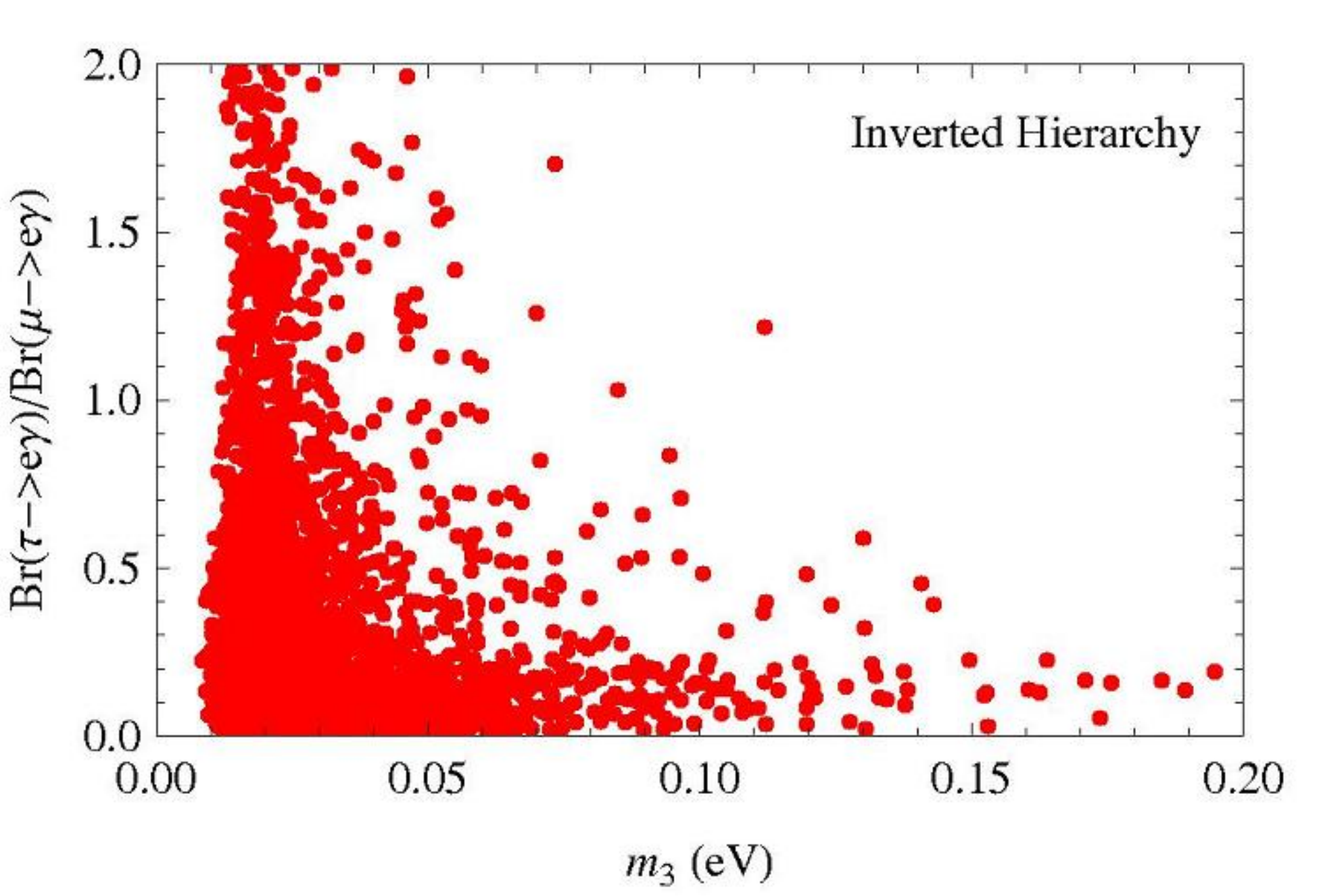}
\end{tabular}
\caption{\label{fig:AF_ratio}Scatter plots of ${Br(\tau\rightarrow
e\gamma)}/{Br(\mu\rightarrow e\gamma)}$ against the lightest
neutrino mass for both NH and IH in AF model.}
\end{center}
\end{figure}

\section{Predictions for LFV in $S_4$ model}

Similar to section \ref{sec:LFVA4}, we first study the predictions
for the LFV branching ratios at LO of the model \cite{Ding:2009iy},
then present the NLO corrections. From the discussion in section
\ref{sec:S4_model}, we learn that the charged lepton mass matrix is
diagonal, and the heavy right handed neutrinos are degenerate at LO
. In the base where both the right handed neutrino and charged
lepton mass matrices are diagonal and real, the neutrino Yukawa
coupling matrix is $\hat{\mathbf{Y}}_{\nu}=U^{T}\mathbf{Y}_{\nu}$,
where the unitary matrix $U$ is given by Eq. (\ref{40}).
Straightforwardly we can calculate the hermitian matrix
$\hat{{\mathbf Y}}^{\dagger}_{\nu}{\mathbf L}{\mathbf Y}_{\nu}$ as
follows
\begin{eqnarray}
\nonumber\hat{{\mathbf Y}}^{\dagger}_{\nu}{\mathbf L}\hat{{\mathbf
Y}}_{\nu}&=&\left(\begin{array}{ccc} 2|a|^2+6|b|^2-4|a||b|\cos\Phi&
|a|^2-3|b|^2+2|a||b|\cos\Phi &
|a|^2-3|b|^2+2|a||b|\cos\Phi\\
|a|^2-3|b|^2+2|a||b|\cos\Phi & 2|a|^2+6|b|^2+2|a||b|\cos\Phi
&|a|^2-3|b|^2-4|a||b|\cos\Phi\\
|a|^2-3|b|^2+2|a||b|\cos\Phi & |a|^2-3|b|^2-4|a||b|\cos\Phi &
2|a|^2+6|b|^2+2|a||b|\cos\Phi
\end{array}\right)\ln\frac{M_G}{|M|}\\
\label{55}&&=\left(\begin{array}{ccc}2m_1+m_2
&m_2-m_1&m_2-m_1\\
m_2-m_1&\frac{1}{2}(m_1+2m_2+3m_3)&~\frac{1}{2}(m_1+2m_2-3m_3)\\
m_2-m_1&\frac{1}{2}(m_1+2m_2-3m_3)&~\frac{1}{2}(m_1+2m_2+3m_3)
\end{array}\right)\frac{|M|}{3v^2_u}\ln\frac{M_G}{|M|}
\end{eqnarray}
where $\Phi$ is the relative phase between $a$ and $b$, and
$m_{1,2,3}$ are the light neutrino masses given in Eq.(\ref{43}).
The off-diagonal elements of $\hat{{\mathbf
Y}}^{\dagger}_{\nu}{\mathbf L}\hat{{\mathbf Y}}_{\nu}$ can be read
out directly
\begin{eqnarray}
\nonumber&&(\hat{{\mathbf Y}}^{\dagger}_{\nu}{\mathbf
L}\hat{{\mathbf Y}}_{\nu})_{12}=(\hat{{\mathbf
Y}}^{\dagger}_{\nu}{\mathbf L}\hat{{\mathbf
Y}}_{\nu})_{21}=(m_2-m_1)\frac{|M|}{3v^2_u}\ln\frac{M_G}{|M|}=\frac{\Delta m^2_{21}}{3(m_1+m_2)v^2_u}|M|\ln\frac{M_G}{|M|}\\
\nonumber&&(\hat{{\mathbf Y}}^{\dagger}_{\nu}{\mathbf
L}\hat{{\mathbf Y}}_{\nu})_{13}=(\hat{{\mathbf
Y}}^{\dagger}_{\nu}{\mathbf L}\hat{{\mathbf
Y}}_{\nu})_{31}=(m_2-m_1)\frac{|M|}{3v^2_u}\ln\frac{M_G}{|M|}=\frac{\Delta m^2_{21}}{3(m_1+m_2)v^2_u}|M|\ln\frac{M_G}{|M|}\\
\label{56}&&(\hat{{\mathbf Y}}^{\dagger}_{\nu}{\mathbf
L}\hat{{\mathbf Y}}_{\nu})_{23}=(\hat{{\mathbf
Y}}^{\dagger}_{\nu}{\mathbf L}\hat{{\mathbf
Y}}_{\nu})_{32}=(m_1+2m_2-3m_3)\frac{|M|}{6v^2_u}\ln\frac{M_G}{|M|}
\end{eqnarray}
It is obvious that the off-diagonal elements of $\hat{{\mathbf
Y}}^{\dagger}_{\nu}{\mathbf L}\hat{{\mathbf Y}}_{\nu}$ are closely
related to the light neutrino mass difference and the right handed
neutrino mass $|M|$, and they increase with $|M|$. Interestingly we
notice $(\hat{{\mathbf Y}}^{\dagger}_{\nu}{\mathbf L}\hat{{\mathbf
Y}}_{\nu})_{12}=(\hat{{\mathbf Y}}^{\dagger}_{\nu}{\mathbf
L}\hat{{\mathbf Y}}_{\nu})_{13}$, which turns out to be true in AF
model at LO as well. The relation $(\hat{{\mathbf
Y}}^{\dagger}_{\nu}{\mathbf L}\hat{{\mathbf
Y}}_{\nu})_{12}=(\hat{{\mathbf Y}}^{\dagger}_{\nu}{\mathbf
L}\hat{{\mathbf Y}}_{\nu})_{13}$ seems universal in models which
reproduce TB mixing at LO. Consequently we also have exactly
$Br(\tau\rightarrow e\gamma)/Br(\mu\rightarrow
e\gamma)\simeq17.84\%$ at LO, which is not satisfied anymore after
including the subleading corrections. Since the factor $\frac{\Delta
m^2_{21}}{3(m_1+m_2)}$ is rather small for IH spectrum, the
branching ratios of $\mu\rightarrow e\gamma$, $\tau\rightarrow
e\gamma$, $\mu\rightarrow3e$, $\tau\rightarrow 3e$ and $\mu-e$
conversion in Al and Ti should be small substantially at leading
order. In the limit of degenerate light neutrino mass spectrum, the
off-diagonal elements tend to be zero, the LFV processes would be
highly suppressed.

It is interesting to estimate the order of magnitude for the right
handed neutrino mass $|M|$. Since the parameters $a$ and $b$ are
expected to be of order $\lambda^2_c$, with this and using
$\sqrt{\Delta m^2_{31}}\simeq0.05$ eV as the light neutrino mass
scale in the Seesaw formula, we obtain
\begin{equation}
\label{57}|M|\sim10^{12}-10^{13}{\rm GeV}
\end{equation}
We set $|M|$ to be equal to $1.0\times10^{13}$ GeV, the LO
predictions for the branching ratios of LFV processes are plotted in
Fig.\ref{fig:Ding_NH_LO} and Fig.\ref{fig:Ding_IH_LO} for NH and IH
spectrum respectively, where the SSB parameters are chose to be the
SPS3 point as well. We see that the LFV branching ratio is smaller
the corresponding one of AF model, all the LFV processes are below
the planned experiment sensitivity except $\mu-e$ conversion in Ti.
Consequently the next generation experiment of $\mu-e$ conversion in
Ti is very important to test this $S_4$ model, and the parameter
space of the model could be tightly constrained by $CR(\mu-e,Ti)$.
The contour plot of $CR(\mu-e,Ti)$ in the $m_1-|M|$ ($m_3-|M|$)
plane is showed in Fig.\ref{fig:Ding_CS}, where both NH and IH mass
spectrum are considered. It is obvious that the contour curves move
toward the left with the increase of the sensitivity. We notice that
in this $S_4$ model the signal of $\mu-e$ conversion in Ti could be
observed in a large part of the allowed parameter space. As we will
demonstrate that in case of IH spectrum, $CR(\mu-e,Ti)$ receives
relative large correction from the NLO contributions, as a result,
the contour plot for IH should be taken with a grain of salt.

\begin{figure}[hptb]
\begin{center}
\begin{tabular}{c}
\includegraphics[scale=1,width=3.75cm]{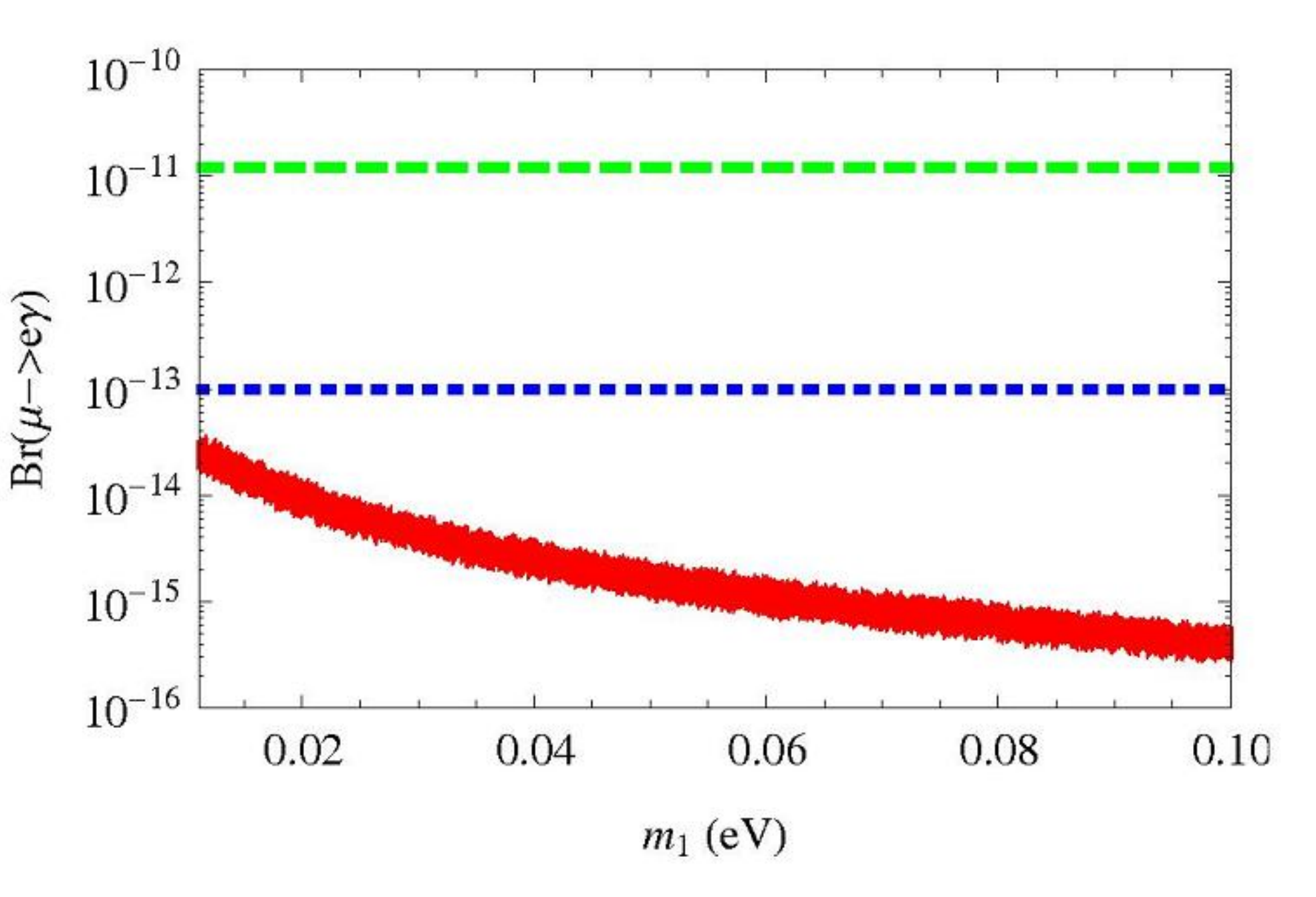}
\includegraphics[scale=1,width=3.75cm]{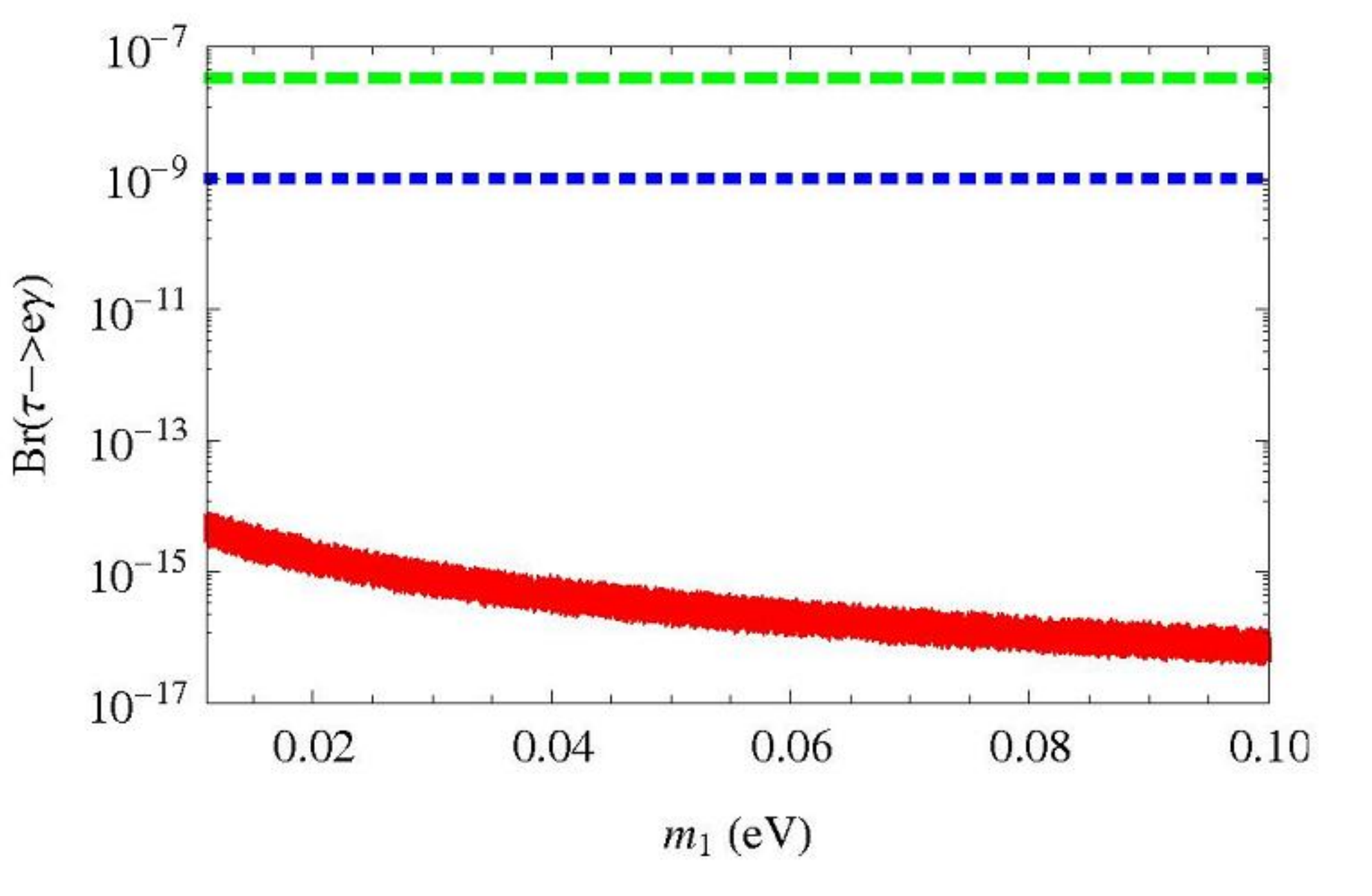}
\includegraphics[scale=1,width=3.75cm]{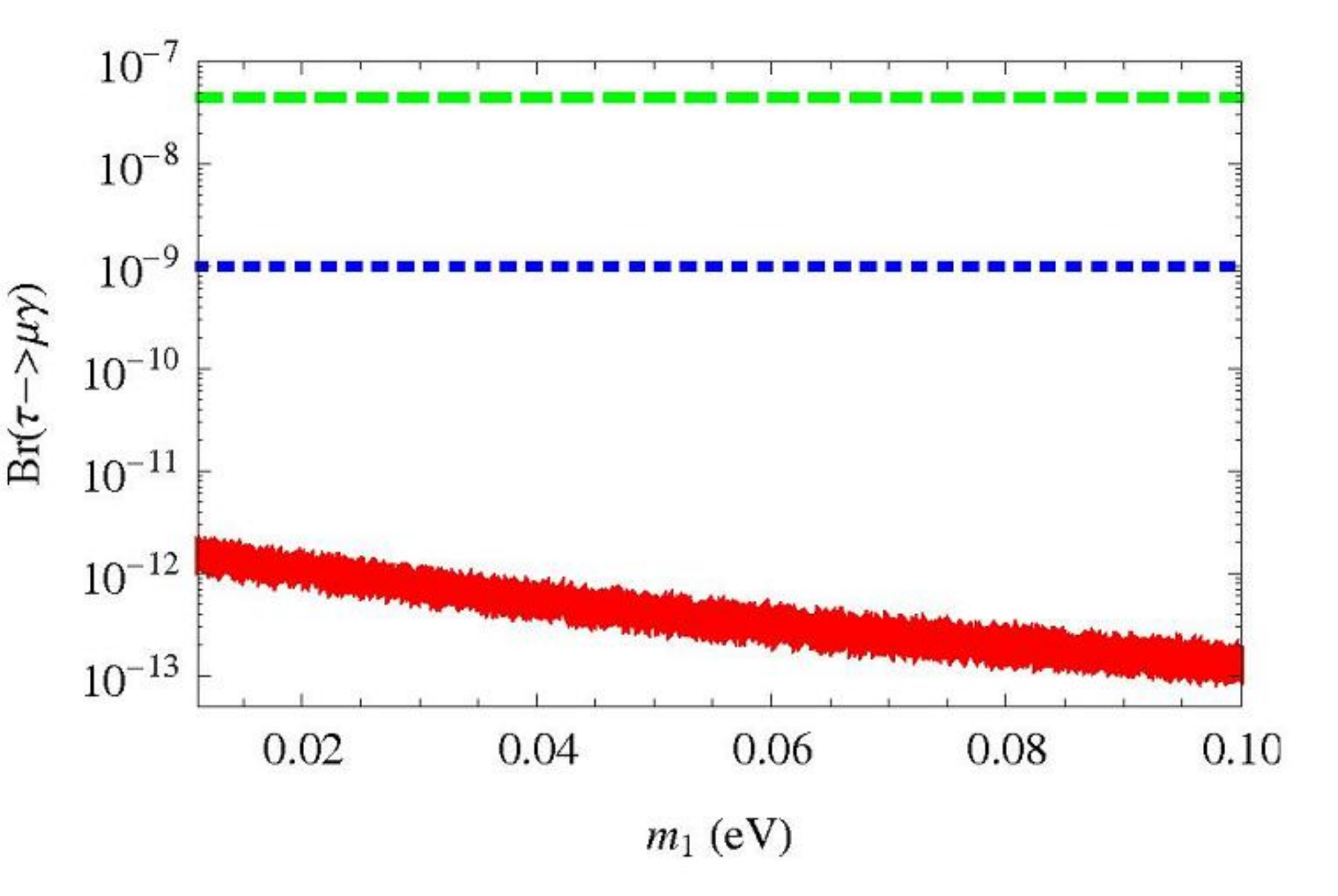}
\includegraphics[scale=1,width=3.75cm]{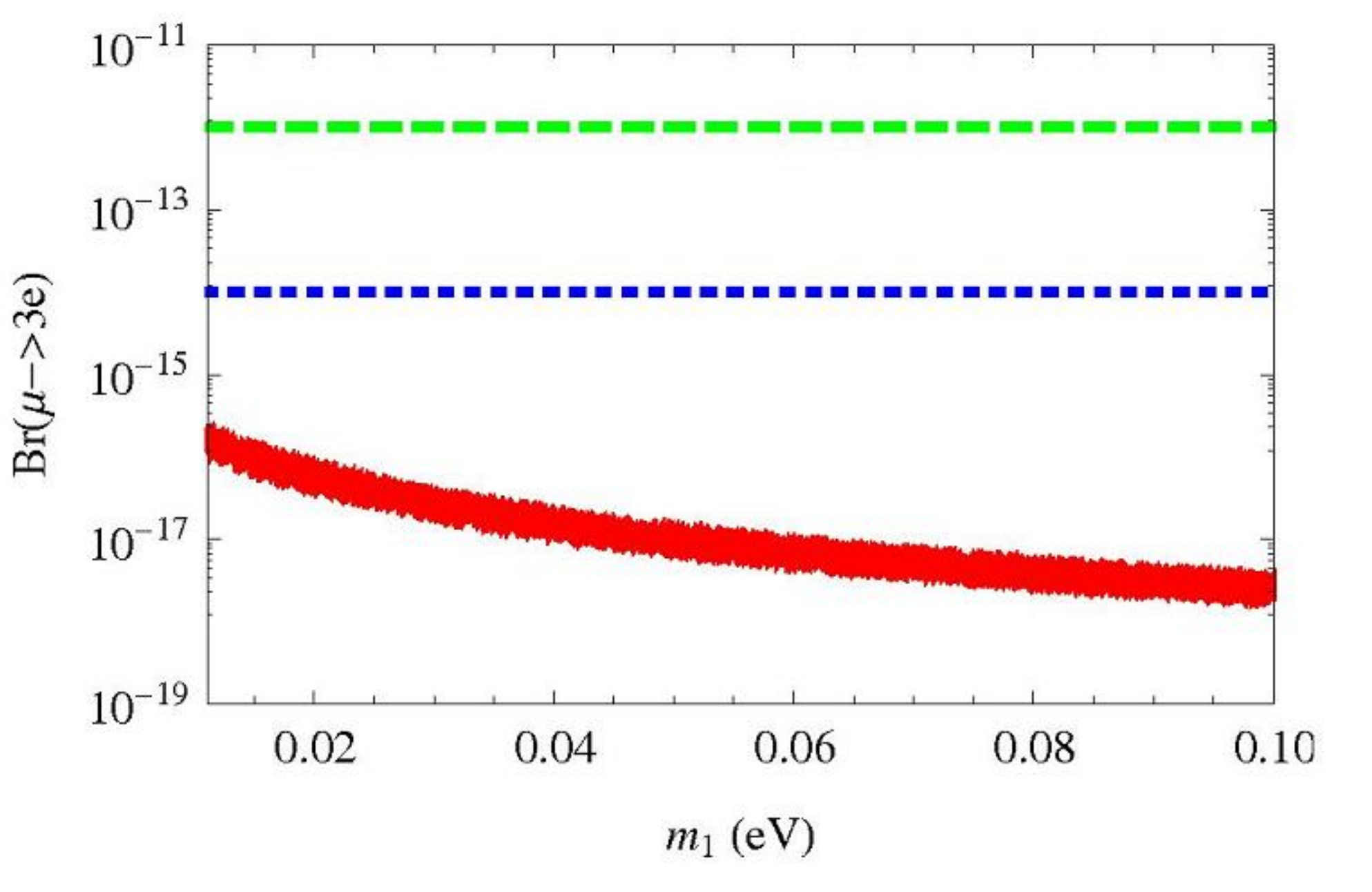}\\\\
\includegraphics[scale=1,width=3.75cm]{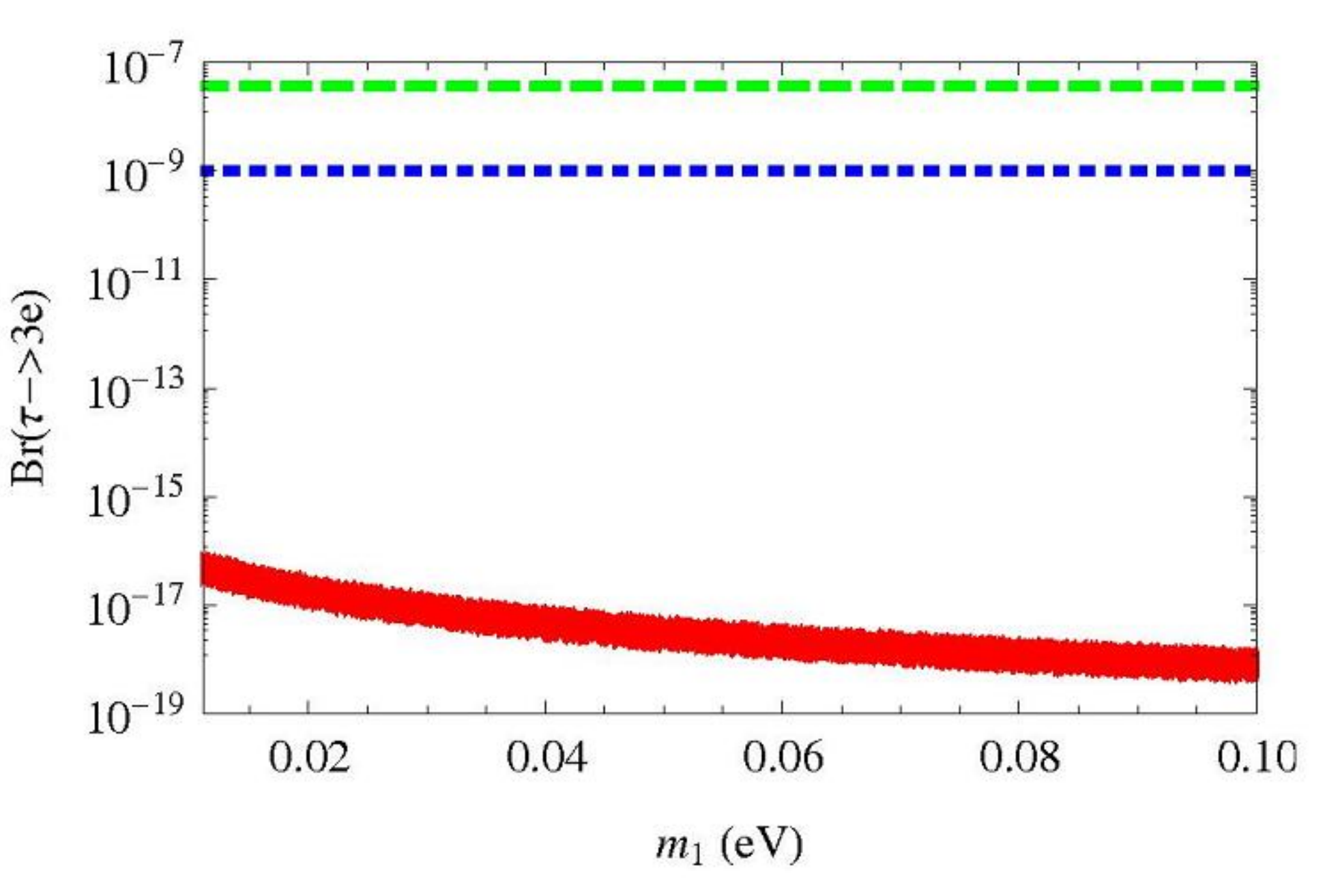}
\includegraphics[scale=1,width=3.75cm]{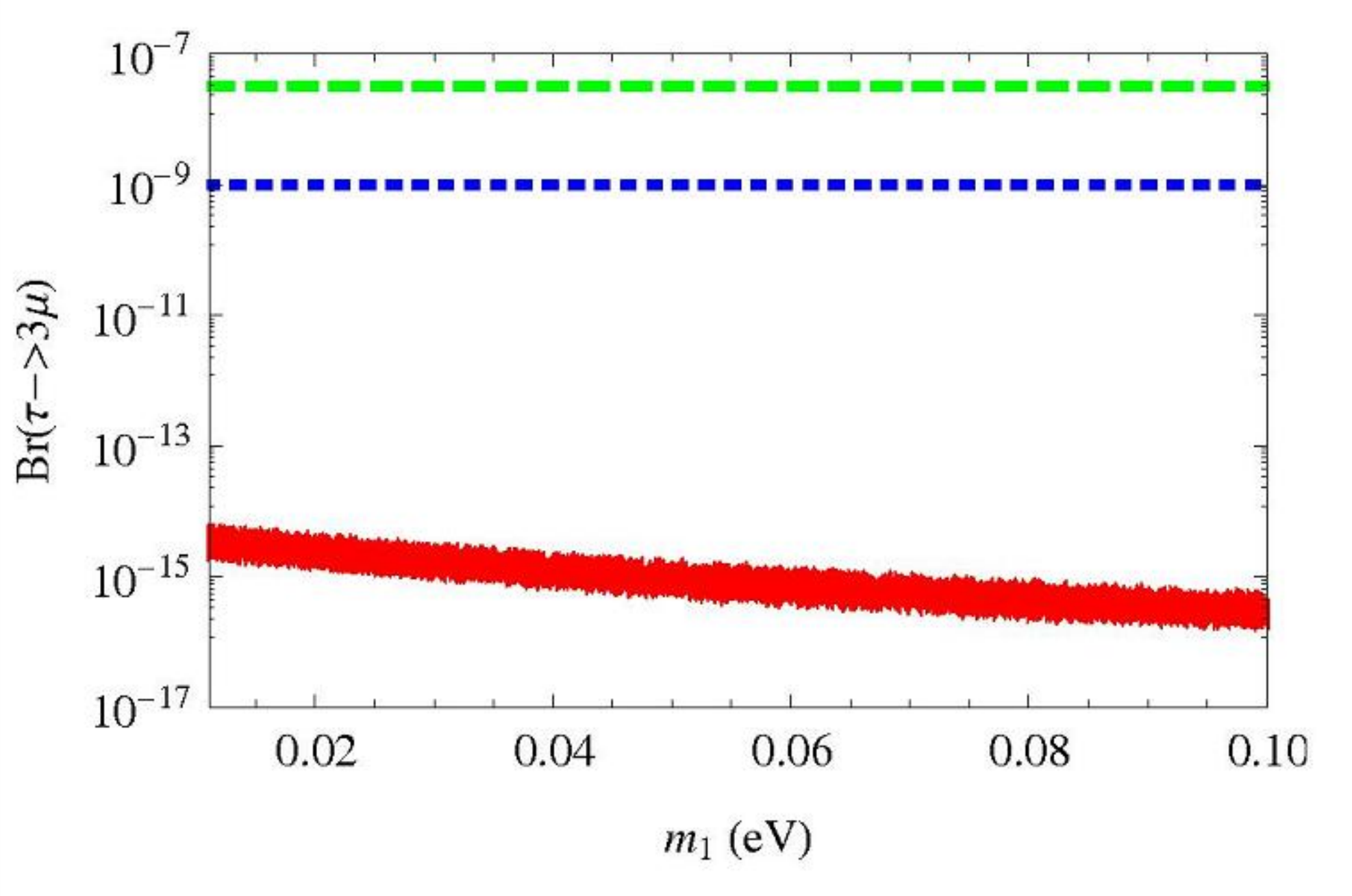}
\includegraphics[scale=1,width=3.75cm]{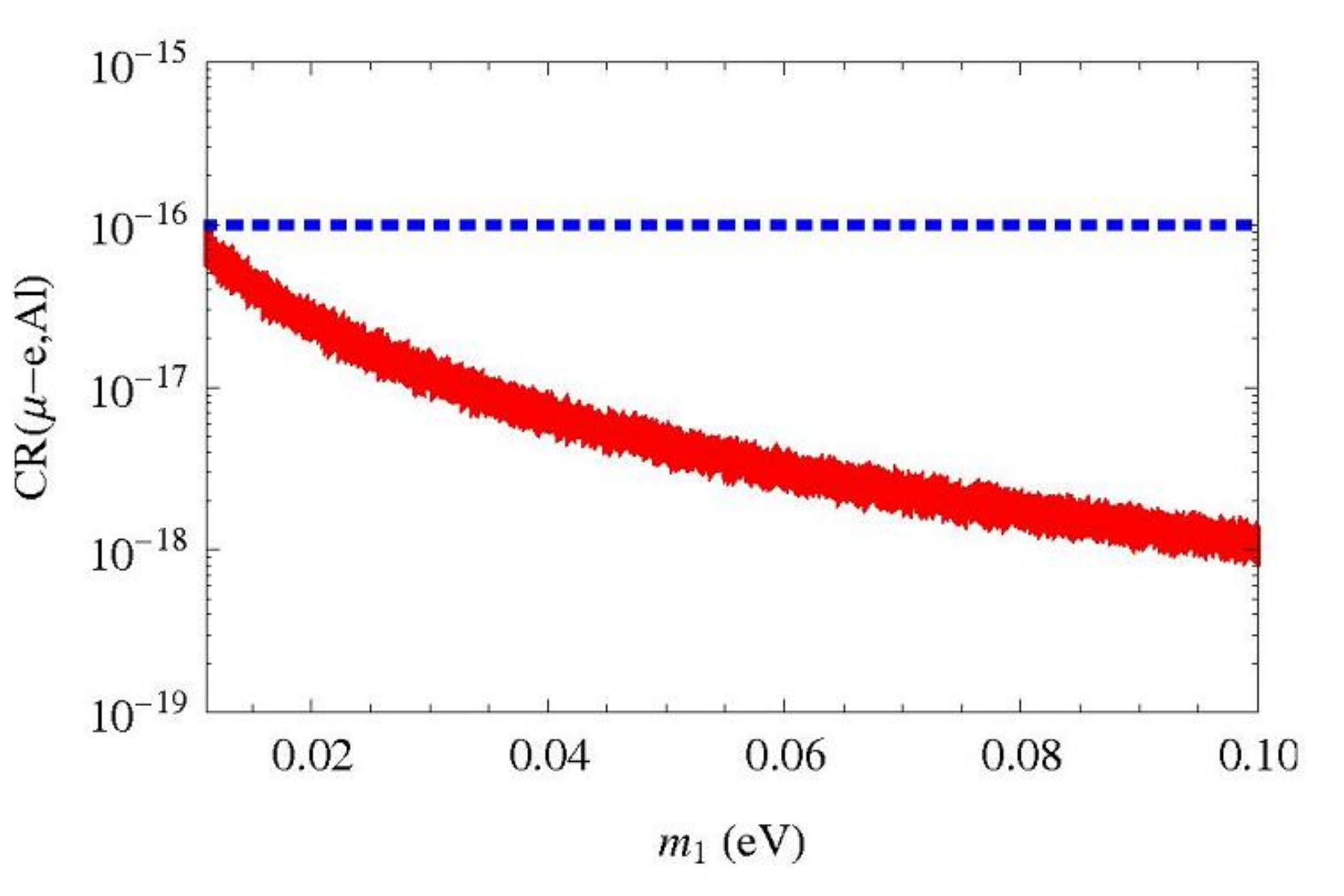}
\includegraphics[scale=1,width=3.75cm]{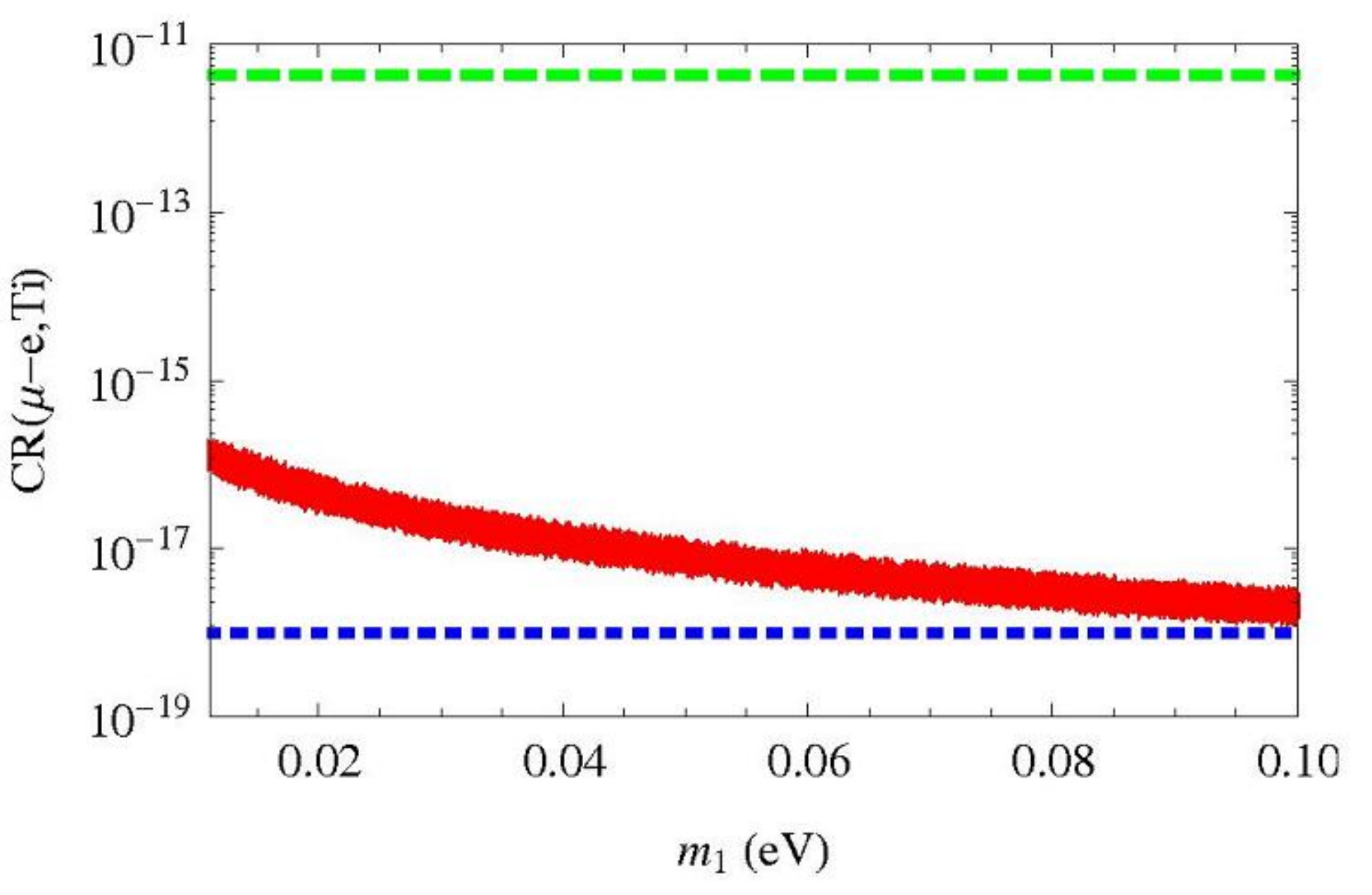}
\end{tabular}
\caption{\label{fig:Ding_NH_LO}Dependence of
$Br(\ell_i\rightarrow\ell_j\gamma)$, $Br(\ell_i\rightarrow3\ell_j)$,
$CR(\mu-e,Al)$ and $CR(\mu-e,Ti)$ on the lightest neutrino mass
$m_1$ in the Ding 's $S_4$ model for normal hierarchy spectrum. The
bands have been obtained by varying $\Delta m^2_{21}$ and $\Delta
m^2_{31}$ in their $3\sigma$ allowed range. The dashed and dotted
lines represent the present and future experimental sensitivity
respectively. There is still no upper bound for $CR(\mu-e,Al)$ at
present.}
\end{center}
\end{figure}

\begin{figure}[hptb]
\begin{center}
\begin{tabular}{c}
\includegraphics[scale=1,width=3.75cm]{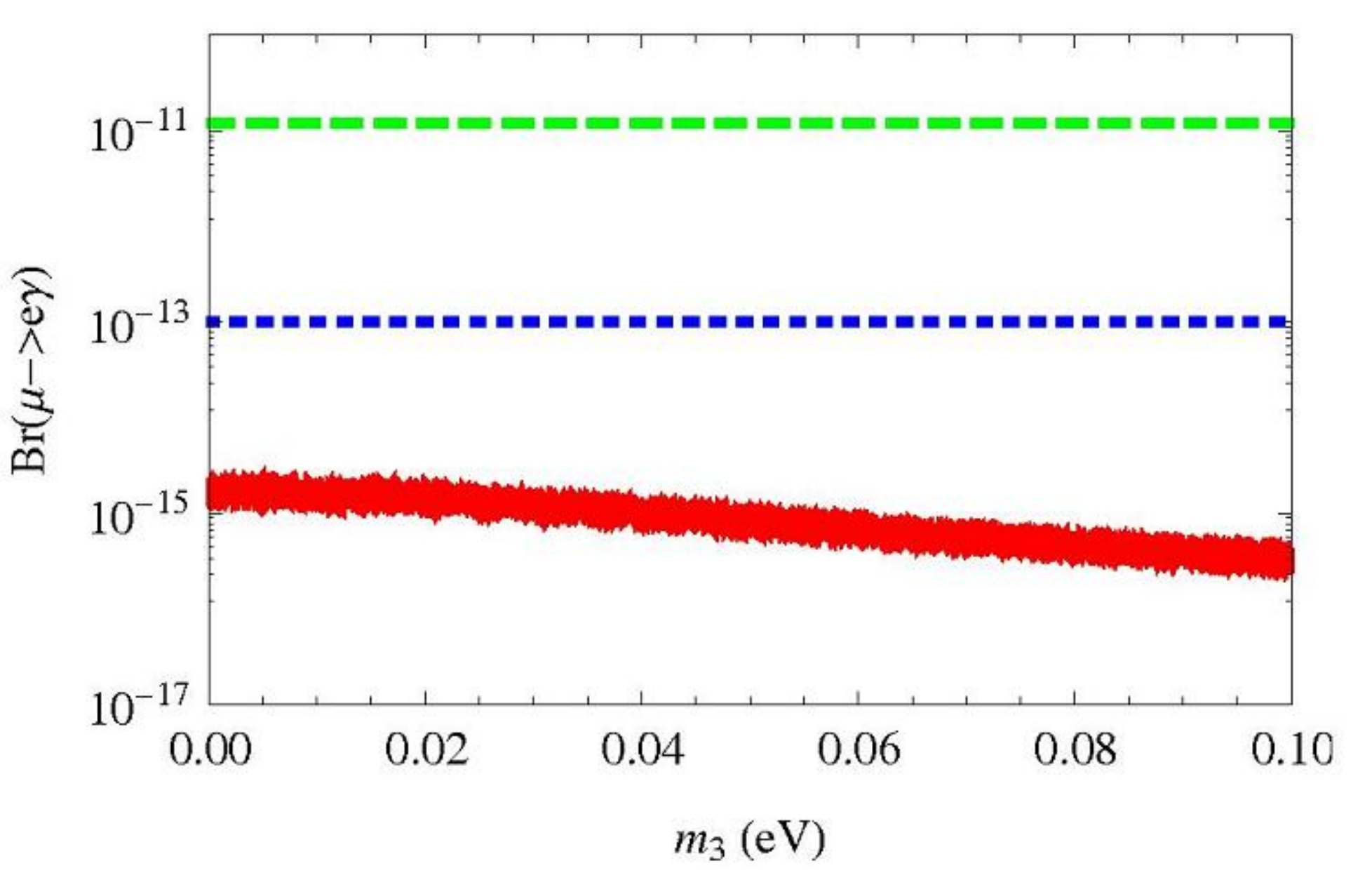}
\includegraphics[scale=1,width=3.75cm]{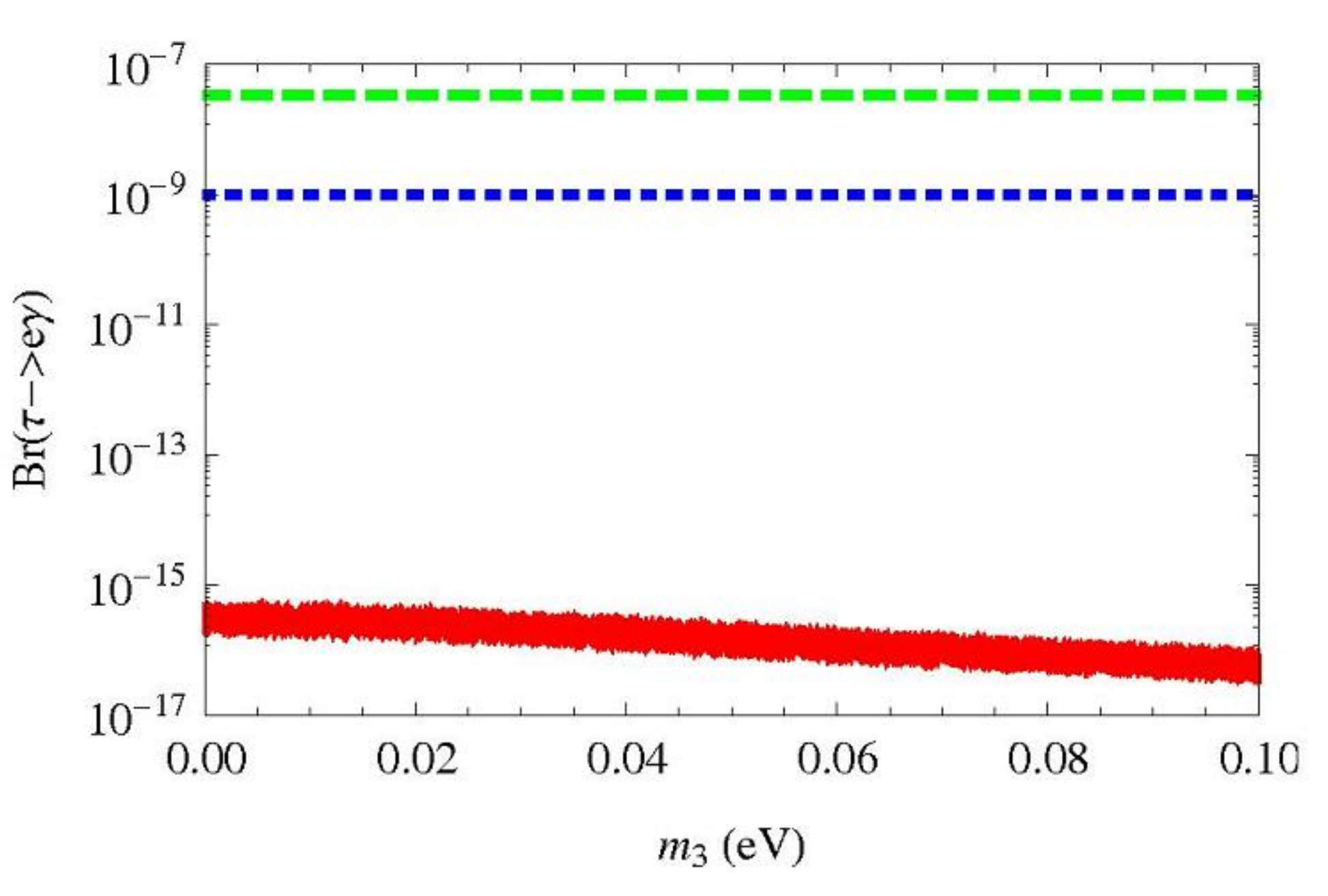}
\includegraphics[scale=1,width=3.75cm]{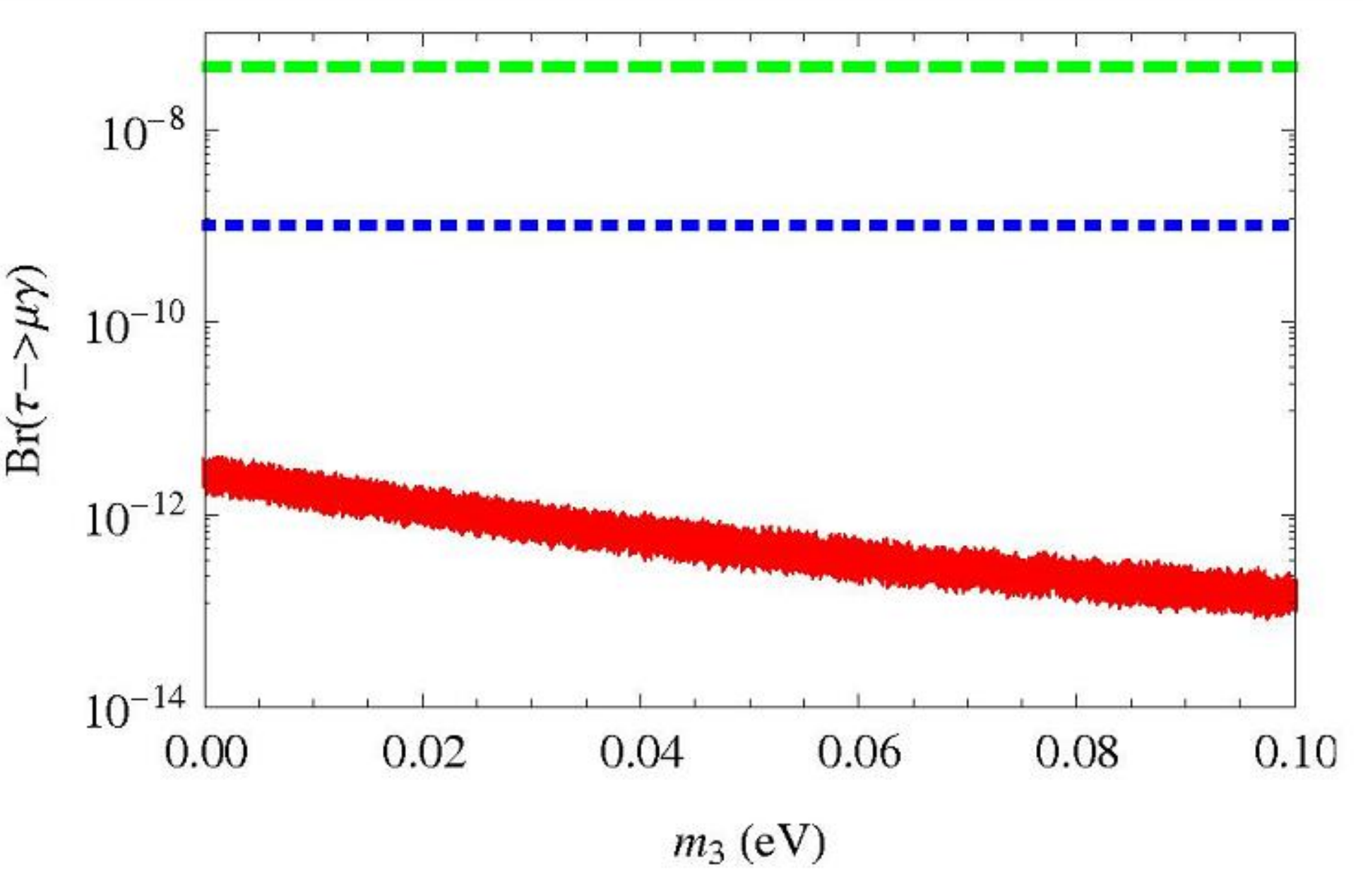}
\includegraphics[scale=1,width=3.75cm]{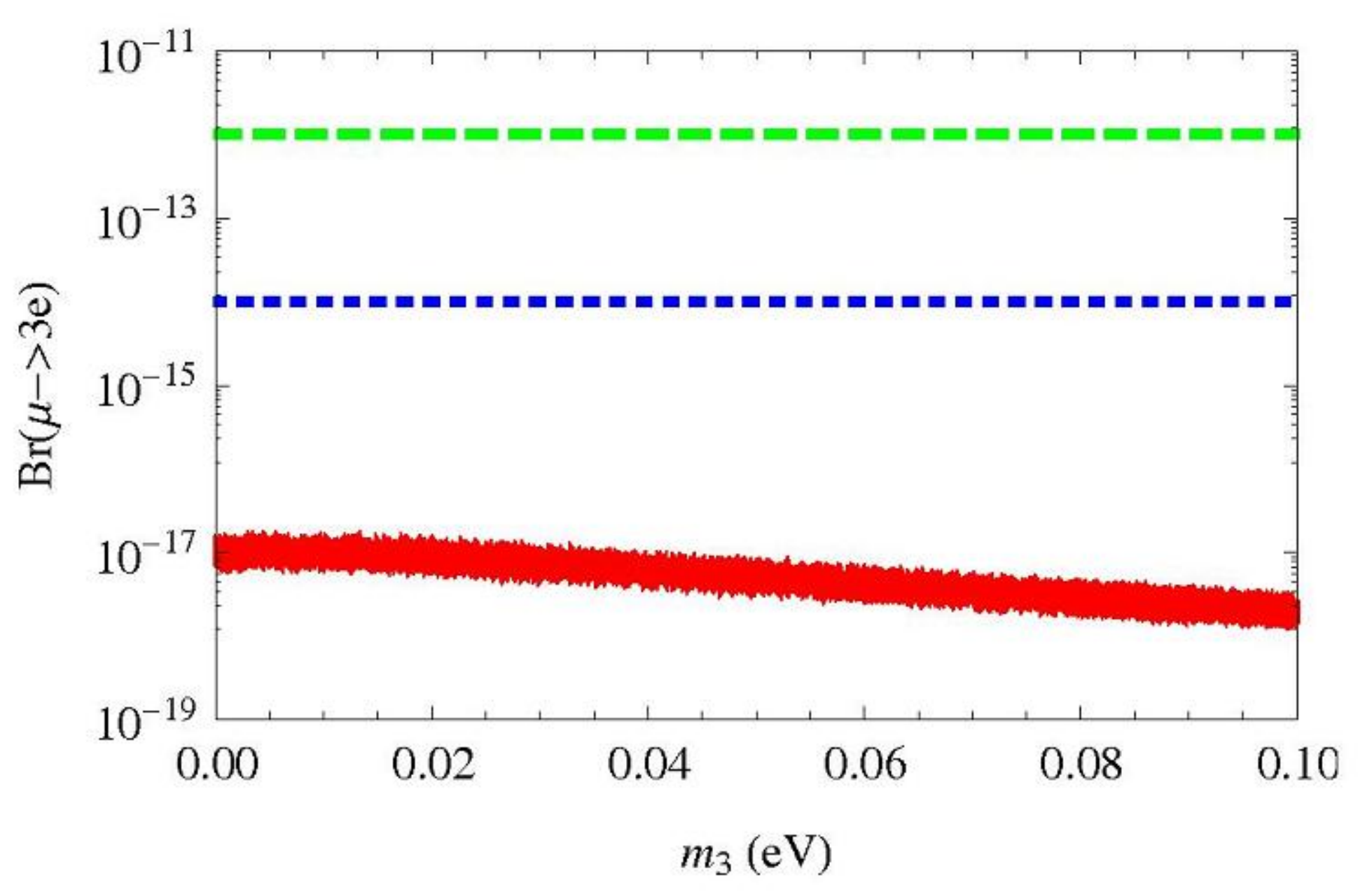}\\\\
\includegraphics[scale=1,width=3.75cm]{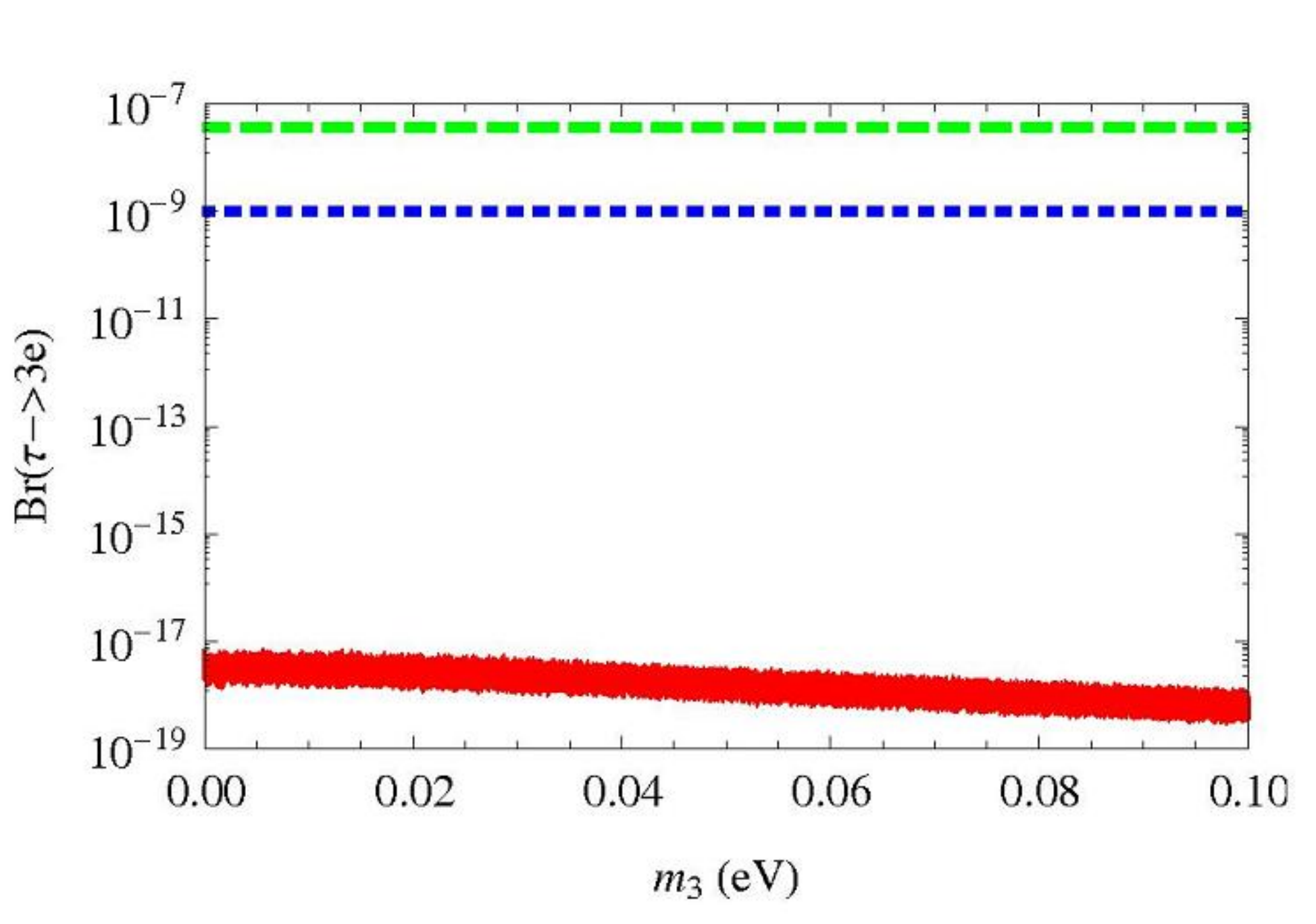}
\includegraphics[scale=1,width=3.75cm]{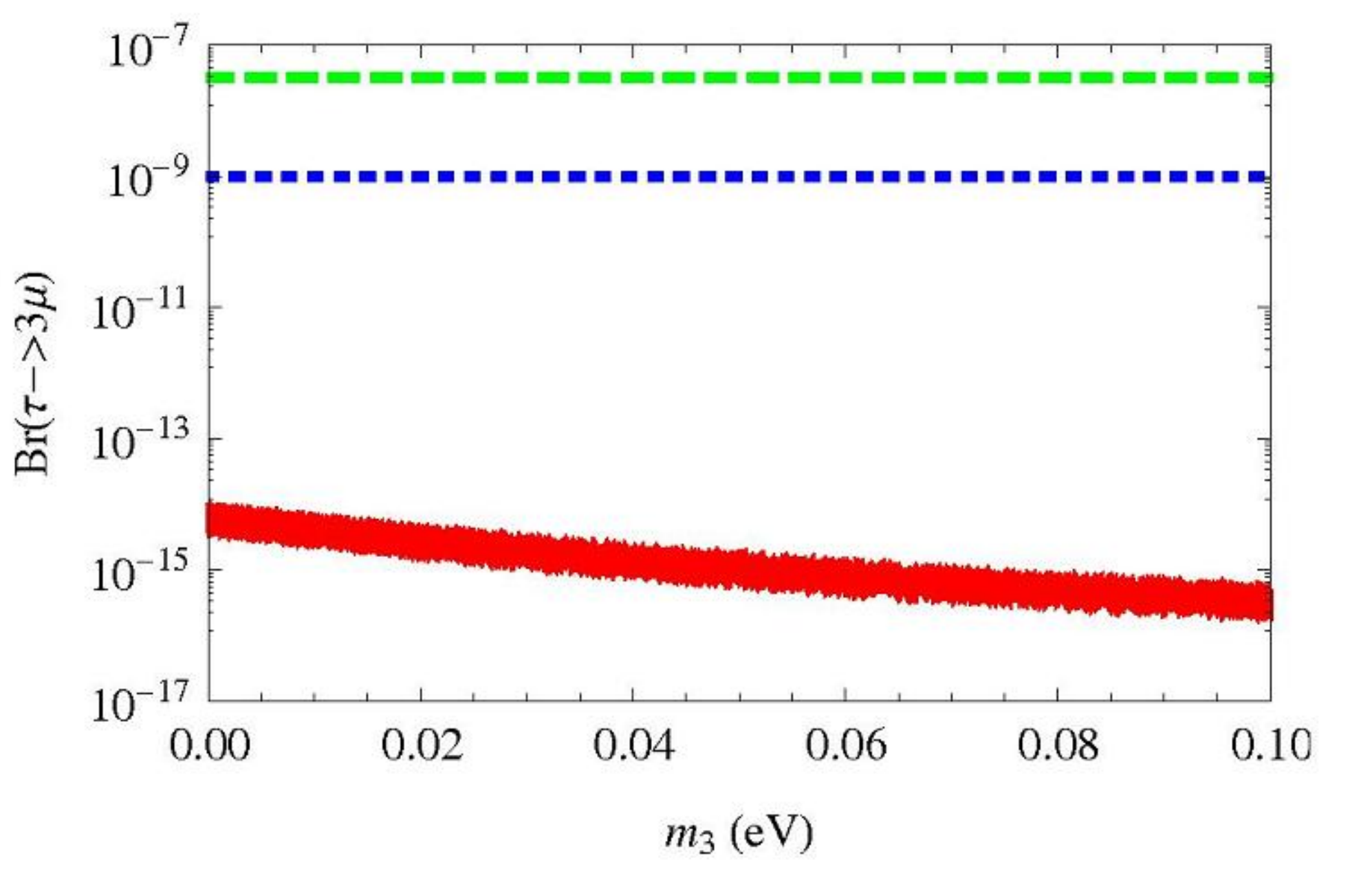}
\includegraphics[scale=1,width=3.75cm]{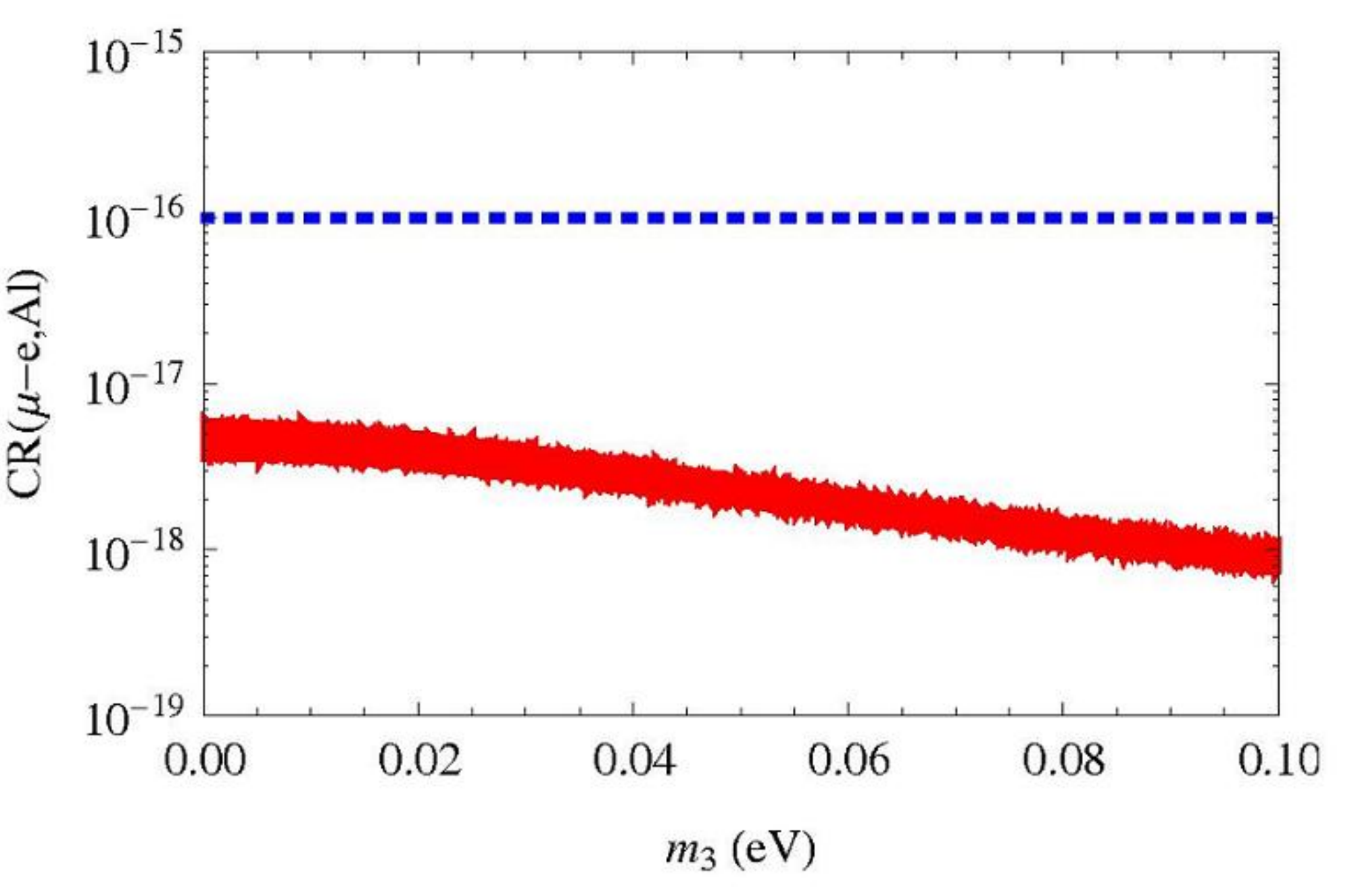}
\includegraphics[scale=1,width=3.75cm]{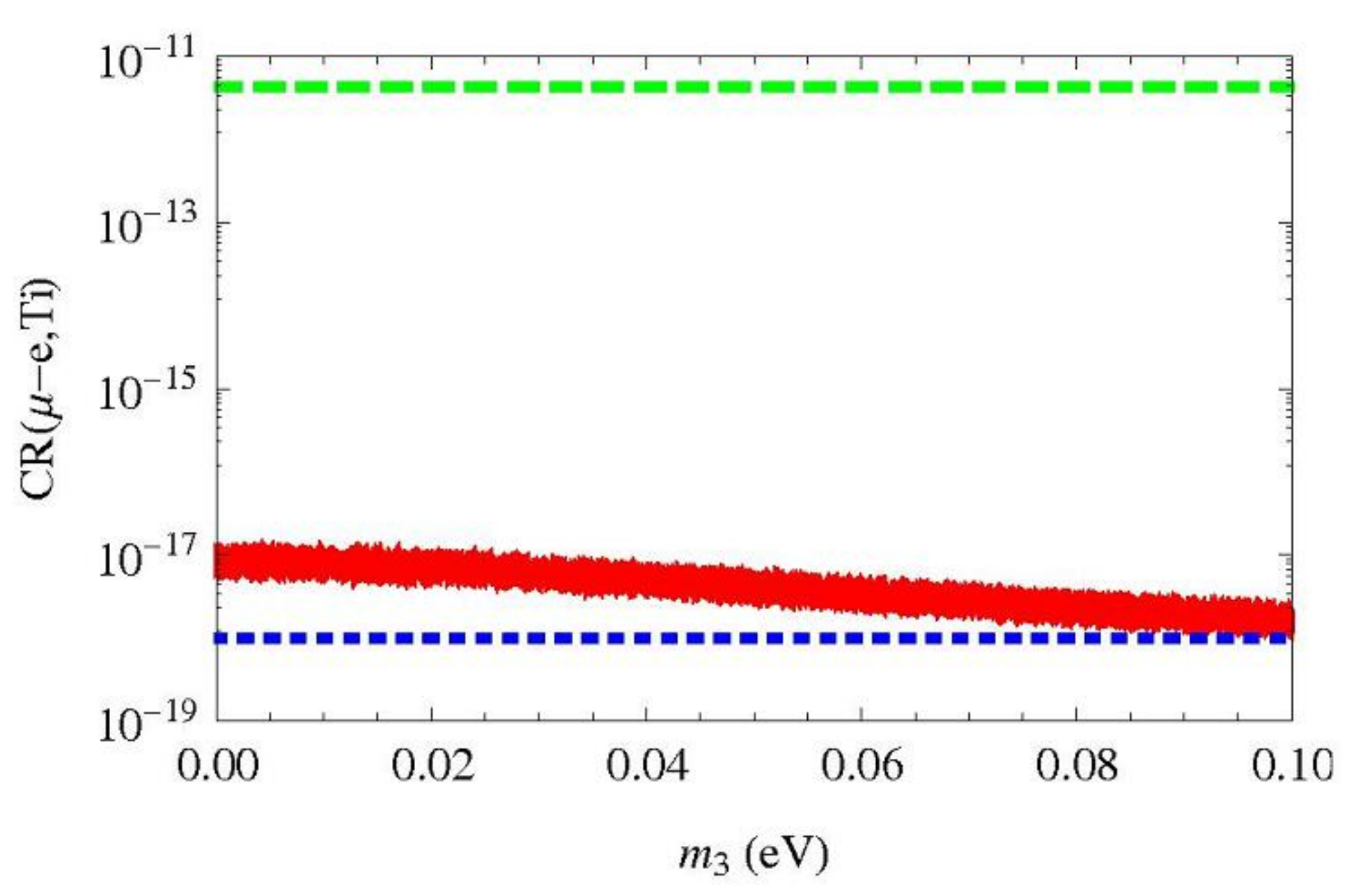}
\end{tabular}
\caption{\label{fig:Ding_IH_LO}Dependence of
$Br(\ell_i\rightarrow\ell_j\gamma)$, $Br(\ell_i\rightarrow3\ell_j)$,
$CR(\mu-e,Al)$ and $CR(\mu-e,Ti)$ on the lightest neutrino mass
$m_3$ in the Ding's $S_4$ model for inverted hierarchy spectrum. The
bands have been obtained by varying $\Delta m^2_{21}$ and $\Delta
m^2_{31}$ in their $3\sigma$ allowed range. The dashed and dotted
lines represent the present and future experimental sensitivity
respectively.}
\end{center}
\end{figure}

\begin{figure}[hptb]
\begin{center}
\begin{tabular}{c}
\includegraphics[scale=1,width=5.0cm]{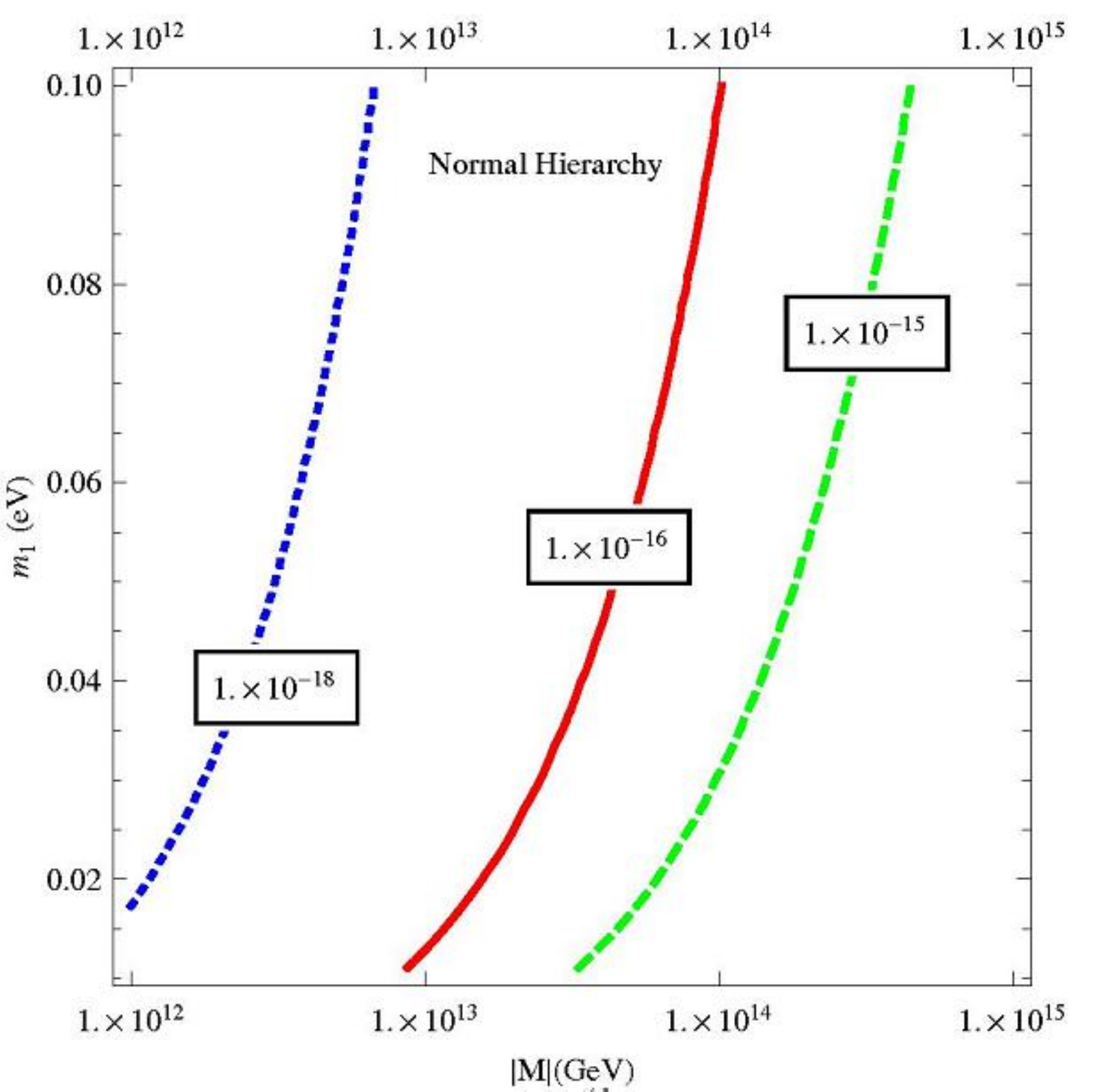}
\hspace{1cm}
\includegraphics[scale=1,width=5.0cm]{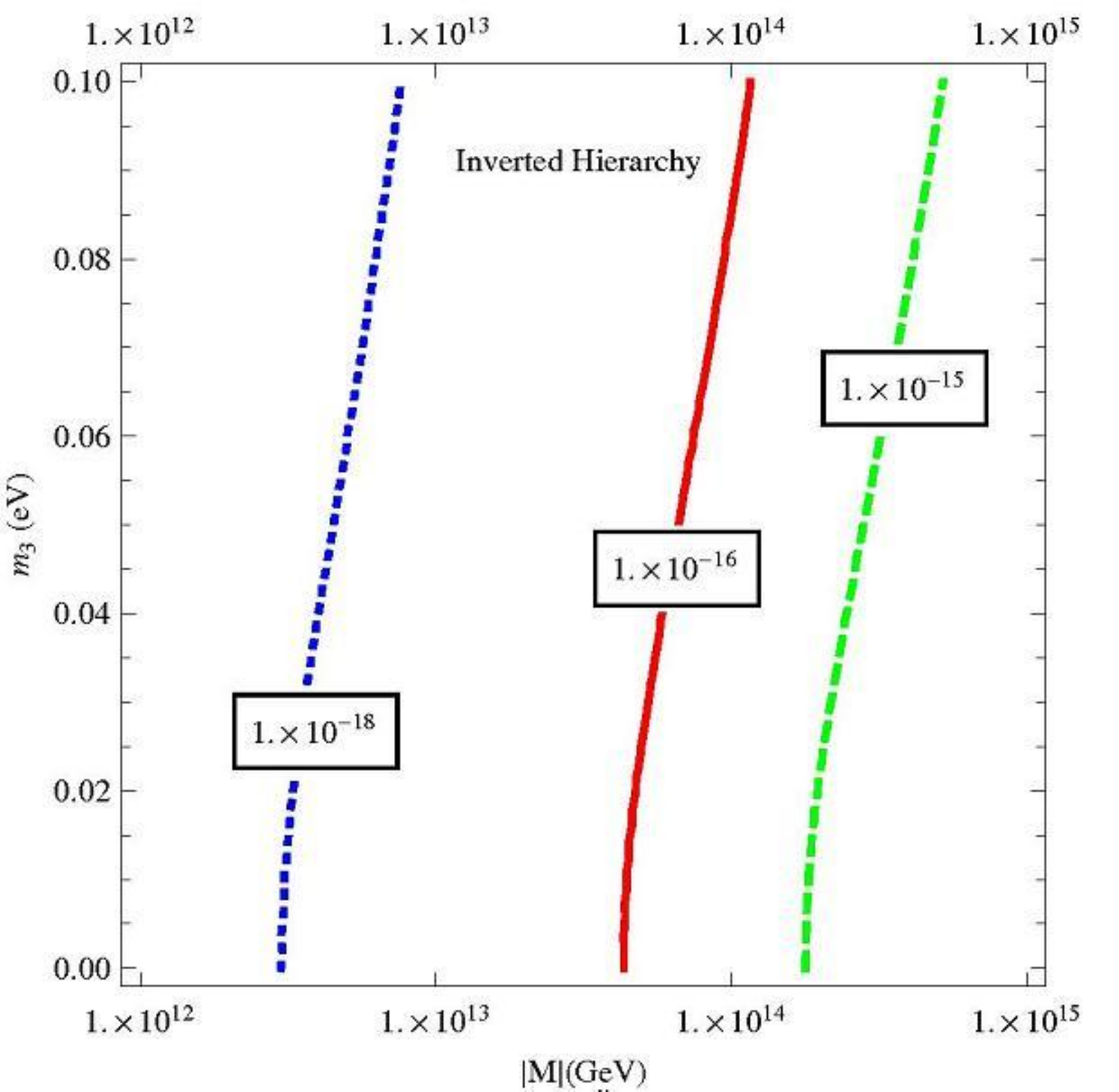}
\end{tabular}
\caption{\label{fig:Ding_CS}The contour plot for $CR(\mu-e,Ti)$ in
the parameter space of the lightest neutrino mass and right handed
neutrino mass $|M|$ in the Ding's $S_4$ model, where $\Delta
m^2_{21}$ and $\Delta m^2_{31}$ are chose to be the best fit values
$7.65\times10^{-5}{\rm\, eV^2}$ and $\pm2.40\times10^{-3}{\rm\,
eV^2}$ respectively.}
\end{center}
\end{figure}

Since the $\mu e$ and $\tau e$ involved LFV processes are predicted
to be suppressed at LO in the case of IH spectrum, the NLO
contributions may be comparable to the leading ones. Therefore we
should take into account the NLO contributions in order to reach a
much more solid conclusion. In the following numerical analysis, the
NLO effective parameters $m^{\ell}_{21}/m^{\ell}_{22}$,
$m^{\ell}_{31}/m^{\ell}_{33}$ and $m^{\ell}_{32}/m^{\ell}_{33}$ in
Eq.(\ref{45}) are treated as random complex numbers with absolute
value between 0 and 2. $a$, $b$, $\tilde{a}$ and $\tilde{b}$ are
taken to be complex random number with absolute value in the range
of 0.01-0.1. The heavy neutrino mass $|M|$ is fixed at
$1.0\times10^{13}$ GeV in order to compare with LO results, and the
expansion parameters $\epsilon$ is set equal to 0.04. The scatter
plots of LFV branching ratios for the NH and IH spectrum are
displayed in Fig.\ref{fig:Ding_NH_NLO} and Fig.\ref{fig:Ding_IH_NLO}
respectively.

\begin{figure}[hptb]
\begin{center}
\begin{tabular}{c}
\includegraphics[scale=1,width=3.75cm]{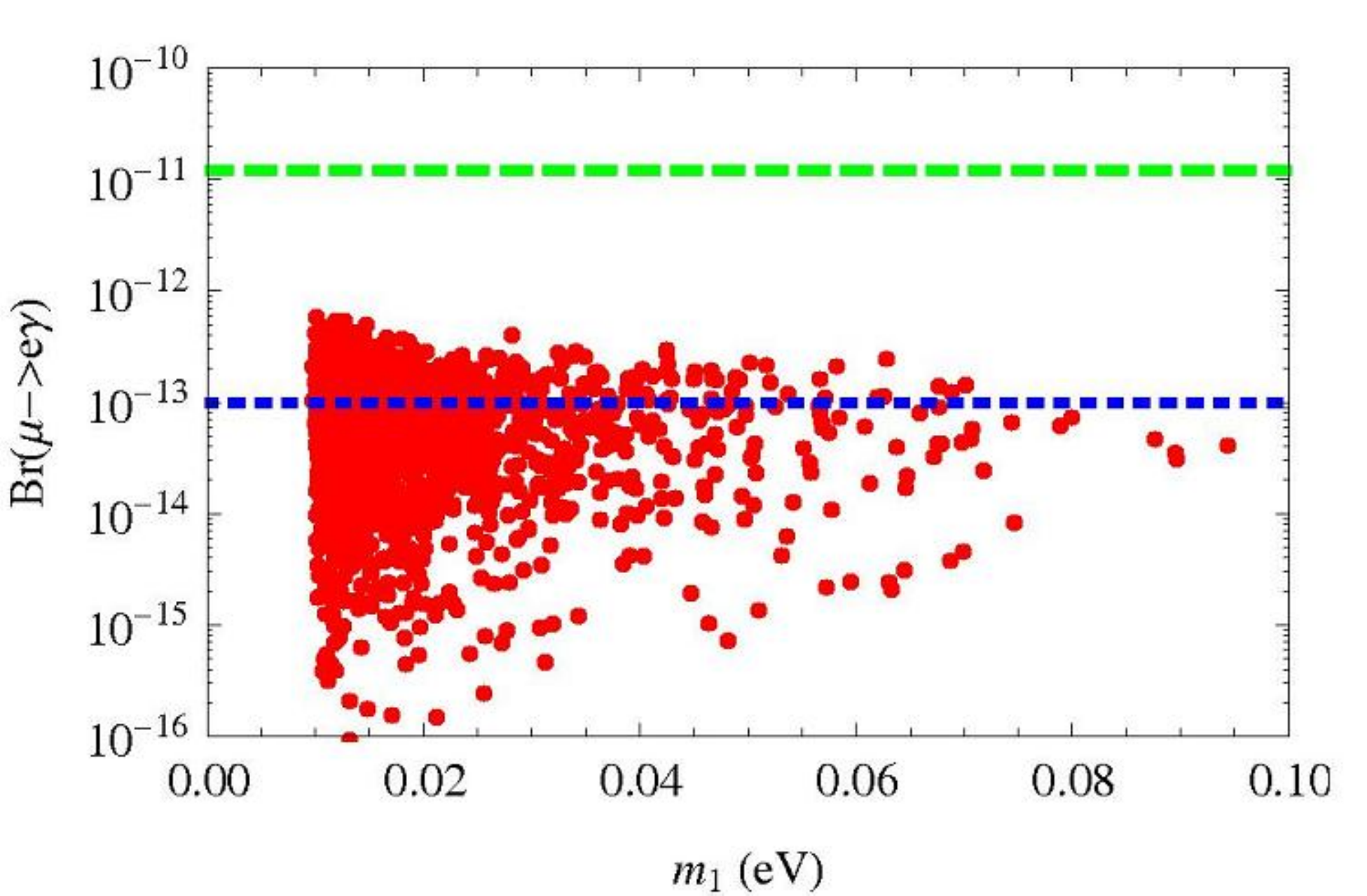}
\includegraphics[scale=1,width=3.75cm]{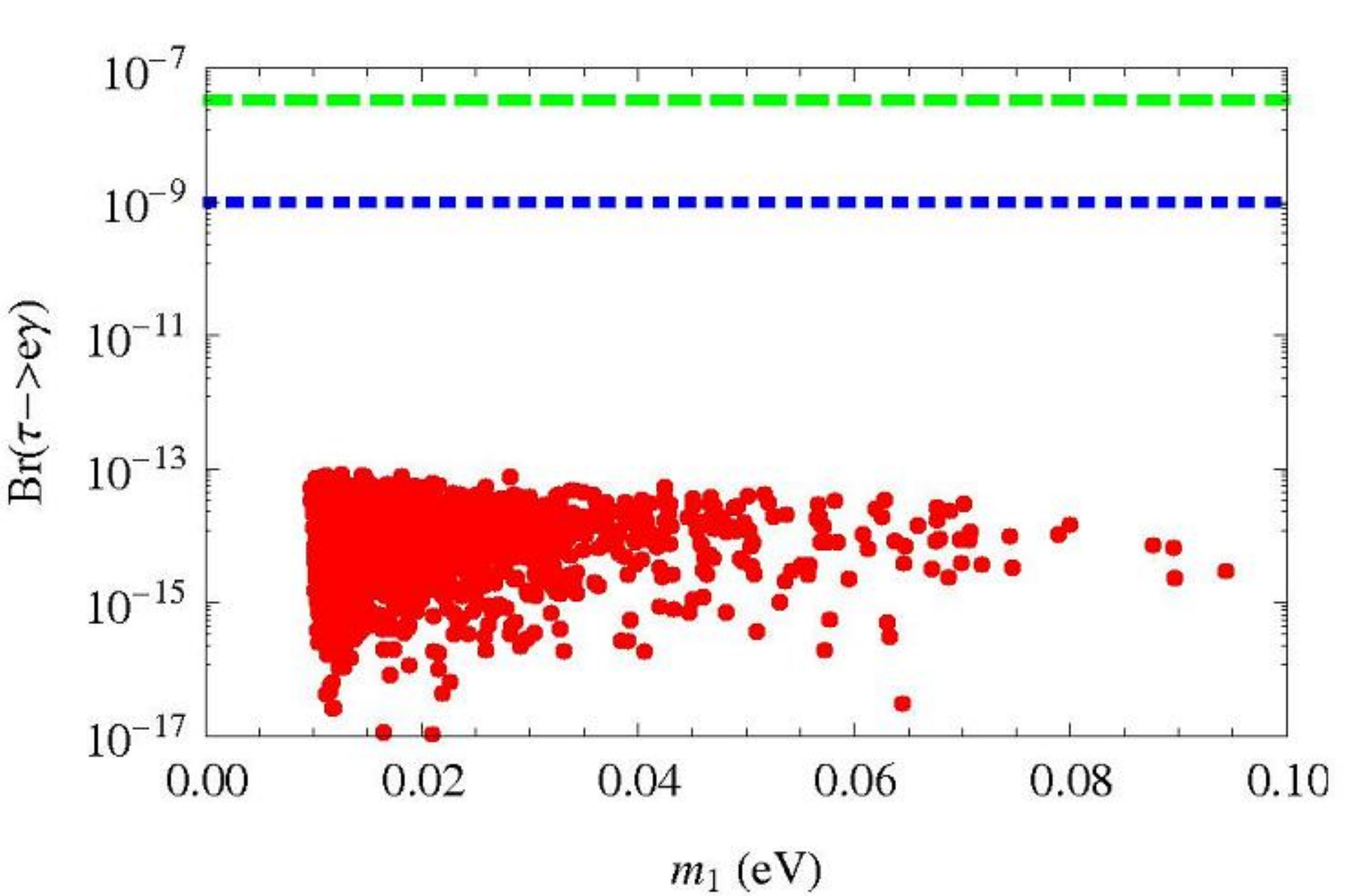}
\includegraphics[scale=1,width=3.75cm]{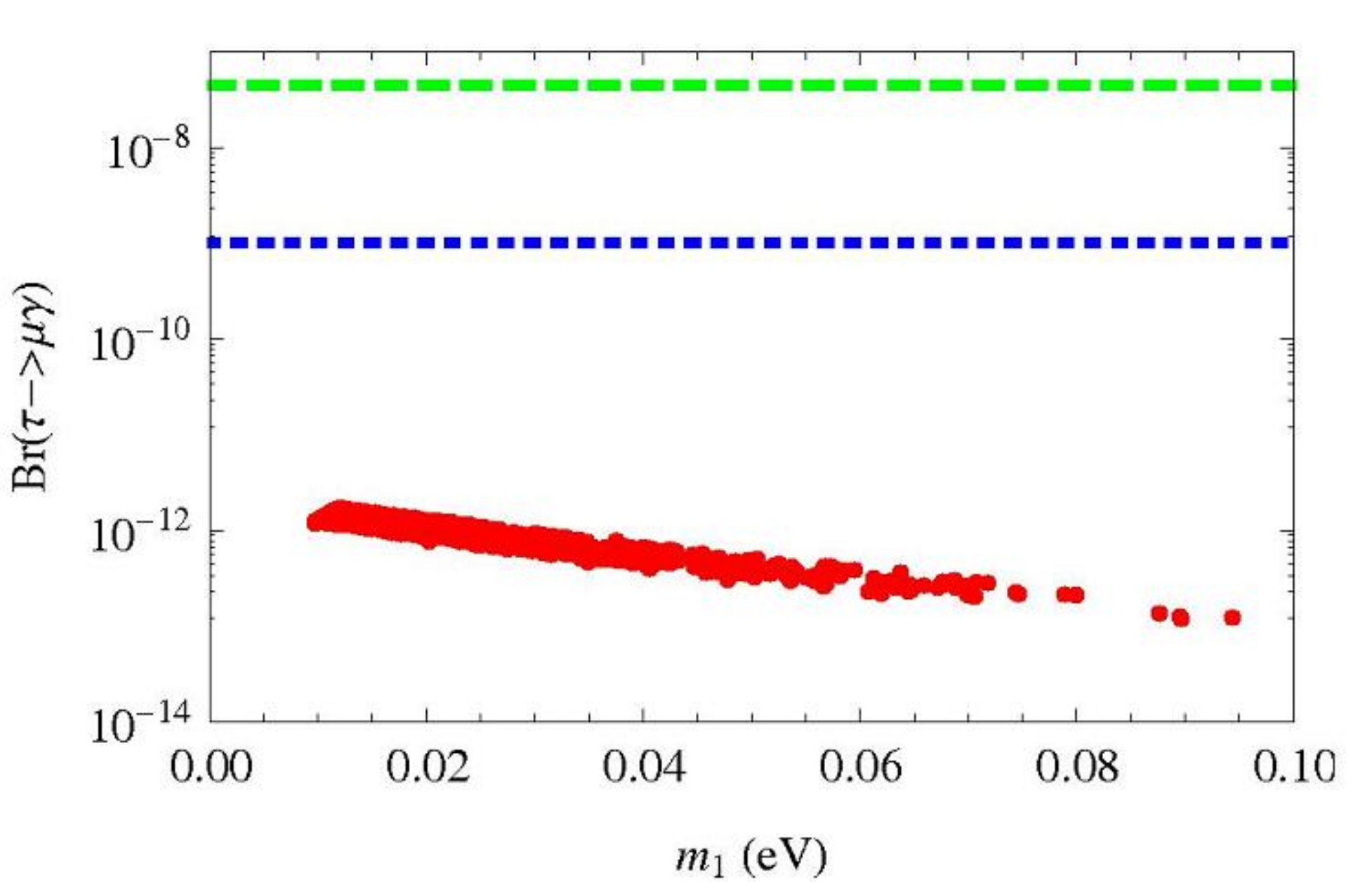}
\includegraphics[scale=1,width=3.75cm]{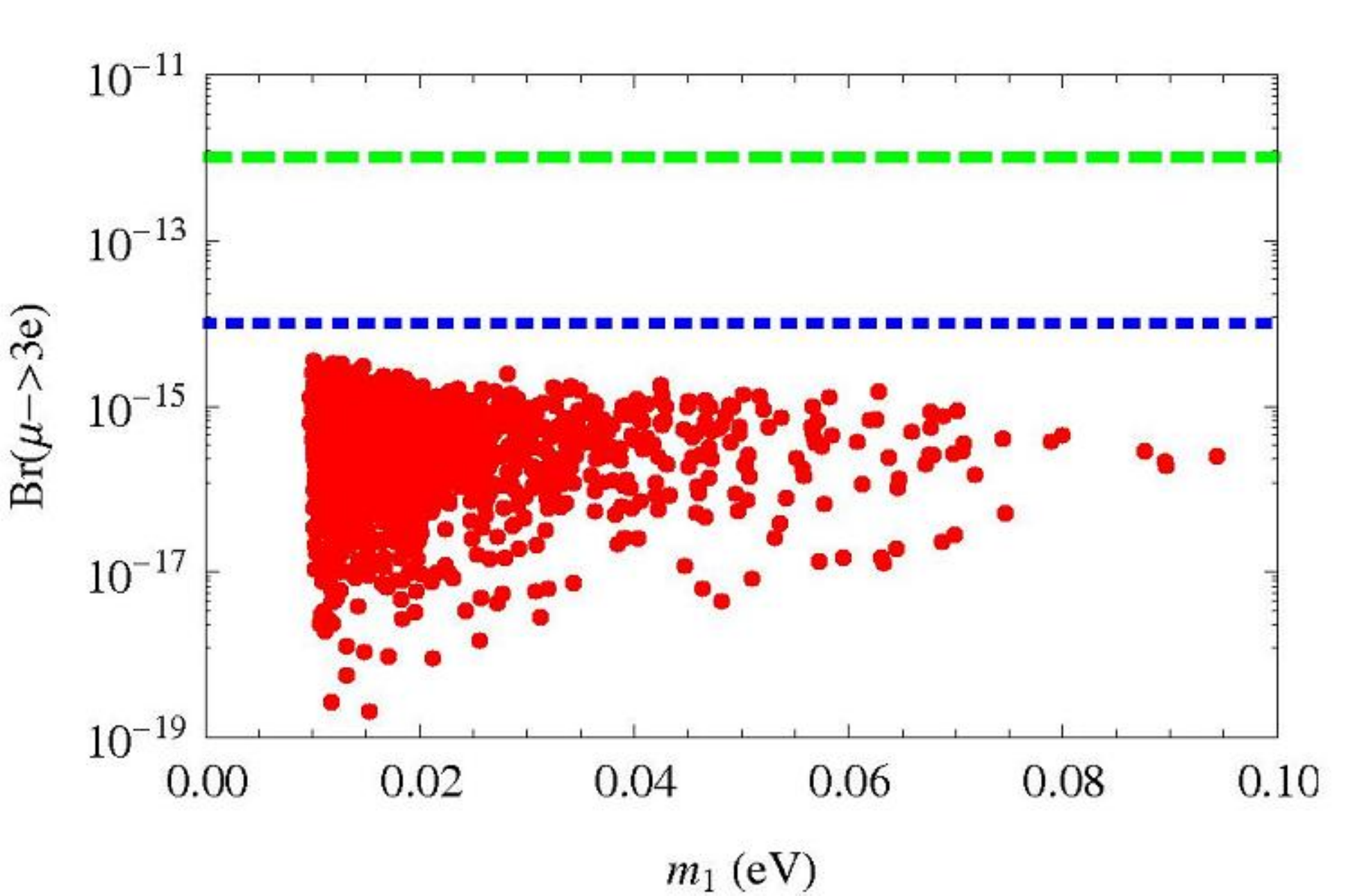}\\\\
\includegraphics[scale=1,width=3.75cm]{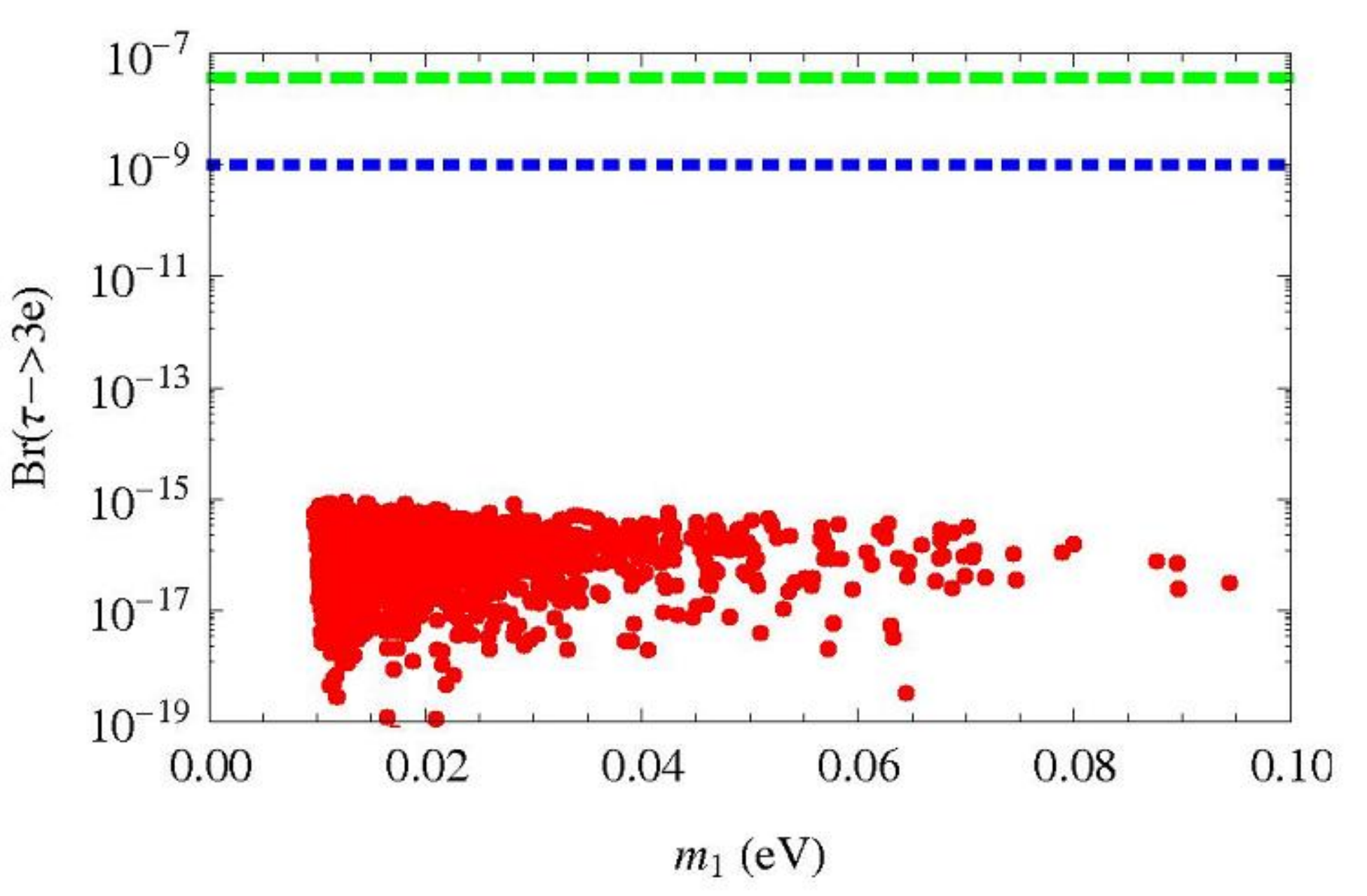}
\includegraphics[scale=1,width=3.75cm]{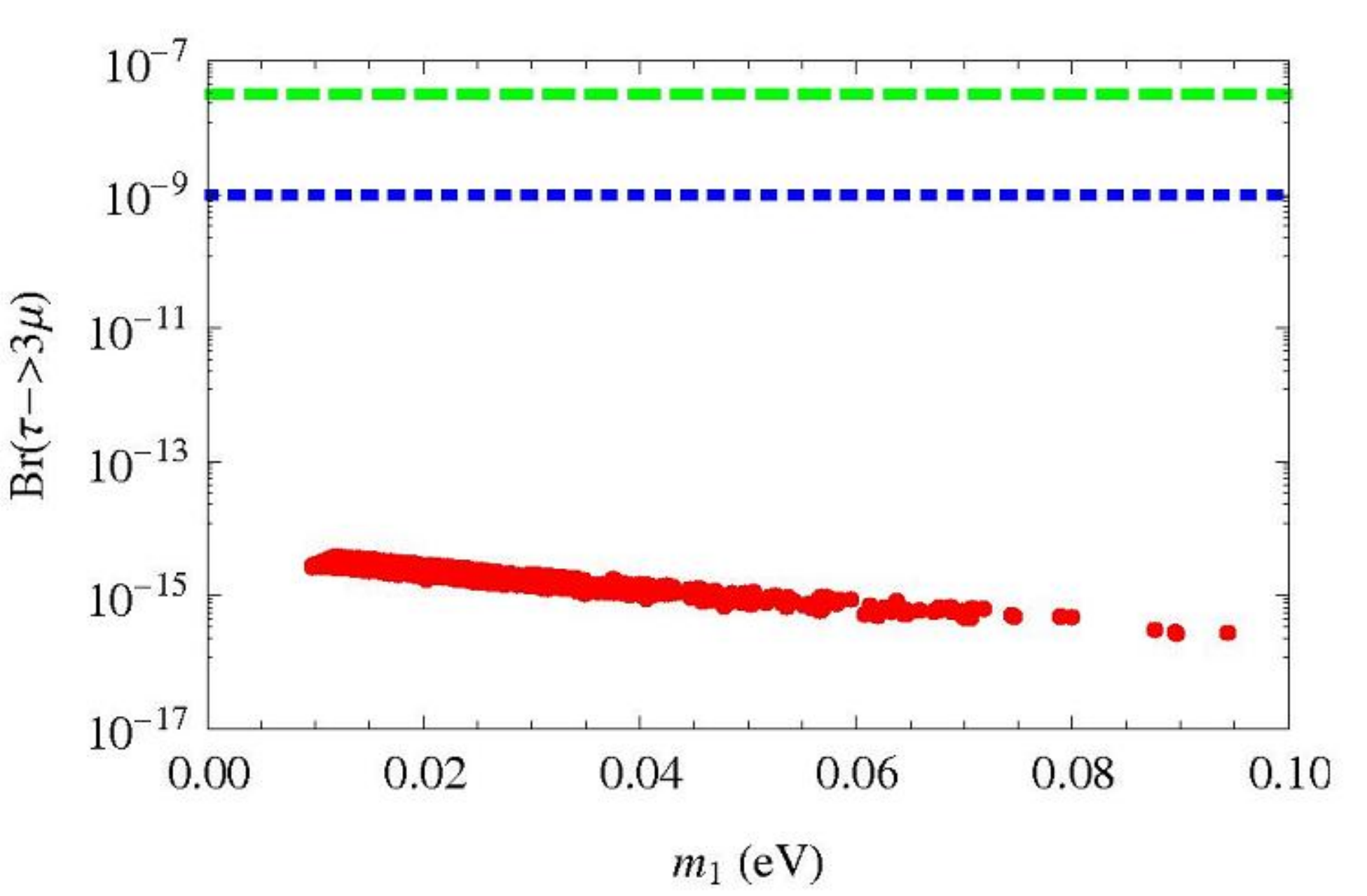}
\includegraphics[scale=1,width=3.75cm]{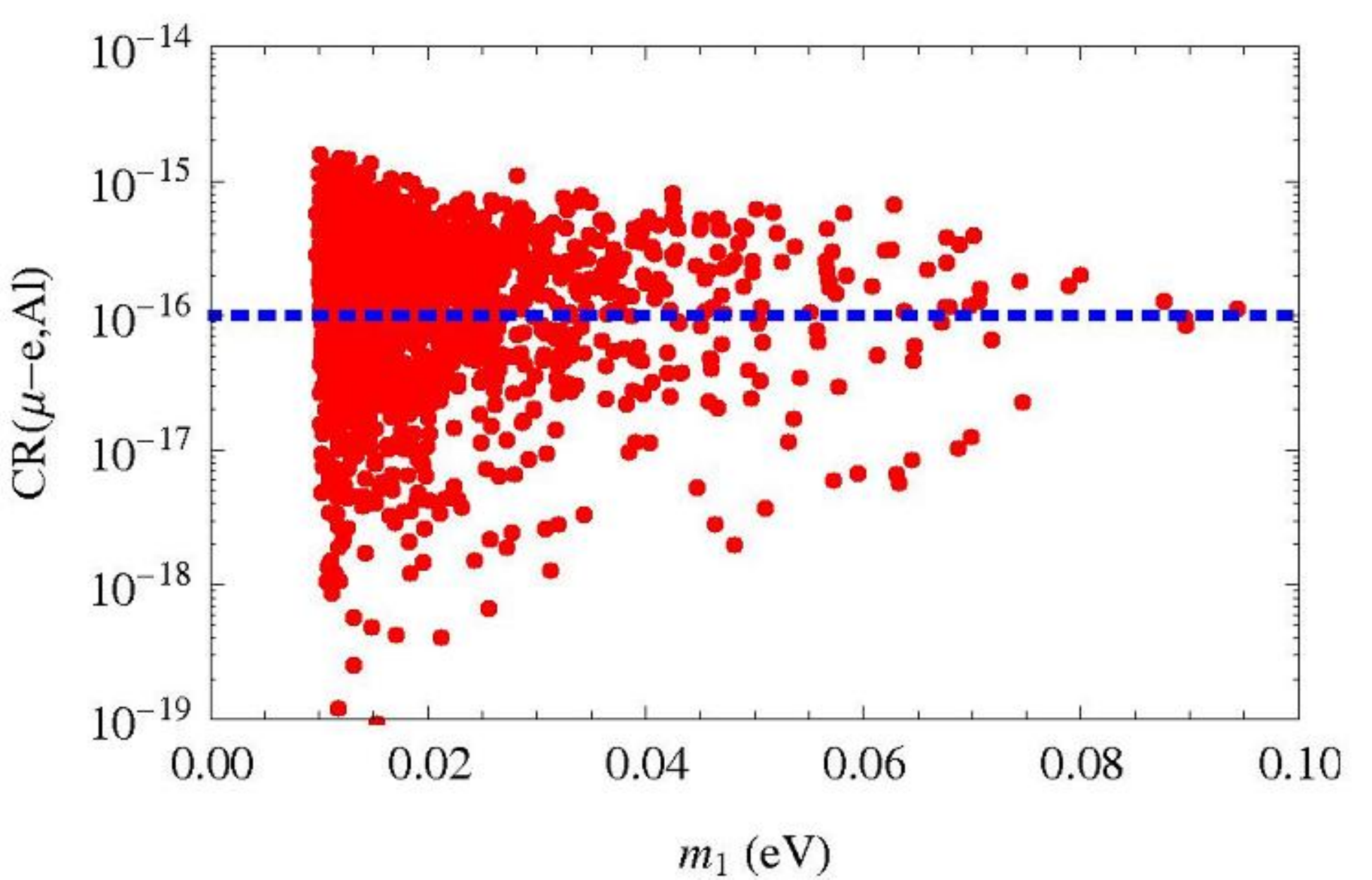}
\includegraphics[scale=1,width=3.75cm]{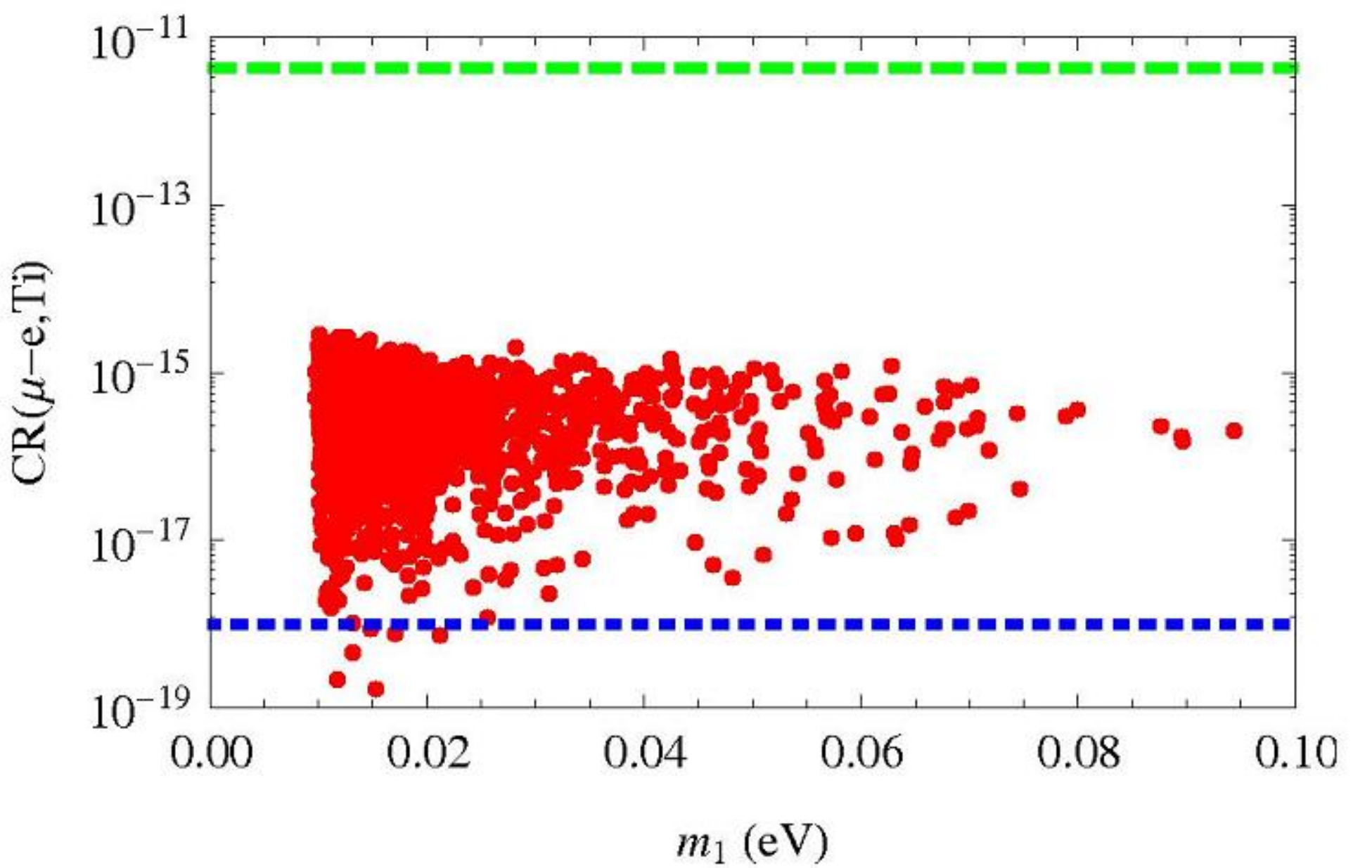}
\end{tabular}
\caption{\label{fig:Ding_NH_NLO} Scatter plot of
$Br(\ell_i\rightarrow\ell_j\gamma)$, $Br(\ell_i\rightarrow3\ell_j)$,
$CR(\mu-e,Al)$ and $CR(\mu-e,Ti)$ against the lightest neutrino mass
$m_1$ in Ding's $S_4$ model for normal hierarchy spectrum. The
dashed and dotted lines represent the present and future
experimental sensitivity respectively. There is no upper bound for
$CR(\mu-e,Al)$ so far.}
\end{center}
\end{figure}

\begin{figure}[hptb]
\begin{center}
\begin{tabular}{c}
\includegraphics[scale=1,width=3.75cm]{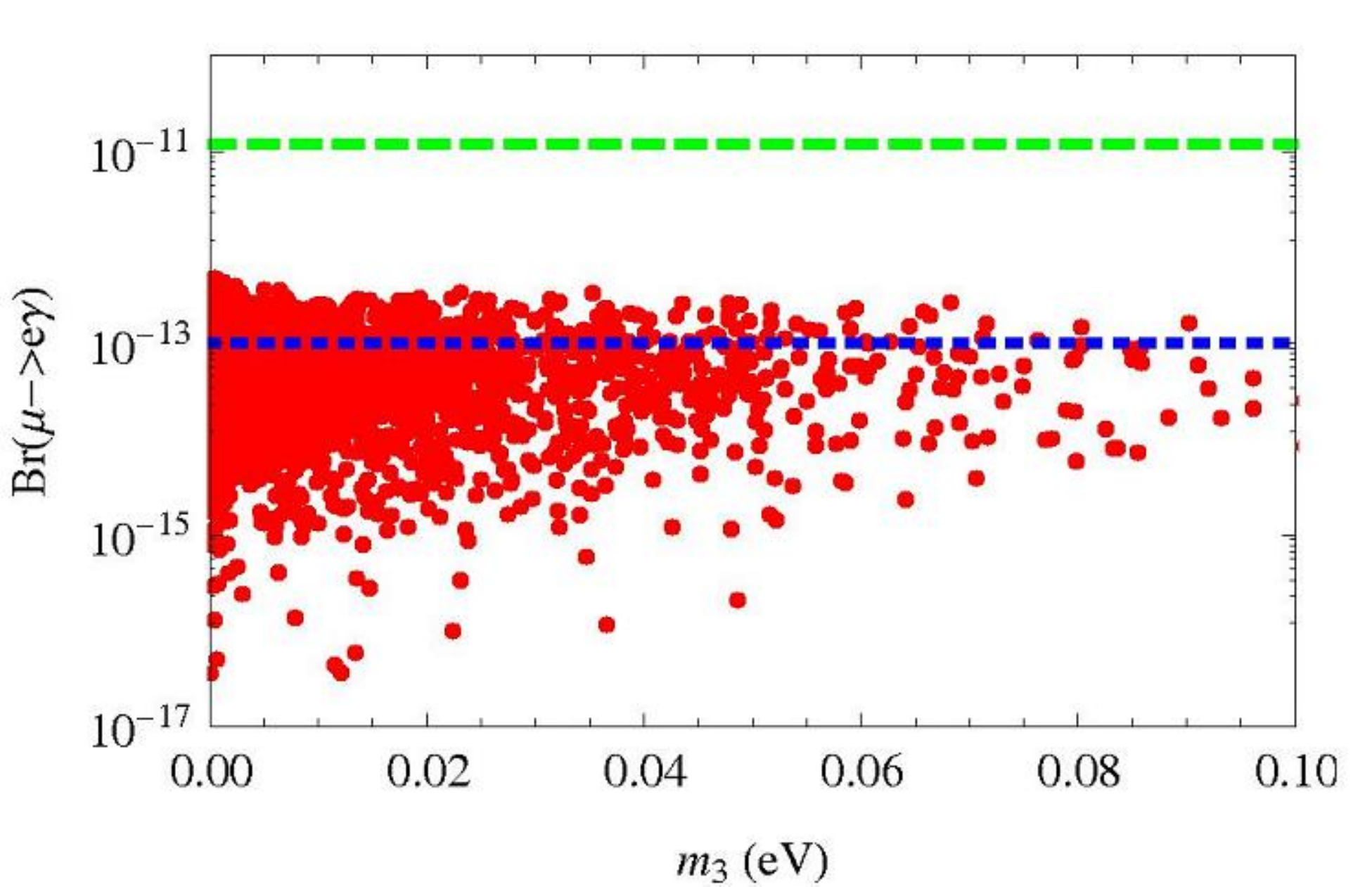}
\includegraphics[scale=1,width=3.75cm]{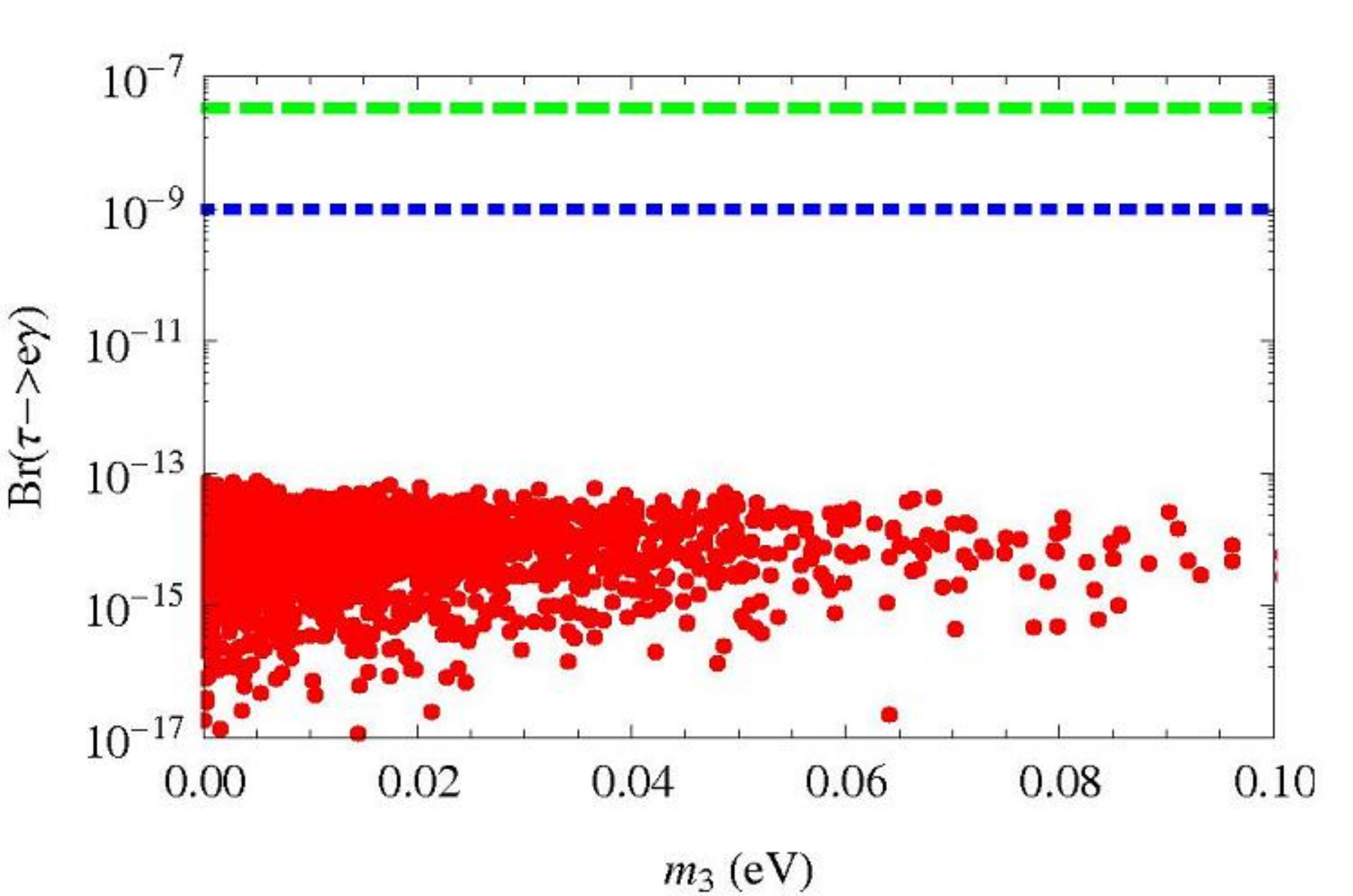}
\includegraphics[scale=1,width=3.75cm]{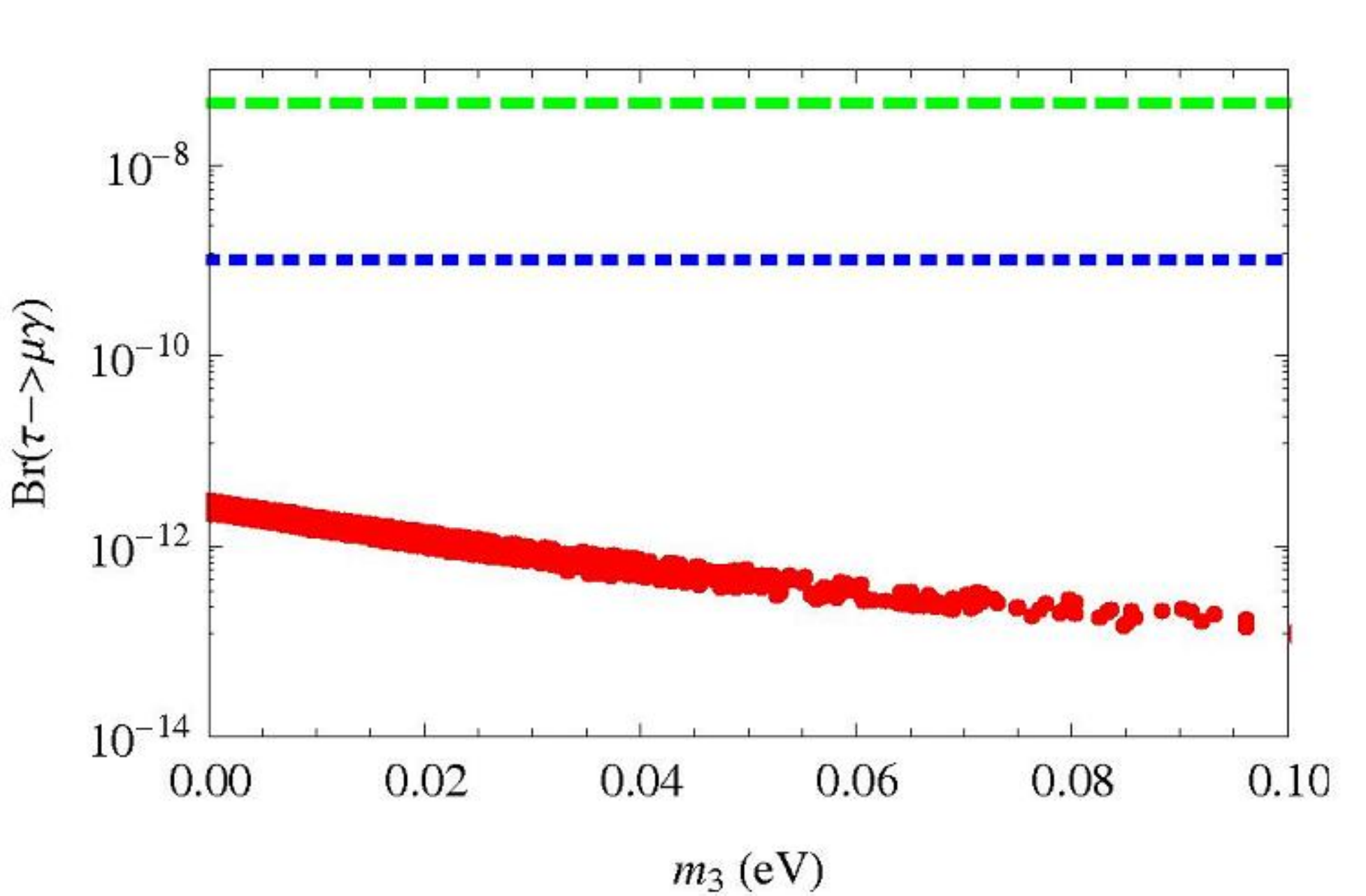}
\includegraphics[scale=1,width=3.75cm]{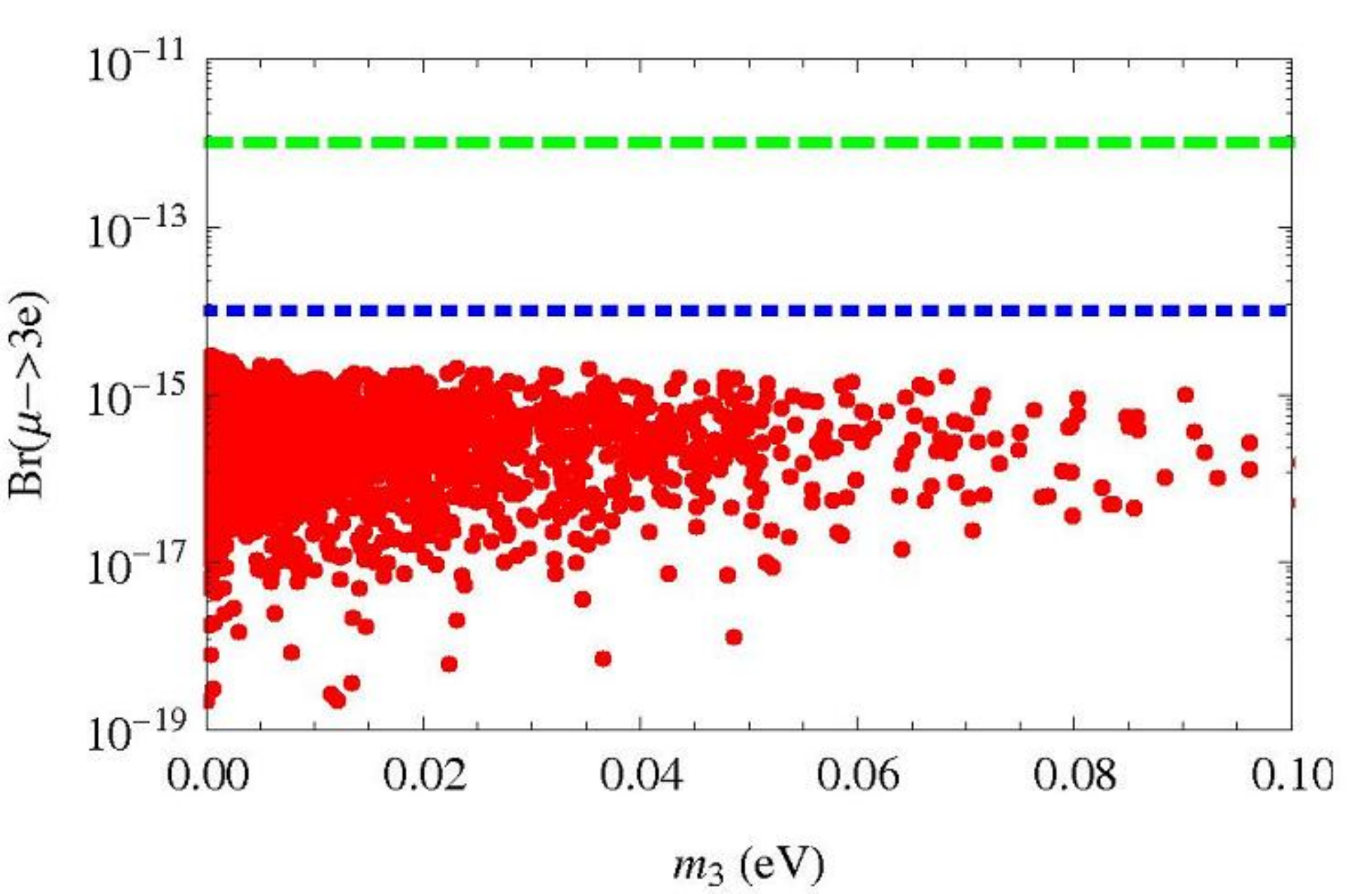}\\\\
\includegraphics[scale=1,width=3.75cm]{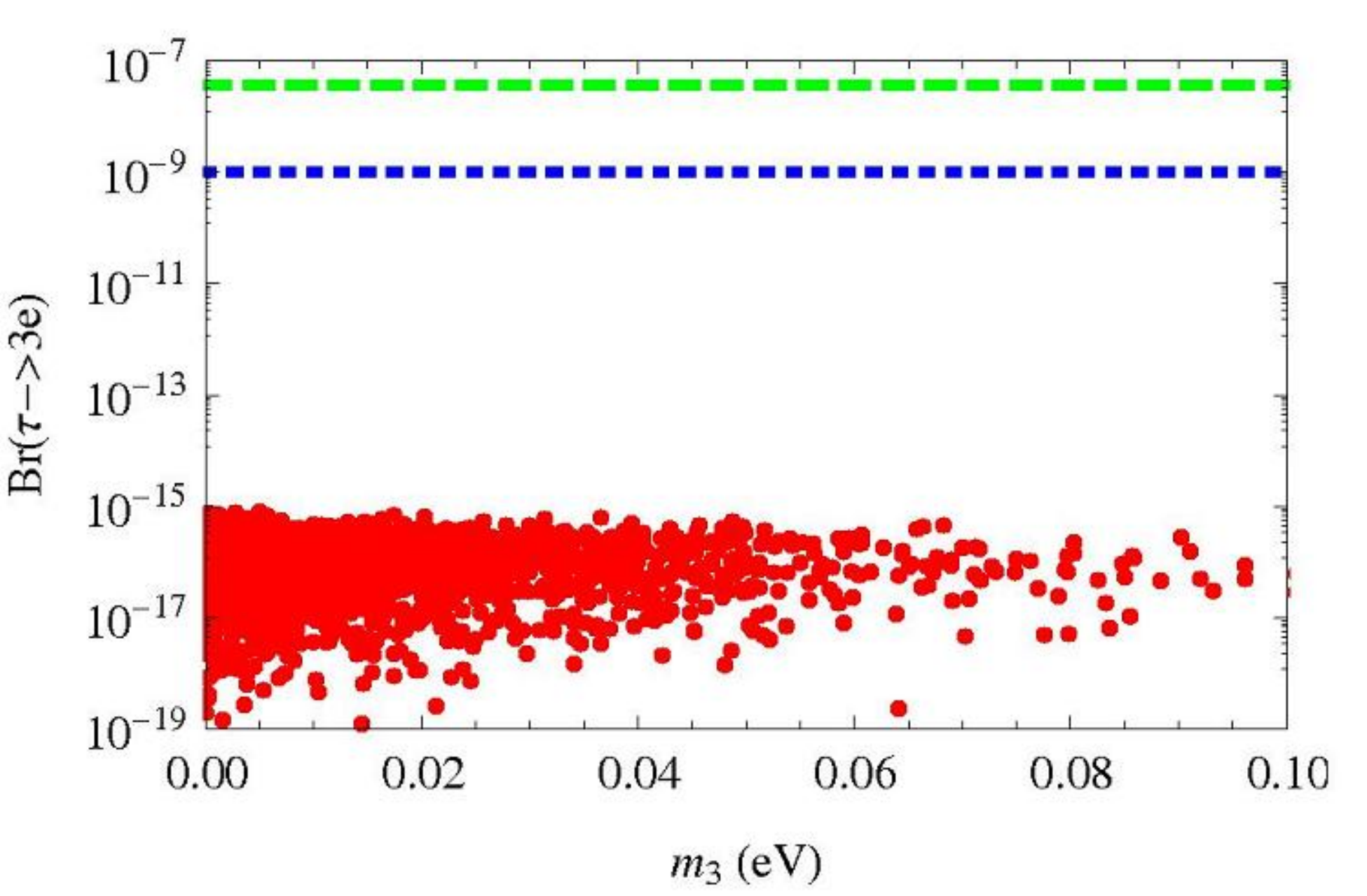}
\includegraphics[scale=1,width=3.75cm]{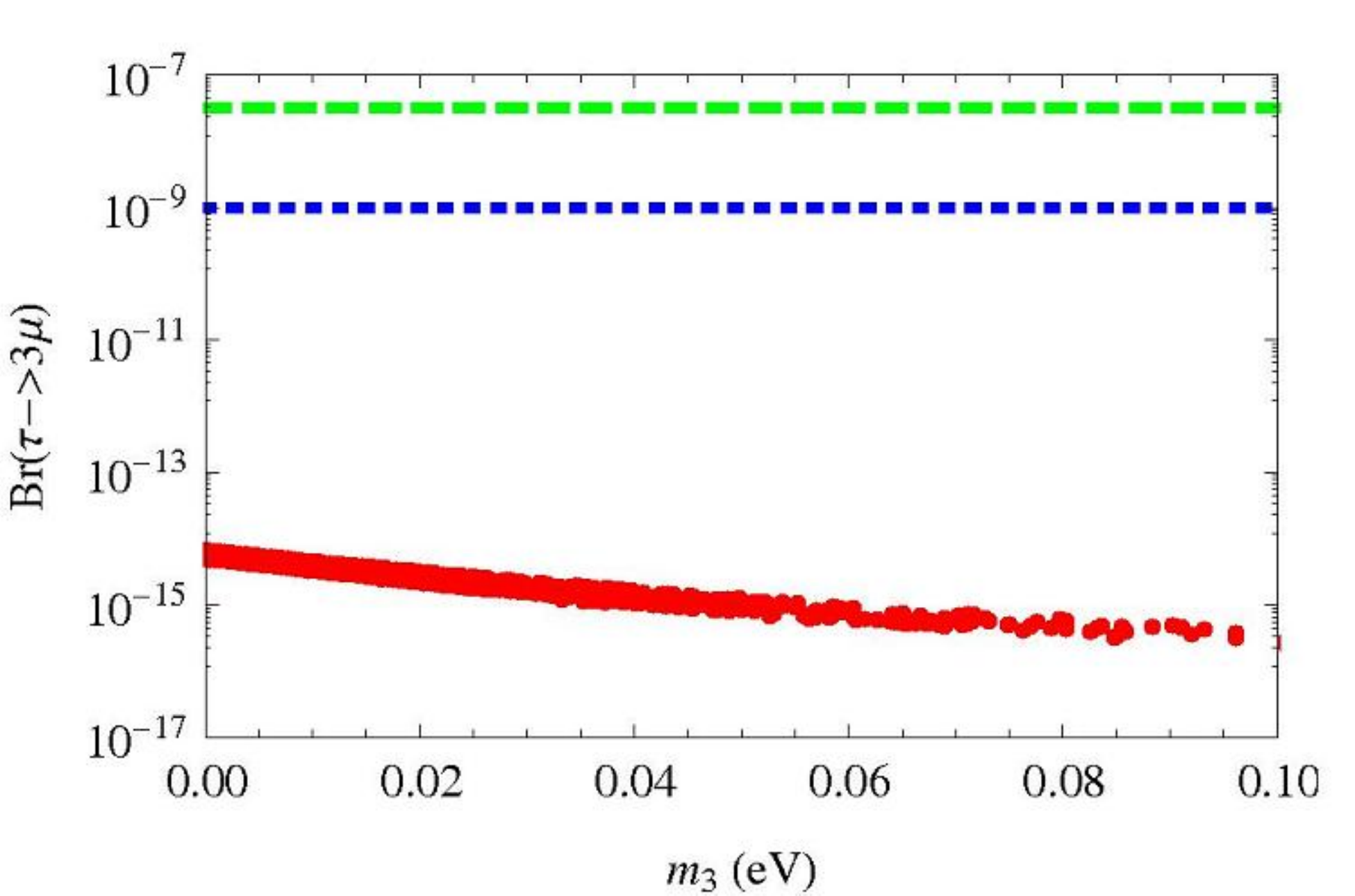}
\includegraphics[scale=1,width=3.75cm]{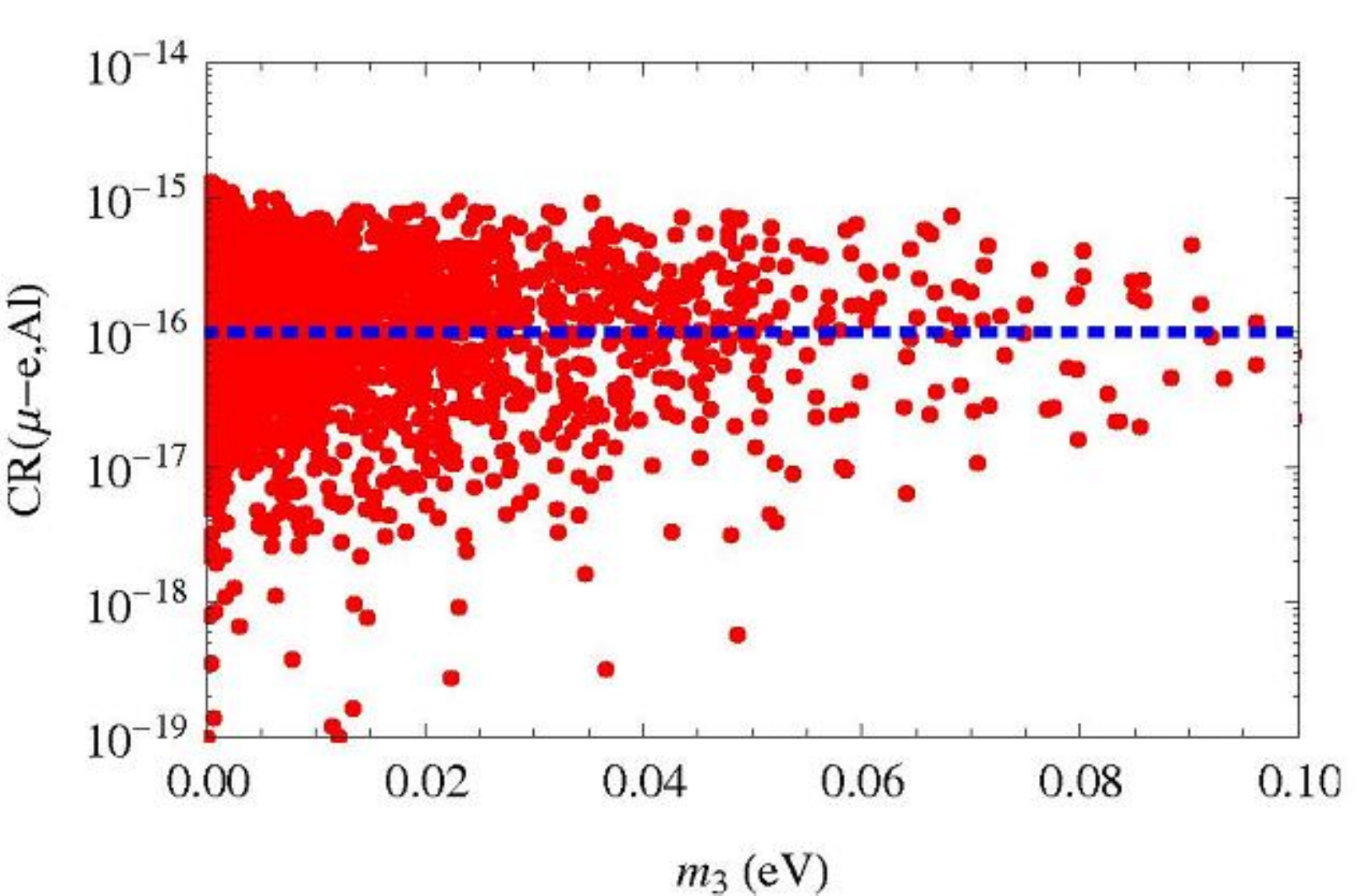}
\includegraphics[scale=1,width=3.75cm]{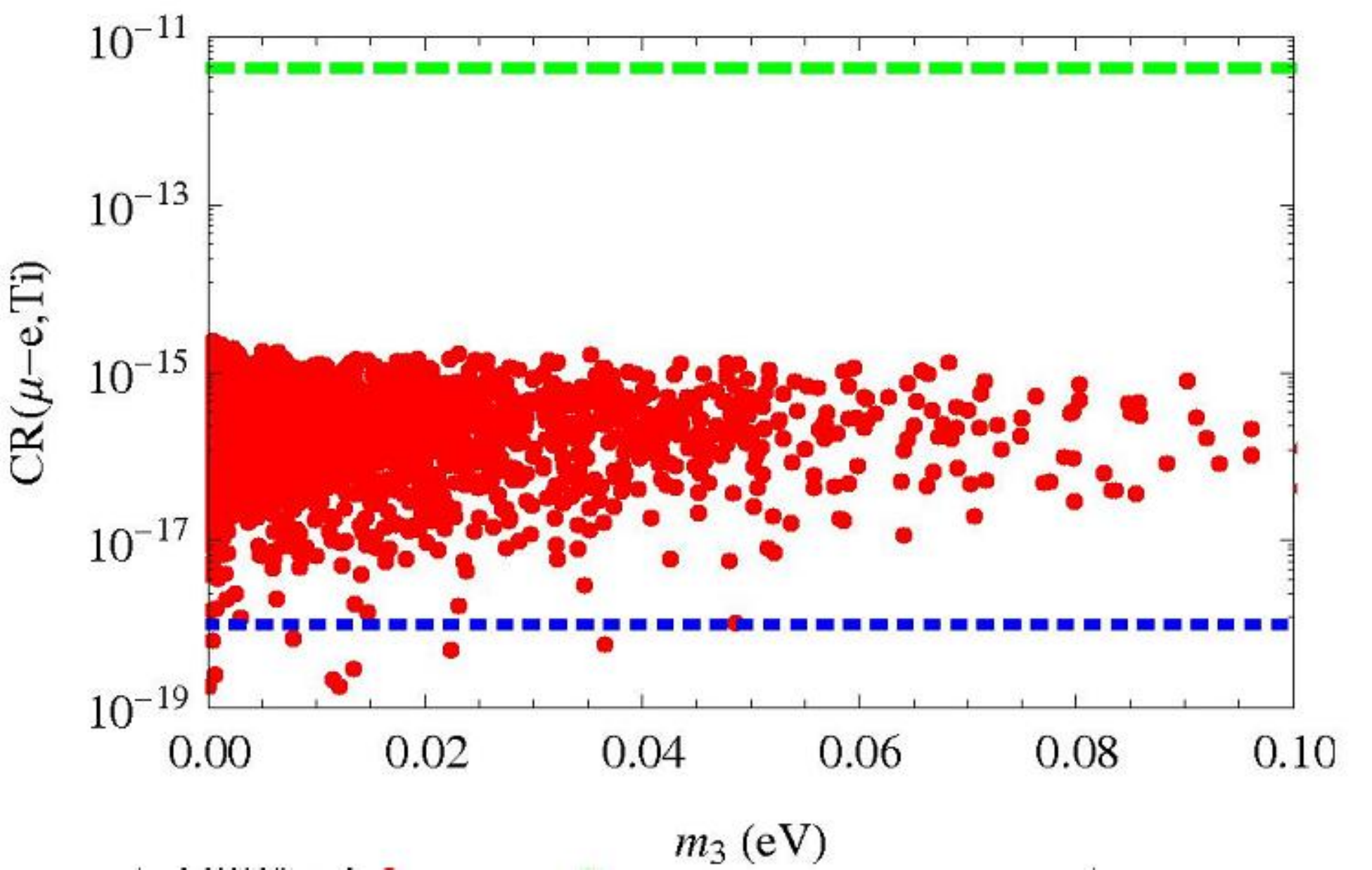}
\end{tabular}
\caption{\label{fig:Ding_IH_NLO} Scatter plot of
$Br(\ell_i\rightarrow\ell_j\gamma)$, $Br(\ell_i\rightarrow3\ell_j)$,
$CR(\mu-e,Al)$ and $CR(\mu-e,Ti)$ against the lightest neutrino mass
$m_3$ in Ding's $S_4$ model for inverted hierarchy spectrum. The
dashed and dotted lines represent the present and future
experimental sensitivity respectively.}
\end{center}
\end{figure}

We see that the LFV branching ratios can vary within a larger region
than the LO results in Fig.\ref{fig:Ding_NH_LO} and
Fig.\ref{fig:Ding_IH_LO}, and the rates for $\mu e$ and $\tau e$
involved processes could be enhanced by a factor a 10-100 for the IH
case. However, $Br(\tau\rightarrow\mu\gamma)$ and
$Br(\tau\rightarrow3\mu)$ are only modified slightly by NLO
contributions. It is remarkable that in this model the $\mu-e$
conversion in Ti is completely within the precision of the next
generation experiment, the $\mu-e$ conversion in Al should be
observable in a large part of the parameter space, the radiative
decay $\mu\rightarrow e\gamma$ can only be observed in a marginal
part of the parameter space at near future experiment, and
$\tau\rightarrow e\gamma$, $\tau\rightarrow\mu\gamma$,
$\mu\rightarrow3e$, $\tau\rightarrow3e$ and $\tau\rightarrow3\mu$
are below the expected future sensitivities. This fact can be used
to distinguish this $S_4$ model from the AF model.


\section{Conclusions and discussions}

Recently some models with discrete flavor symmetry such as $A_4$ and
$S_4$ have been showed to be able to naturally produce TB mixing at
leading order. Since only few parameters are involved at LO in these
models, they are rather predictive. In this work, we have studied
the LFV $\mu\rightarrow e\gamma$, $\tau\rightarrow e\gamma$,
$\tau\rightarrow\mu\gamma$, $\mu\rightarrow 3e$,
$\tau\rightarrow3e$, $\tau\rightarrow 3\mu$ and $\mu-e$ conversion
in Al and Ti in the AF $A_4$ model \cite{Altarelli:2005yx} and the
$S_4$ model of Ding \cite{Ding:2009iy} within the framework of
mSUGRA.

At LO the branching ratio for LFV process is closely related to the
light neutrino mass. For inverted hierarchy neutrino mass spectrum,
$\mu\rightarrow e\gamma$, $\tau\rightarrow e\gamma$,
$\mu\rightarrow3e$, $\tau\rightarrow3e$ and $\mu-e$ conversion in Al
and Ti are highly suppressed at LO. After taking into account the
NLO contributions, the LFV branching ratios for NH are corrected
properly, and the allowed region becomes larger. Whereas the
branching ratios of the $\mu e$ and $\tau e$ involved LFV are
enhanced drastically in the case of IH spectrum, and
$Br(\tau\rightarrow\mu\gamma)$ and $Br(\tau\rightarrow3\mu)$ are
approximately the same as the LO predictions.

Our predictions for the rare LFV processes in the two models are
summarized in Table \ref{LFV_sum}. From detailed numerical analysis,
we find that for NH spectrum of AF model, $\mu\rightarrow e\gamma$,
$\mu\rightarrow3e$ and $\mu-e$ conversion in Al and Ti are within
the reach of next generation experiments (Fig. \ref{fig:AF_NH_NLO}).
While for IH spectrum $\mu-e$ conversion in Ti is above the expected
future sensitivity in all the parameter space considered,
$\mu\rightarrow e\gamma$ and $\mu-e$ conversion in Al could be
observed by near future experiments in a very significant proportion
of the parameter space, nevertheless the signal of
$\mu\rightarrow3e$ could be only detected in a marginal part of the
parameter space (Fig. \ref{fig:AF_IH_NLO}). Then we turn to the
Ding's $S_4$ model, for both NH and IH spectrum, we find that only
$\mu-e$ conversion in Ti is within the precision of the future
experiment, $\mu-e$ conversion in Al should be observed in a large
part of the parameter space, while $\mu\rightarrow e\gamma$
observable in a marginal part of the parameter space (Fig.
\ref{fig:Ding_NH_NLO} and Fig. \ref{fig:Ding_IH_NLO}). In both AF
model and the $S_4$ model of Ding, $\tau\rightarrow e\gamma$,
$\tau\rightarrow\mu\gamma$, $\tau\rightarrow3e$ and
$\tau\rightarrow3\mu$ are below the sensitivity of next generation
experiments. We conclude that $\mu-e$ conversion in Ti is a
particularly robust test to the AF model and the Ding's $S_4$ model.
If it is really observed in future, it would be a great support to
these discrete flavor models.

\begin{table}
\begin{center}
\begin{tabular}{|c|cc||cc|}\hline\hline
\multicolumn{1}{|c}{}&\multicolumn{2}{|c||}{AF}&\multicolumn{2}{c|}{$S_4$}\\\cline{2-5}
 & NH&IH&NH&IH\\\hline
$\mu\rightarrow e\gamma$&$\surd$&$\checkmark$&?&?\\

$\tau\rightarrow e\gamma$&$\times$&$\times$&$\times$& $\times$\\

$\tau\rightarrow\mu\gamma$&$\times$&$\times$&$\times$& $\times$\\

$\mu\rightarrow3e$&$\surd$&?&$\times$& $\times$\\

$\tau\rightarrow 3e$&$\times$&$\times$&$\times$& $\times$\\

$\tau\rightarrow 3\mu$&$\times$&$\times$&$\times$& $\times$\\

$\mu-e$ {\rm
in\,Al}&$\surd$&$\checkmark$&$\checkmark$&$\checkmark$\\

$\mu-e$ {\rm in\,Ti}&$\surd$&$\surd$&$\surd$&$\surd$\\\hline\hline
\end{tabular}
\caption{\label{LFV_sum}Summary of predictions for LFV processes in
AF model \cite{Altarelli:2005yx} and Ding's $S_4$ model
\cite{Ding:2009iy}. The symbol $\surd$ denotes the rare process is
above the sensitivity of next generation experiments for the whole
parameter space considered, $\checkmark$ represents the LFV process
should be observed in a very large part of parameter space in near
future, $?$ for the process only observable in a marginal part of
parameter space, and $\times$ denotes the process is below the
sensitivity of next generation experiments.}
\end{center}
\end{table}

We note that our consideration are not at all restricted to the AF
model and Ding 's $S_4$ model, but could apply to a much wider class
of theories. Models with discrete flavor symmetry can be strongly
constrained or be excluded by existing or future LFV bounds, a
combined analysis of LFV processes provide us a way to distinguish
different models.

{\it Note Added: near the completion of this work, papers that
address similar issues appear
\cite{Hagedorn:2009df,Feruglio:2009hu}. }

\section{Acknowledgments}
We are grateful to Prof. Mu-Lin Yan for stimulating discussion, and
we acknowledge Dr. Jia-Wei Zhao and Wei Li for their help on
numerical calculations. This work is supported by the National
Natural Science Foundation of China under Grant No.10905053, Chinese
Academy KJCX2-YW-N29 and the 973 project with Grant No.
2009CB825200. Jia-Feng Liu is supported in part by the National
Natural Science Foundation of China under Grant No.10775124.

\end{document}